\documentclass[journal]{IEEEtran}

\usepackage{amsmath,amsfonts}
\usepackage{algorithmic}
\usepackage{algorithm}
\usepackage{array}
\usepackage[caption=false,font=normalsize,labelfont=sf,textfont=sf]{subfig}
\usepackage{textcomp}
\usepackage{stfloats}
\usepackage{url}
\usepackage{verbatim}
\usepackage{graphicx}
\usepackage{cite}
\usepackage{mathrsfs}
\usepackage{threeparttable}
\usepackage{booktabs}
\usepackage{nicematrix}
\usepackage{multirow}
\usepackage{refcount}

\usepackage{xr}
\usepackage{threeparttable}
\usepackage{color}
\usepackage{amssymb}
\newtheorem{thm}{Theorem}

\newtheorem{lemma}{Lemma}
\newtheorem{claim}{Claim}
\newtheorem{defi}{Definition}
\newtheorem{prop}{Proposition}

\newcommand{\mS}{\mathcal S}

\newcommand{\mE}{\mathcal E}

\newcommand{\pri}{\text{pri}}

\begin{document}

\title{Alternating Subspace Method for Sparse Recovery of Signals}

\author{Xu Zhu, Yufei Ma, Xiaoguang Li, and Tiejun Li
        % <-this % stops a space
% <-this % stops a space
\thanks{X. Zhu, Y. Ma, and T. Li are with Laboratory of Mathematics and Applied Mathematics, School of Mathematical Sciences,  Center for Machine Learning Research, Peking University, Beijing 100871, P.R. China}
\thanks{Email: xuzhu@pku.edu.cn (X. Zhu),  1800010737@pku.edu.cn (Y. Ma), tieli@pku.edu.cn (T. Li)}
\thanks{X. Li is with Department of Mathematics, Hunan Normal University, P.R. China}
\thanks{Email: lixiaoguang@hunnu.edu.cn (X. Li)}
\thanks{Corresponding author: Xiaoguang Li, Tiejun Li}
\thanks{This work is a thorough revision of our preprint~\cite{ASAMP}, with revisions focused solely on enhancing clarity and rigor.}
}

% The paper headers
\markboth{Journal of \LaTeX\ Class Files,~Vol.~14, No.~8, August~2021}%
{Shell \MakeLowercase{\textit{et al.}}: A Sample Article Using IEEEtran.cls for IEEE Journals}

\IEEEpubid{%0000--0000/00\$00.00~\copyright~2021 IEEE
pubid\ \ \ \ \ \ \ }
% Remember, if you use this you must call \IEEEpubidadjcol in the second
% column for its text to clear the IEEEpubid mark.

\maketitle

\begin{abstract}

Numerous renowned algorithms for tackling the compressed sensing problem employ an alternating strategy, which typically involves data matching in one module and denoising in another. 
%Based on an in-depth analysis of the connection between the message passing and operator splitting, 
We present a novel approach, the Alternating Subspace Method (ASM), which integrates the principles of the greedy methods (e.g., the orthogonal matching pursuit type methods) and the splitting methods (e.g., the approximate message passing type methods). Crucially, ASM enhances the splitting method by achieving fidelity in a subspace-restricted fashion. \textcolor{black}{We reveal that such a restriction strategy guarantees global convergence via proximal residual control and establish its local geometric convergence on the LASSO problem.} Numerical experiments on the LASSO, channel estimation, and dynamic compressed sensing problems demonstrate its high convergence rate and its capacity to incorporate different prior distributions. Overall, the proposed method is promising in terms of efficiency, accuracy, and flexibility, and has the potential to be competitive in different sparse recovery applications.

\end{abstract}

\begin{IEEEkeywords}
Compressed sensing, alternating subspace method, operator splitting, LASSO problem, approximate message passing, channel estimation, dynamic compressed sensing
\end{IEEEkeywords}

\section{Introduction}\label{sec:intro}

\IEEEPARstart{T}{he} linear inverse problem aims to recover a signal $x^\dag\in\mathbb R^N$ from the following observation model
\begin{equation}
y=Ax^\dag+w, \label{problem:LinearInverse}
\end{equation}
where $A\in \mathbb{R}^{M\times N}$ is the measurement matrix, $y\in\mathbb R^M$ is the received signal and $w\sim \mathcal N(\boldsymbol 0, \sigma^2_w I)$ is Gaussian noise. When $M<N$ and $x^\dagger$ is known to be sparse, the problem becomes the well-known compressed sensing problem \cite{CS, CSTao}. Given that the system is underdetermined, the sparsity of $x^\dag$ plays a vital role in narrowing down the solution space. Conceptually, successful recovery involves two key components: identifying an active set $\mE$ (with $|\mE|\leq M$) that contains the ground truth support $\mE^\dagger$, and performing data fidelity within $\mE$ to estimate the coefficients. Although this ideal procedure cannot be accomplished in a single step, various iterative methods emulate this routine by alternating between support determination and subspace fidelity.

For instance, \emph{greedy methods} such as Orthogonal Matching Pursuit (OMP) \cite{OMP} implement this framework by iteratively expanding the support set. Specifically, at each iteration $k$, OMP selects new elements based on their correlation with the current residual to form $E^k$, then performs subspace fidelity within $E^k$ to obtain coefficient estimates. This process continues until the support size reaches a predefined number $L$, yielding a sparse solution. Variants including StOMP \cite{StOMP} and CoSaMP \cite{CoSaMP} follow similar expansion strategies.

To achieve sparsity without manually determining the support size, one can resort to optimization models such as basis pursuit denoising, which minimizes $\|x\|_1$ while enforcing data consistency through $\|y-Ax\|^2\leq \delta$. This formulation has the advantages of being convex and admitting an equivalent unconstrained version~\cite{BPDN}:
\begin{equation}
\text{(LASSO)}\quad \mathop{\min}_{x\in\mathbb R^N} \frac12\|y-Ax\|^2+\lambda\|x\|_1,
\quad \lambda>0
\label{problem:LASSO}
\end{equation}
which is the well-known LASSO problem. The composite structure of~\eqref{problem:LASSO} makes it amenable to solution by the Alternating Direction Method of Multipliers (ADMM) \cite{ADMM:Boyd}, which naturally decomposes into an alternation between data fidelity (the regularized least squares, RLS, step) and support determination (the soft-thresholding step).

Although composite optimization offers advantages over greedy methods (e.g., guaranteed convergence and the flexibility to leverage more prior information), greedy methods remain popular due to their significantly lower computational cost from explicit sparsity exploitation. Specifically, OMP performs data fidelity by solving least-squares within a low-dimensional subspace $E^k$, while ADMM executes RLS in the full $\mathbb{R}^N$ space despite sparsity introduced by the soft-thresholding step. Since the RLS step requires solving a linear system, its computational expense becomes a primary concern in large-scale problems. Given this shared alternating structure, a compelling question arises: can we integrate \textit{subspace fidelity} into the ADMM framework to enhance computational efficiency, while preserving its advantages? This constitutes the central question addressed in this paper.

\begin{figure}
\centering
\includegraphics[width=0.45\textwidth]{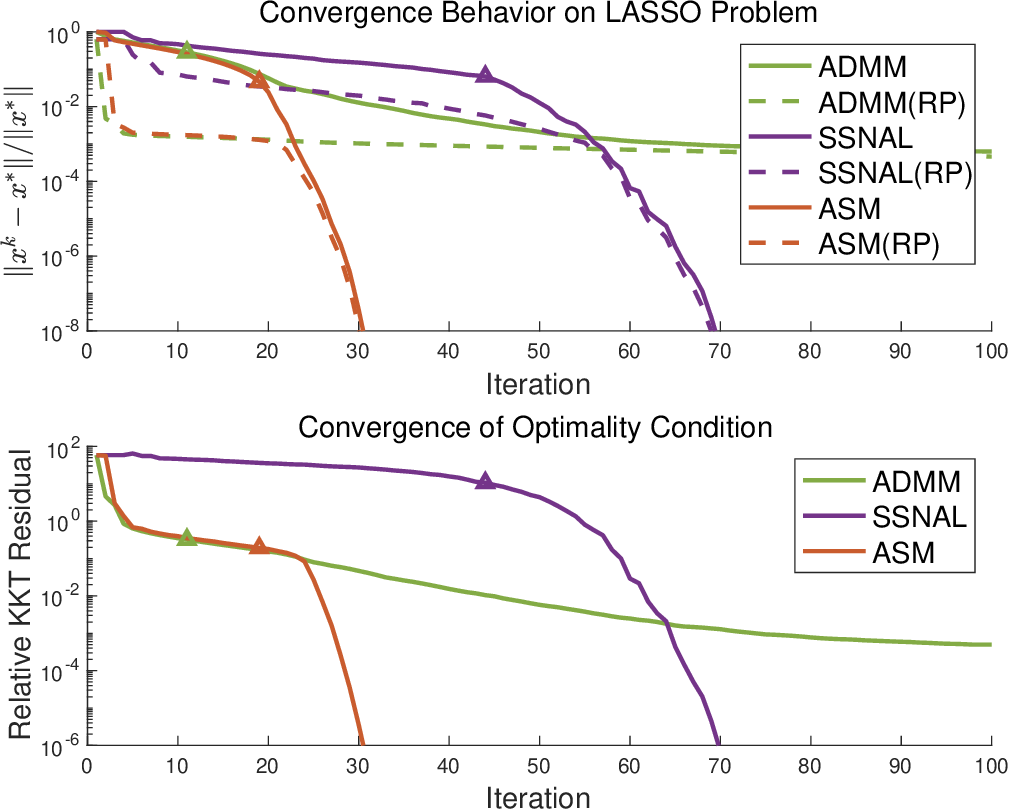}
\caption{Median convergence behaviors of various algorithms for the LASSO problem with $M=200$, $N=500$, and $\lambda=10^{-3}$, evaluated over $500$ independent realizations. The term ``RP'' (row projection) denotes the projection of the error $e^k=x^k - x^*$ onto the row space of $A$, which decays rapidly while the null space component  is actually the dominant part of $e^k$. The triangle marks the transition point beyond which $|E^k| \leq M$; note the accelerated convergence of ASM near this transition.}\label{fig:BehaveCom}
\end{figure}

Surprisingly, numerical experiments demonstrate that subspace fidelity not only reduces the computational cost but also accelerates ADMM's convergence. This subspace variant, combined with careful update of the multiplier, is named by us as the Alternating Subspace Method (ASM) and evaluated in Fig.~\ref{fig:BehaveCom} for solving the LASSO problem~\eqref{problem:LASSO} (see experimental setup in Section~\ref{sec:4A}). The results show that ASM maintains the fast initial convergence of standard ADMM while sustaining acceleration for high-accuracy solutions, ultimately achieving convergence behavior comparable to the asymptotically superlinear Semi-Smooth Newton Augmented Lagrangian (SSNAL) method \cite{SSNAL}.

\textcolor{black}{
Since ASM retains the RLS structure of ADMM, it remains fully applicable even when $|E^k| > M$, without reverting to a full-space fallback. Although the acceleration is reduced in this regime, operating strictly within the restricted subspace still substantially cuts the computational overhead.
}

\textcolor{black}{
Theoretically, establishing the global convergence of ASM requires resolving two critical challenges inherent to subspace methods. First, a naive restriction may lead to the permanent exclusion of certain indices that are supposed to be maintained. Second, introducing subspace fidelity risks compromising ADMM's well-established convergence properties. We resolve these issues by revealing the intrinsic proximal gradient structure within ADMM and splitting the update into two formally independent \emph{operator splitting}~\cite{LSCOA} schemes. Because ASM preserves the full-space proximal gradient evaluation, we show that restricting the data fidelity subproblem to the chosen subspace will not break global convergence, provided that the averaging operation is appropriately applied to control the proximal residual.
}

The detailed iteration of ASM is summarized in Algorithm~\ref{algo:ASframework}. A key component is the \textit{denoising module}, so named for its capacity to incorporate general denoisers beyond the soft-thresholding operator. This module can determine the support set using richer prior information than mere sparsity. This flexibility is inherited from the plug-and-play (PnP) framework prevalent in ADMM-like methods \cite{PnP1, PnP2, PnP3}, where denoisers can be substituted in a plug-and-play manner or even modeled as learnable neural networks \cite{ADMM-NET1, ADMM-NET2}. Furthermore, from a Bayesian perspective, replacing the Laplace prior (associated with the LASSO's MAP estimate) with more sophisticated priors (e.g., modeling variable correlations via a hidden Markov chain \cite{STCS}) allows the module to pursue minimum mean-squared error (MMSE)-type estimates. We will demonstrate this generalizability in Sec.~\ref{sec:ASM_Structured}.

In summary, ASM, as a subspace fidelity enhancement of ADMM, offers the following advantages for sparse linear inverse problems: 1) \textbf{Cost reduction}: While ADMM-like methods may become prohibitive when solving a large-scale regularized least squares subproblem, ASM leverages sparsity and restricts the subproblem to a low-dimensional subspace. \textcolor{black}{Besides, we demonstrate that low-rank update techniques can be integrated to further enhance computational efficiency.} 2) \textbf{Fast convergence}: Sharing the rapid initial convergence of ADMM-like methods, ASM additionally avoids their eventual decelaration, leading to fast convergence even to high-accuracy solutions. In fact, for the LASSO problem, ASM demonstrates competitive performance with SSNAL, which is proven to be asymptotically superlinear. 3) \textbf{Flexibility to incorporate different priors}: The framework of ASM facilitates integration with structured denoisers. This capability, which it shares with ADMM-like methods, is challenging for SSNAL.

We remark that the ASM idea has already inspired the alternating estimation framework in~\cite{AE-SC-VBI}, which demonstrates its strong generalizability.

The remainder of this paper is organized as follows. In Sec.~\ref{sec:derivating_asm_from_admm_on_LASSO}, we derive ASM by introducing subspace fidelity to ADMM. Sec.~\ref{sec:Sub4Accelerate} \textcolor{black}{illustrates the asymptotic acceleration results via subspace restriction. }
%gives an intuitive explanation of the acceleration mechanism via subspace restriction. 
\textcolor{black}{In Sec.~\ref{sec:averaging_and_the_proximal_residual_control}, we establish the global convergence of ASM for the LASSO problem. 
%by investigating the proximal residual descent and illustrating the impact of the averaging operation. 
Practical strategies, including low-rank updates and parameter tuning, are discussed in Sec.~\ref{sec:practical_strategies_for_asm},}
and the integration with general denoisers is presented in Sec.~\ref{sec:ASM_Structured}. Numerical experiments on LASSO, channel estimation, and dynamic compressed sensing are provided in Sec.~\ref{sec:experiment}. %Finally, Sec.~\ref{sec:convergence} proves the local convergence of ASM-L1 for the LASSO problem and analyzes its convergence rate.

\begin{algorithm}[t]
\caption{Alternating Subspace Method (ASM)} 
{\bf Input:} 
Initialize $x^0_{\text{ave}}=\boldsymbol0$ and set up   
an updating strategy for $\hat v^{k}, v^{k}$ (cf. Sec.~\ref{sec:ave_strategies}).\\
 {\bf Output:} 
$x^{\text{Maxit}}$
\begin{algorithmic}[1]\label{algo:ASframework}
\FOR{$k=1,2,\ldots,\text{Maxit}$}
    \STATE \% {\it  Denoising}\\
    \STATE $\mu^{k}=x^{k-1}_{\text{ave}}+v^{k}A^T(y-Ax^{k-1}_{\text{ave}})$\label{line:GD}\\
    \STATE $z^{k}=\mathcal D(\mu^{k}, v^{k};p^\pri)$\\
    \STATE Figure out $E^{k}$ with respect to $z^{k}$\\
    \STATE $\hat z^{k}=z^{k}|_{E^{k}}$; $\hat\mu^{k}=\mu^{k}|_{E^{k}}$ \% {\it Restriction}
    \STATE \% {\it  Subspace Fidelity}\\
    \STATE $\hat \nu^{k}=\hat z^{k}-(\hat v^{k}/v^{k})(\hat\mu^{k}-\hat z^{k})$
    \STATE $\hat x^{k+1}=\mathop{\arg\min}_{\hat u}\frac{1}{\sigma^2_w}\|y-\hat A^{k}\hat u\|^2+\frac1{
    \hat v^k}\|\hat u-\hat\nu^k\|^2$\\
    \STATE $x^{k}|_{E^k}=\hat x^{k}$; $x^{k}|_{(E^k)^c}=\boldsymbol0$ \% {\it Extension}
    \STATE \% {\it Averaging}\\
    \STATE $x^{k}_{\text{ave}}= d\cdot x^{k}+(1-d)\cdot x^{k-1}_{\text{ave}}$
\ENDFOR
\end{algorithmic}
\end{algorithm}

\section{Notations}

The following notational conventions are adopted throughout this paper. We denote $\{1,\ldots,N\}$ as $[N]$ and the $\ell_2$ norm $\|\cdot\|_2$ as $\|\cdot\|$. If $\mathcal{C}$ is a subset of $[N]$ and $x\in\mathbb R^N$, then $x_{\mathcal{C}}$ or $x|_{\mathcal{C}}$ denotes the restriction of $x$ to $\mathcal{C}$. For a matrix $A$, we use $\text{Null}(A)$, $\text{Row}(A)$ to denote the null and row space of $A$, respectively. We define $\delta_{\mathrm B}$ as $1$ if the Boolean expression $\mathrm{B}$ is true and $0$ otherwise. $\delta_{ij}$ is the shorthand for $\delta_{\{i=j\}}$, and $I_k\in\mathbb R^{k\times k}$ is the identity matrix.
%and $\mathcal{I}(\cdot)$ to denote the identity mapping. 
We use $\odot$ to denote the entry-wise multiplication and $\partial(\cdot)$ to denote the sub-gradient. The symbol $\mathrm{diag}(\boldsymbol a)$ means a diagonal matrix with diagonal entries corresponding to $\{a_i\}_{i=1}^N$. $\mathrm{eig}_{\max}(A)$ and $\mathrm{eig}_{\min}(A)$ denote the maximum and the minimum eigenvalue, respectively. \textcolor{black}{For any sequence $u^k$ generated during the iterations, we use $\Delta u^k := u^{k+1} - u^k$ to denote its successive difference}.

In general, we use italic font letters to represent vectors or matrices (e.g., $x$ or $A$), though in some cases, vectors may appear in boldface like $\boldsymbol v=(v_i)_i$ for clarity. %Considering the frequent subspace restriction in this paper, 
We adopt the convention that the symbol $\hat{x}$ means the restriction of vectors or matrices to the subspace/subcolumns (e.g., $\hat{x}^k:=x|_{E^k}$, $\hat{A}^k:=A|_{E^k}$, etc.). 
%which corresponds to the essentially non-zero components of $x$. 
However, this convention does not apply to scalar parameter $\hat{v}^{k}$, in which the hat symbol is used only for notational consistency. 
%We denote by $\mathcal{P}$ the projection operator. In particular, 
We use $\mathcal{P}_{\text{Row}(A)}(x)$ and $\mathcal{P}_{\text{Null}(A)}(x)$ to denote the projection of $x$ onto $\text{Row}(A)$ and $\text{Null}(A)$, respectively, and $\mathcal{P}_{[a,b]}(x)$ to denote the projection of each component of $x$ onto the closest point in the interval $[a,b]$.

\section{Deriving ASM from ADMM on LASSO}\label{sec:derivating_asm_from_admm_on_LASSO}

\subsection{Preliminaries of ADMM on Composite Optimization}

The composite structure in the LASSO problem naturally calls for a solver that optimizes in an alternating manner, such as ADMM. To apply this, we define an additional variable $z$ and reformulate the problem into the following version:
\begin{equation}
    \min_{x,z\in\mathbb R^N}\ \ h(x)+\lambda g(z),\ \ \mathrm{s.t.}\ x=z,\label{problem:ADMM}
\end{equation}
where $h(x):=\frac12\|y-Ax\|^2$ and $g(x):=\|x\|_1$. 
The standard ADMM procedure begins by writing its augmented Lagrangian function $L(x,z)$ and then minimizing over $x$ and $z$ alternately, where $L(x,z)$ is defined as
\begin{equation}\label{eq:Lxz}
    L(x,z):=h(x)+\lambda g(z)+s^T(x-z)+\frac1{2v}\|x-z\|^2
\end{equation}
\textcolor{black}{for a fixed $v>0$}. If we define $\mathcal L_{v}$ and $\mathcal S_{\lambda v}$ as follows:
\begin{align}
    \mathcal L_{v}(u)&:=\mathop{\arg\min}_x \left\{h(x)+\frac1{2v}\|x-u\|^2\right\},\label{eq:prox_h}\\
    \mathcal S_{\lambda v}(u)&:=\mathop{\arg\min}_x \left\{\lambda g(x)+\frac1{2v}\|x-u\|^2\right\},\label{eq:prox_g}
\end{align}
then the ADMM iterations for the primal variables $x$, $z$ and the dual variable (multiplier) $s$ are:
\begin{equation}
\begin{aligned}
    x^{k+1}&=\mathcal L_{v}(z^k-vs^k),\\
    z^{k+1}&=\mathcal S_{\lambda v}(x^{k+1}+vs^k),\\
    s^{k+1}&=s^k+(x^{k+1}-z^{k+1})/v.
\end{aligned}\label{eq:DRS_ADMM}
\end{equation}

The operators $\mathcal L_{v}$ and $\mathcal S_{\lambda v}$ are the so-called proximal operators with respect to $h$ and $g$, respectively \cite{FISTA,ADMM:Boyd,LSCOA}. When $\mathcal L_{v}$ and $\mathcal S_{\lambda v}$ have closed-form expressions, which is the case for $h$ and $g$ defined in~\eqref{problem:ADMM}, the iteration described above typically exhibits excellent numerical stability. However, even if $\mathcal L_v$ has an explicit expression, it may still involve significant computational cost. When $h$ represents the data fidelity term in~\eqref{problem:ADMM}, which is common in modern data science, we have
\begin{equation}
    \mathcal L_{v}(u)=\left(I_N+vA^TA \right)^{-1}\left(u+vA^Ty \right).\label{eq:V_inverse}
\end{equation}
When dealing with large-scale problems or under strict computational constraints, repeatedly applying this operator may become prohibitive.

Fortunately, when we focus on sparse regularization with $g(x)=\|x\|_1$, 
we find that $\mathcal S_{\lambda v}$ satisfies
\begin{equation}
    \mathcal S_{\lambda v}(u)=\mathrm{sgn}(u)\odot\max(|u|-\lambda v,0).\label{eq:STvOP}
\end{equation}
% \textcolor{black}{We also remark that $z=\mathcal S_{\lambda v}(u)$ is equivalent to $z=u-\lambda v p(u)$ where $p(u):=\mathcal{P}_{[-1, 1]}(u/(\lambda v))$.} 
\textcolor{black}{We also remark that $z=\mathcal S_{\lambda v}(x-v\nabla h(x))$ is equivalent to $z=x-v\nabla h(x)-\lambda v p(x)$ where $p(x):=\mathcal{P}_{[-1, 1]}[(x-v\nabla h(x))/(\lambda v)]$.}
As established in~\cite{2008-Elaine}, a key advantage of $\mathcal S_{\lambda v}$ is its capability to yield a support set $E^k$ covering or almost covering the optimal $\mE^*$ after a moderate number of iterations. This selector capability, also leveraged by methods like FPC\_AS~\cite{FPC_AS}, motivates a natural question: \textit{Can this subspace restriction be introduced into ADMM to reduce fidelity cost, without impairing its fast initial convergence?}

\subsection{Introducing Subspace Fidelity to ADMM}\label{sec:admm_sub_fidelity}

We now incorporate this subspace restriction into the ADMM iteration. The core idea leverages the fact that $\mathcal S_{\lambda v}$ produces a sparse support set $E^k$, which can be exploited to fix $x^{k+1}|_{(E^k)^c} = 0$, thus bypassing the computational cost of processing these components through $\mathcal L_v$. Within the subspace $E^k$, we solve the reduced subproblem defined by $\hat{\mathcal L}_v^k$:
\begin{equation}
    \hat{\mathcal L}_v^k(\hat u) := \mathop{\arg\min}_{\hat x\in\mathbb R^{|E^k|}} \left\{ \frac{1}{2} \| y - \hat A^k \hat x \|^2 + \frac{1}{2v} \|\hat x - \hat u\|^2 \right\},\label{eq:mL_def}
\end{equation}
where the columns of $\hat A^k$ and the elements of $\hat{u}^k$ and $\hat{x}^k$ are restricted to the subspace $E^k$.

Although this would appear to be a natural modification, one soon realizes that once $x_i^k = z_i^k = 0$, the multiplier $s_i^k$ ceases to update, and $x_i$ becomes permanently fixed at $0$, regardless of the optimal value $x^*_i$. This failure stems from an inappropriate restriction of the multiplier. To demonstrate this, we consider the gradient of the objective in~\eqref{eq:prox_h}. From the first-order optimality condition, the following relationship holds for $u = z^k - v s^k$:
\begin{equation}
    v\nabla h(x^{k+1})+x^{k+1}-z^k+vs^k=0.\label{eq:1st_OPT_L}
\end{equation}
Substitute this into \eqref{eq:DRS_ADMM}, we get 
\begin{equation}
    z^{k+1}=\mathcal S_{\lambda v}(z^k-v\nabla h(x^{k+1})).\label{eq:new_z_update}
\end{equation}
In this form, enforcing $x^{k+1}|_{(E^k)^c} = 0$ no longer implies a permanent restriction, as the gradient $\nabla h(x^{k+1})$ has global support and can transform all coordinates.

Obviously, without the subspace restriction, updating $z^k$ according to~\eqref{eq:DRS_ADMM} or~\eqref{eq:new_z_update} is equivalent, and standard ADMM employs the former for its computational efficiency due to being gradient-free. However, this simplification also conceals the proximal gradient structure inherent in~\eqref{eq:new_z_update}, which can be seen by performing a change of variable ${x}_\text{ave}^{k+1} := (x^{k+1} + x^k_\text{ave})/2$ and verifying that $z^k - v\nabla h(x^{k+1}) = {x}_\text{ave}^{k+1} - v\nabla h({x}_\text{ave}^{k+1})$. \textcolor{black}{This equivalence stems from the connection between ADMM and Douglas-Rachford Splitting (DRS) for composite optimization\cite{ADMM:Boyd}, where strict equality holds provided that certain initialization consistency conditions are met (e.g., when $z^1=vA^Ty$ and $s^1=x^1_{\mathrm{ave}}=0$; see Supplementary Material (SM)~\ref{sec:equivalent_expressions_of_the_admm_iteration} for details), while it presents a new scheme in general case.}
This structure itself can serve as a solver for the LASSO problem and is well-known as ISTA~\cite{ISTA1}.

Recognizing this intrinsic proximal gradient structure, we argue that ADMM can be interpreted as an accelerated ISTA variant. Our key insight is that ISTA with subspace-restricted data fidelity (ASM) can outperform full-space ADMM in both computational efficiency and convergence rate. We will elaborate on this claim in Sec.~\ref{sec:Sub4Accelerate}.

\subsection{Modifications for Consistency}\label{sub:modifications_for_consistency}

Having revealed the proximal gradient structure in ADMM, we now address the consistency maintenance when adopting subspace fidelity. If we formally assume that $g$ is differentiable, we can derive an equation analogous to~\eqref{eq:1st_OPT_L} for $\nabla g$. Specifically, applying the first-order optimality condition to~\eqref{eq:prox_g} with $u = x^{k} + v s^{k-1}$, we obtain $v\lambda\nabla g(z^{k})+z^{k}-x^{k}-vs^{k-1}=0$.
Applying this to $s^{k+1}$ in~\eqref{eq:DRS_ADMM}, we obtain:
\begin{equation}
    v s^{k} = v s^{k-1} + x^{k} - z^{k} = \textcolor{black}{v \lambda \nabla g(z^{k})}.\label{eq:mul_equal}
\end{equation}
This relation allows us to reformulate the ADMM iteration in~\eqref{eq:DRS_ADMM} without explicit reference to the multiplier:
\begin{equation}\hspace*{-1cm}
\begin{aligned}
 &   z^{k}=\mathcal S_{\lambda v}(x_\text{ave}^{k}-v\nabla h(x_\text{ave}^{k})),\\
\text{(ADMM)}\qquad\quad  &  x^{k+1}=\mathcal L_{v}(z^{k}-v\lambda\nabla g(z^{k})),\\
   &  x_\text{ave}^{k+1}=x^{k+1}/2+x_\text{ave}^k/2.
\end{aligned}\label{eq:cplx_L}
\end{equation}
%Despite its formal derivation and the omission of the averaging term ${x}_\text{ave}^{k+1}$, 
This formulation reveals a symmetric structure that alternates between applying the proximal gradient operators for $g$ and $h$.

Based on this derivation, we now claim that if we restrict the $x^{k+1}$ update in~\eqref{eq:cplx_L} to the subspace $E^k$, the iteration becomes rigorous, and exactly corresponds Algorithm~\ref{algo:ASframework} except that $\mathcal D$ is substituted by $\mathcal S_{\lambda v^k}$. Specifically, when $g(x) = \|x\|_1$, %it is non-differentiable, and 
we must instead consider the subgradient $\partial g$ defined as:
\begin{equation*}
    \partial g(x) = \mathrm{sign}(x),\ \text{if } x \neq 0; \quad 
    \partial g(x) = [-1, 1],\ \text{if } x = 0.
\end{equation*}
However, when restricted to the subspace $E^k$, where $\hat{z}^k := z^k|_{E^k}$ is non-zero by definition, we have $\partial g(z^k)|_{E^k} = \nabla g(z^k)|_{E^k} := \partial g(\hat{z}^k)$ exactly.

As a summary of this section, we present the condensed ASM-L1 algorithm, omitting intermediate variables and adjustable parameters for clarity:
\begin{equation}
\begin{aligned}
    &z^{k}=\mathcal S_{\lambda v}(x_\text{ave}^{k}-v\nabla h(x_\text{ave}^{k})),\\
\text{(ASM-L1)}\quad\qquad    &{x}^{k+1}|_{E^k}\leftarrow\hat{\mathcal L}^k_{v}(\hat{z}^{k}-v\lambda\partial g(\hat{z}^k)),\\
    &x_\text{ave}^{k+1}=x^{k+1}/2+x_\text{ave}^{k}/2,
\end{aligned}\label{eq:core_ASM}
\end{equation}
where the notation `$\leftarrow$' indicates updating $z^k$ restricted to $E^k$ to obtain $x^{k+1}$, and $\hat{\mathcal{L}}^k_v$ is defined in~\eqref{eq:mL_def}. 
This formulation encapsulates the \textit{core ASM philosophy}:  (i) subspace data fidelity for acceleration, (ii) averaging for stability, and (iii) full-space proximal gradient for consistency.

\section{Asymptotic Acceleration of ASM}\label{sec:Sub4Accelerate}

In this section, we will provide intuitive and theoretical explanations of why the subspace fidelity accelerates convergence.  First, we compare the convergence behavior of ASM and ADMM under a moderately conditioned scenario, demonstrating how subspace restriction fundamentally alters the convergence behavior. Then we discuss the intuitive and theoretical explanations of ASM acceleration.
%then provide an intuitive interpretation of this phenomenon. 
\textcolor{black}{Practical implementation strategies, including low-rank updates and parameter selection, are discussed in Sec.~\ref{sec:practical_strategies_for_asm}.} 

\textcolor{black}{
For simplicity, we focus on the real-domain LASSO. To analyze the asymptotic behavior of the subspace, we define
\begin{equation}\label{eq:mu(x)}
\mu(x) := x - v \nabla h(x),\quad p(x) := \mathcal P_{[-1,1]}[\mu(x)/(v\lambda)]
\end{equation}
and $\mu^k:=\mu(x_\text{ave}^k),\ p^k:=p(x_\text{ave}^k)$. As $v$ remains constant in this section and Sec.~\ref{sec:averaging_and_the_proximal_residual_control}, the dependence of $\mu(\cdot)$ on $v$ is omitted for brevity. Besides, we introduce the following definition to characterize a class of well-behaved LASSO problems (which are shown to be generic in \cite{tibshirani2012LASSO}):
\begin{defi}[Generic Conditions]\label{defi:genetic}
    The LASSO problem \eqref{problem:LASSO} is said to satisfy the \textbf{generic conditions} if: 1) its optimal solution $x^*$ and support set $\mE^*$ are unique, and 2) the strong complementarity condition holds, i.e., $\omega_0>0$ where $\omega_0:=\min_{i\in(\mE^*)^c}\{1-|\mu_i(x^*)|/(v\lambda)\}$. Any $M$ columns of $A$ are linearly independent; furthermore, for any index set $S$ with $|S|>M$ and $e \in \{1, -1\}^{|S|}$, $e \notin \mathrm{Row}(A_S)$. 
\end{defi}
}

\subsection{Behavior of the Subspace Fidelity Strategy}\label{sec:4A}

To demonstrate the core benefits of subspace fidelity, we first compare the performance of ASM (i.e., ASM-L1), ADMM, and SSNAL \cite{SSNAL} on LASSO problem~\eqref{problem:LASSO}. We fix the parameter $v$ for ADMM and ASM to ensure they strictly adhere to the iterations defined in~\eqref{eq:cplx_L} and~\eqref{eq:core_ASM}, respectively. The elements of $A$ and $x^\dagger$ are drawn i.i.d. from Gaussian distributions: $A_{ij} \sim \mathcal{N}(0, 1/\sqrt{M})$ and $x_k^{\dagger} \sim \mathcal{N}(0, 1)$ for $k \in \mathcal{E}^\dagger$. We set $\lambda = 10^{-3}$ without introducing noise (noisy settings are presented in Sec.~\ref{sec:experiment}).

Using $v = 100$ for both ADMM and ASM, we perform 500 independent trials and plot the median error evolution in Fig.~\ref{fig:BehaveCom}. The relative KKT residual, derived from first-order optimality conditions, provides an accuracy metric independent of the optimal solution $x^*$ to the LASSO problem:
\begin{equation}
    \mathrm{Res}_{\mathrm{KKT}}(x^k) := \frac{\|x^k - \mathcal{S}_{\lambda}(x^k - \nabla h'(x^k))\|}{1 + \|x^k\| + \|y' - A' x^k\|},\label{eq:ResKKT}
\end{equation}
where $y' := y / \sqrt\lambda$, $A' := A / \sqrt\lambda$, and $h'(x) := \|y' - A' x\|^2 / 2$ are normalized quantities. We run ADMM until $\mathrm{Res}_{\mathrm{KKT}}(x^k) \leq 10^{-8}$ and use the output as $x^*$ to compute $\|x^k - x^*\|$. Fig.~\ref{fig:BehaveCom} shows both $\|x^k - x^*\|$ and $\|\mathcal{P}_{\text{Row}(A)}(x^k - x^*)\|$ (labeled as RP for row projection). Triangle markers indicate the transition to $|E^k| \leq M$, which persists thereafter.

Based on the results in Fig.~\ref{fig:BehaveCom}, three key observations can be made:
1) ASM maintains the rapid initial convergence of ADMM while eventually achieving a local convergence rate comparable to that of SSNAL, a method known for its asymptotic superlinear convergence properties.
2) The null space error dominates the convergence behavior across all methods.
3) A distinct behavioral transition occurs near the marked step, after which $|E^k|$ remains below $M$. These phenomena are consistently observed in our subsequent experiments.

\color{black}
\subsection{Acceleration via Null Space Shrinking}\label{sec:sub_accelerate}

To understand the acceleration effect of subspace fidelity, we decompose the error $e^k:=x^k - x^*$ into the null space and row space components of $A$, denoted as $e_{\text{Null}}^k$ and $e_{\text{Row}}^k$, respectively. The averaging operation is omitted here as it preserves the error evolution's structural form. In sparse recovery with small $\lambda$, $e_{\text{Null}}^k$ typically dominates the convergence bottleneck. As shown in Fig.~\ref{fig:BehaveCom}, $e_{\text{Row}}^k$ (dashed line) decays much faster than the total error.

As established in Sec.~\ref{sub:modifications_for_consistency}, both ADMM and ASM can be interpreted as enhanced ISTA variants. However, ASM introduces an additional acceleration mechanism beyond that of ADMM. Specifically, ADMM incorporates a full-space RLS step, which breaks the strict step-size restriction required for ISTA's stability ($v \leq 2 / \lambda_{\max}(A^\top A)$). Nevertheless, as derived in SM~\ref{app:derivation_null_space}, the null space error $e_{\text{Null}}^k$ for both methods shares an identical structural evolution:
\begin{equation}\label{eq:eNull_ISTA}
    e^{k+1}_{\text{Null}} = e^{k}_{\text{Null}} - \beta v \lambda \cdot \mathcal{P}_{\text{Null}(A)}(p^k),
\end{equation}
where $\beta=1$ for ISTA, $\beta=2$ for ADMM, and $p^k := p(x^k)$. This implies that although ADMM accelerates convergence by relaxing the restriction on $v$, as the iteration proceeds and $p^k \to p^* := A^\top(y - Ax^*)$, its projection onto $\text{Null}(A)$ inherently diminishes. Consequently, the marginal gain from merely utilizing a larger step size becomes insignificant.

Beyond similarly relaxing the restriction on $v$ (as detailed in Sec.~\ref{sec:averaging_and_the_proximal_residual_control} and Sec.~\ref{sec:Adap_sub_with_LR}), ASM fundamentally accelerates convergence by directly shrinking the null space itself. Specifically, when $E^k \supseteq \mE^*$ and $x^k|_{(E^k)^c} = 0$, the null space error is strictly confined to $\text{Null}(\hat A^k)$. In particular, when $|E^k| \leq M$, $\text{Null}(\hat A^k)$ degenerates to the trivial zero space $\{0\}$, triggering a strict local contraction. We have the following local convergence theorem for the general averaging step 
\begin{equation}
    {x}_\text{ave}^{k+1} = d^k\cdot x^{k+1} + (1 - d^k)\cdot {x}_\text{ave}^k,\quad d^k \in (0, 1],\label{eq:ave_dk}
\end{equation}
by extending the averaging factor $1/2$ in \eqref{eq:core_ASM} to $d^k$.
\begin{thm}[Local geometric convergence]\label{thm:Local_geometric_convergence}
Consider the sequence $\{x_{\text{ave}}^k\}$ generated by the ASM-L1 iteration \eqref{eq:core_ASM} with subspace sequence $\{E^k\}$ and averaging factor sequence $\{d^k\}$. Suppose $x_{\text{ave}}^k \to x^*$, $v$ is fixed, and $I - v(A|_{\mE^*})^T A|_{\mE^*}$ is non-singular. Then, under the \textit{generic conditions}, there exists an integer $k_0 > 0$ such that for any $k \geq k_0$,
\begin{equation}
    \|x_{\text{ave}}^{k+1} - x^*\|_{\mE^*} \leq \left[ d^k \cdot C_{v} + (1 - d^k) \right] \|x_{\text{ave}}^{k} - x^*\|_{\mE^*},\label{eq:acceleration_thm}
\end{equation}
where $C_{v} < 1$ is defined as the maximum of $|1 - v\tau| / |1 + v\tau|$ over the eigenvalues $\tau$ of $(A|_{\mE^*})^T A|_{\mE^*}$, and $\|\cdot\|_{\mE^*}$ is the vector norm defined by $\|x\|_{\mE^*} := \|x|_{\mE^*} - v(A|_{\mE^*})^T Ax\|_2 + \|x|_{(\mE^*)^c}\|_2$.
\end{thm}
\begin{IEEEproof}
See Appendix~\ref{app:Local_geometric_convergence}.
\end{IEEEproof}

In practical implementations discussed subsequently, we ensure $d^k \geq 1/2$ asymptotically. Consequently, the ultimate geometric convergence rate is predominantly governed by $C_{v}$. Before exactly identifying $\mE^*$, if there exists an integer $k_1$ such that $\mE^* \subseteq E^k$ and $|E^k| \leq M$ for all $k \geq k_1$, a similar contraction holds for the norm $\|\cdot\|_{E^k}$ \textcolor{black}{(defined by replacing the set $\mE^*$ with $E^k$ in $\|\cdot\|_{\mE^*}$), provided the 
the \textit{non-singular condition}, i.e., $I - v(A|_{\mE})^T A|_{\mE}$ is non-singular for $\mE=E^k$, holds.} At this stage, any potential oscillations caused by metric switching across iterations are strictly controlled, decaying geometrically with the factor $1-d^k$.

\textcolor{black}{
\section{Proximal Residual Control, Averaging Strategy and Global Convergence}\label{sec:averaging_and_the_proximal_residual_control}
}

The averaging step, although seemingly simple, plays a pivotal role in ASM. It is well-known that for any non-expansive fixed-point iteration (FPI) in finite dimensions, an additional averaging step can induce convergence \cite{LSCOA} even in the absence of operator contractility. For ASM, this averaging mechanism becomes even more indispensable. While ADMM iteration constitutes an FPI, the subspace restriction in~\eqref{eq:core_ASM} raises doubts about whether ASM inherits this property. In fact, for the LASSO problem, ASM is an FPI only when the optimal support set $\mathcal{E}^*\subseteq E^k$. Fortunately, with the assistance of the averaging strategy, the global convergence of ASM can be guaranteed by controlling the associated proximal residual, even if the condition $\mathcal{E}^*\subseteq E^k$ does not hold. In the following, we consider general averaging factor $d^k$ 
and vary the constant $v$ in the second step of \eqref{eq:core_ASM} to $\hat{v}$, which could be different from $v$ in the first step.
\textcolor{black}{In principle, ASM generalizes to any subspace $\mathcal{E}^k \supseteq E^k$ by replacing $\partial g(z^k)$ with $p^k := (x_{\text{ave}}^k - v\nabla h(x_{\text{ave}}^k) - z^k)/(v\lambda)$ in the second step and, similarly, restricting the update of $x^{k+1}$  to $\mE^k$.}

\textcolor{black}{
\subsection{The Descent of the Proximal Residual and Convergence}
}

Before elucidating the mechanism of the averaging step, we first show how the convergence of ASM can be established by ensuring the descent of a Lyapunov function—the \textit{proximal residual.} We define $R^k$ and $R(x)$ as
\begin{equation}
    R^k := R(x_\text{ave}^k) := \mathcal{S}_{\lambda v}(x_\text{ave}^{k}-v\nabla h(x_\text{ave}^{k})) - x_\text{ave}^k
\end{equation}
for a fixed $v$. As exhibited in \eqref{eq:core_ASM}, \textcolor{black}{$R^k = z^{k} - x_\text{ave}^k$} represents the proximal residual of the intrinsic ISTA step performed in the entire space. We remark that $R^k$ can be characterized via the dual variable as $R^{k}=-v\nabla h(x_{\text{ave}}^{k})-v\lambda p^{k}$ \textcolor{black}{where $p^k=p(x_\text{ave}^{k})$}. If the ASM iteration is adjusted such that $R^k \to 0$, it yields a sequence $x_\text{ave}^k$ converging to the LASSO optimum. For ADMM in~\eqref{eq:cplx_L}, this convergency holds:
\begin{prop}[\textcolor{black}{Convergence of ADMM in terms of $R^k$}]\label{prop:ADMM}
Consider the sequence $\{x_{\mathrm{ave}}^k\}$ generated by the ADMM iteration in \eqref{eq:cplx_L} with $v = \hat{v}$ and a fixed parameter $d \in (0,1)$. The associated residual sequence $\{R^k\}$ is non-expansive. Furthermore, under the \textit{generic conditions}, for any $x \neq x^*$ (the LASSO optimum), \textcolor{black}{let $x^{(n)}$ denote the iterate obtained after $n$ ADMM steps initialized at $x$. Then, there exist $n_s \in \mathbb{Z}_+$ and a constant $\delta_x > 0$ such that $\|R(x^{(n_s)})\| \leq \|R(x)\| - \delta_x$.}
\end{prop}
\begin{IEEEproof}
See Appendix~\ref{app:prop_ADMM}.
\end{IEEEproof}

This transformation of attention from FPI to controlling $R^k$ facilitates the analysis of ASM. First, since $R^k$ no longer depends explicitly on the relationship between $E^k$ and the optimal $\mE^*$, global convergence becomes tractable. Second, the evolution of $R^k|_{E^k}$ throughout the ASM iterations closely resembles that of ADMM. Intuitively, when $R^k|_{(E^k)^c}$ is negligibly small, the behavior of $R^{k+1}|_{E^k}$ is governed by its restricted dynamics on $E^k$, which mirrors the non-expansive property in ADMM.

To establish a rigorous foundation for this intuition, it is necessary to clarify two underlying ambiguities: 1) How to explicitly control the residual's behavior outside the subspace $E^k$? 2) How to ensure that the perturbations arising from a non-zero $R^{k+1}|_{(E^k)^c}$ do not destabilize the non-expansiveness inherited from ADMM within $E^k$? Essentially, resolving these two points provides a sufficient path to convergence:
\begin{thm}[Global convergence framework]\label{thm:convergence_concern12}
Consider the sequence $\{x_{\text{ave}}^k\}$ generated by ASM-L1 \eqref{eq:core_ASM} with subspace sequence $\{\mathcal{E}^k\}$ and averaging factor sequence $\{d^k\}$. \textcolor{black}{Suppose that $\hat v = v$, $E^k \subseteq \mathcal{E}_0^k \subseteq \mathcal{E}^k$, where $E^k := \{i : |\mu_i(x^k_\text{ave})|>v\lambda\}$, which is in fact the support of $z^k$; 
\begin{equation}
\mathcal{E}_0^k := E^k \cup \big\{i : |p_i^k| \geq 1 - \epsilon, \textcolor{black}{\text{or}\ |(x_\text{ave}^k)_i|\geq C} \big\}, \label{eq:ave_epsilon}
\end{equation}
where $\epsilon$ is a small constant \textcolor{black}{while constant $C$ is large}; and there exists $d_1\in (0,1]$ such that $d_1\leq d^k\leq d_2$ with $\mathrm{card}(\{d^k\})<\infty$.} If the associated residual sequence $\{R^k\}$ further satisfies:
    \begin{enumerate}
        \item $\|R^{k+1}|_{(\mathcal{E}^k)^c}\| \leq (1-d^k)\|R^k|_{(\mathcal{E}^k)^c}\|$,
        \item $\|R^{k+1}|_{\mathcal{E}^k}\| \leq \|R^k|_{\mathcal{E}^k}\|$,
    \end{enumerate}
    then under the \textit{generic} conditions, the sequence $\{x_{\text{ave}}^k\}$ converges to the LASSO optimum.
\end{thm}
\begin{IEEEproof}
See Appendix~\ref{sec:convergence_of_asm}.
\end{IEEEproof}

Two further remarks are in order regarding this framework. First, the subspace $E^k$ in \eqref{eq:core_ASM} is generalized here to $\mathcal{E}^k$ to accommodate the flexible screening rules to be discussed later. \textcolor{black}{Second, as shown in the proof, if the proximal residual sequence $R^k$ satisfies the \textit{non-monotonic descent condition} below, 
the separable suppression with respect to $\mE^k$ is not strictly necessary to establish the convergence.}
\begin{defi}[Non-Monotonic Descent Condition]
    Let $\{x^k_\text{ave}\}$ be generated by the ASM-L1 \eqref{eq:core_ASM} with subspaces $\{\mE^k\}$ and averaging factors $\{d^k\}$. If there is a constant $c_0 > 0$ and a sequence $\{a_k\}$ satisfying $a_k \geq 0$ and $\sum a_k < \infty$ such that 
    \begin{equation}
        \begin{aligned}
            \|R^{k+1}\|^2 &\leq (1+a_k)\big[\|R^k\|^2 - c_0 d^k \delta^k_c \big], \\
            \delta^k_c &:= \|R^{k+1}|_{(\mE^k)^c}\|^2 + \|R^k|_{(\mE^k)^c}\|^2,
        \end{aligned}
    \end{equation}
    we say %$\{x^k_\text{ave}\}$ 
    $\{R^k\}$ satisfies the \textit{non-monotonic descent condition}.
\end{defi}

\textcolor{black}{
\subsection{Averaging, Contractility and Subspace Elements Transition}\label{sec:averaging_and_subspace_elements_transition}
}

To fulfill the condition 1) of Theorem~\ref{thm:convergence_concern12}, we first investigate the contractility of the residual outside the support $E^k$. When $i \notin E^{k}$, we have  $z_i^{k}=0$, so  $R_i^k = 0-(x_\text{ave}^k)_i$. This reveals that outside the support, $-R^k$ coincides with $x_\text{ave}^k$ itself.

The control of this residual then bifurcates into two scenarios depending on the subspace evolution. First, consider the stable case where $E^{k+1} \subseteq E^k$. For any index $i \in (E^k)^c$, the thresholding rule ensures \textcolor{black}{$x_i^{k+1} = 0$}, and consequently:
\begin{equation}
    R_i^{k+1} = -(x_\text{ave}^{k+1})_i = -(1-d^k)(x_\text{ave}^k)_i = (1-d^k)R_i^k.\label{eq:Rk_zero}
\end{equation}
In this regime, the averaging factor $d^k$ directly acts as the contraction rate for the residual on $(E^k)^c$, relying on the adaptive property of $p^k$ to maintain the zero-valued entries.

In the second case, this automatic contractility is lost during subspace expansion, i.e., when $E^k\subsetneq E^{k+1}$. For an index $i$ entering the support, its residual $R_i^{k+1}$ is no longer a simple scaled version of $R_i^k$. Since $R_i^{k+1} = -v\nabla h(x_\text{ave}^{k+1})_i - v\lambda p_i^{k+1}$ and the adaptivity of $p_i^{k+1}$ is constrained by $|p_i^{k+1}| \leq 1$, any abrupt change in the gradient $\nabla h(x_\text{ave}^{k+1})_i$ can cause the residual to grow, breaking the contractility.

Instead of attempting to bound the residual growth directly during subspace expansion, we establish a safeguarded mechanism to facilitate a stable re-entry of indices into the support. This idea is embodied in the definition of \eqref{eq:ave_epsilon} and we have the following lemma.
\begin{lemma}\label{lemma:ave_d0}
    Suppose $\{x_{\text{ave}}^k\}$ generated by ASM-L1~\eqref{eq:core_ASM} is bounded and $\mE^k \supseteq \mE^k_0$ holds for all $k$. Then, there exists a constant $d_1 \in (0,1]$ such that each iteration admits an averaging factor $d^k \geq d_1$ satisfying $\|\mu(x_{\text{ave}}^{k+1})|_{(\mE^k)^c}\|_\infty \leq v\lambda$, hence $R^{k+1}|_{(\mE^k)^c} = (1-d^k) R^k|_{(\mE^k)^c}$.
\end{lemma}
\begin{IEEEproof}
    Due to the boundedness of $\{x_{\text{ave}}^k\}$ and the ASM-L1 mapping (regardless of any specific $\mE^k$), there exists a constant $C > 0$ such that $\|\mu(x^{k+1})\| \leq C$ for all $k$. Let $i_0 = \mathop{\arg\max}\{i \in (\mE^k)^c : |\mu_i(x^{k+1})|\}$. We only consider the case where $|\mu_{i_0}(x^{k+1})| > v\lambda$. Then we set $d_1$ such that $C d_1 + (1-d_1)v\lambda(1-\epsilon) = v\lambda$, and we have $d_1 \in (0,1]$ by noting $C \geq |\mu_{i_0}(x^{k+1})| > v\lambda$. Setting $d^k = d_1$ ensures that $|d^k \mu_{i_0}(x^{k+1}) + (1-d^k) v\lambda p_{i_0}^k| \leq v\lambda$ by~\eqref{eq:ave_epsilon}, implying that $\|\mu(x_{\text{ave}}^{k+1})|_{(\mE^k)^c}\|_\infty \leq v\lambda$, and the conclusion follows.
\end{IEEEproof}

\textcolor{black}{
Indeed, at each iteration, if $i_0$ exists, we can always choose $d_{i_0} > 0$ such that the averaged value $|\mu_{i_0}(x^{k+1}_{\text{ave}})|$ equals $v\lambda$:
\begin{equation}
    |d_{i_0} \mu_{i_0}(x^{k+1}) + (1-d_{i_0}) \mu_{i_0}(x^k_{\text{ave}})| = v\lambda, \label{eq:i_0_d_i_0}
\end{equation}
thereby smoothly transitioning $i_0$ into the subspace $\mE^{k+1}$ while obeying the contractility on $(\mE^k)^c$ as required by Theorem~\ref{thm:convergence_concern12}. Specifically, this intuition translates naturally into a practical backtracking rule for $d^k$:
}
\textcolor{black}{
\begin{defi}[Safe Averaging Rule]
Fix an initial step size $d_I \in (0,1)$ (e.g., $d_I = 1/2$) and a decay factor $\alpha \in (0,1)$. At the $k$th iteration of the ASM~\eqref{eq:core_ASM} with a given $\mE^k$, first compute $x^{k+1}$ and $\mu(x^{k+1})$. If $\|\mu(x^{k+1})|_{(\mE^k)^c}\|_\infty \leq v\lambda$, set $d^k = d_I$. Otherwise, compute $d_{i_0}$ from~\eqref{eq:i_0_d_i_0}, find the integer $l \in \mathbb{N}$ such that $\alpha^l d_I \leq d_{i_0} < \alpha^{l-1} d_I$, and set $d^k = \alpha^l d_I$.
\end{defi}}

\begin{lemma}\label{lemma:dk_guarantee}
Consider $\{x_{\text{ave}}^k\}$ generated by ASM-L1~\eqref{eq:core_ASM} with subspaces $\{\mE^k\}$ and averaging factors $\{d^k\}$. Suppose that for all $k$, (i) $\mE^k_0 \subseteq \mE^k$, (ii) the \textit{non-singular condition} holds, and (iii) $d^k$ is determined by the \textit{Safe Averaging Rule}. If the associated residual sequence $\{R^k\}$ satisfies the \textit{non-monotonic descent condition}, then there exists a constant $d_0 > 0$ such that $d^k \geq d_0$ for all $k$, and $\mathrm{card}(\{d^k\}) < \infty$.
\end{lemma}
\begin{IEEEproof}
    The result follows directly by combining Lemma~\ref{lemma:ave_d0} and Lemma~\ref{lemma:ASM_bounded} in Appendix~\ref{sec:convergence_of_asm}.
\end{IEEEproof}

\textcolor{black}{
As long as the ASM iteration converges, $d^k$ asymptotically stabilizes at the initial value $d_I$. Furthermore, once $\epsilon$ in~\eqref{eq:ave_epsilon} falls below the generic threshold $\omega_0$ (Definition~\ref{defi:genetic}), $\mE_0^k$ is guaranteed to identify the optimal support $\mE^*$ in finite steps. This ensures that the safeguarding mechanism does not impede the asymptotic acceleration.
}

\textcolor{black}{
\subsection{Global Convergence of ASM on the LASSO Problem}\label{sub:global_convergence_of_asm}
}

According to \textcolor{black}{the condition 2)} in Theorem~\ref{thm:convergence_concern12}, the global convergence of ASM hinges on ensuring non-expansiveness within the subspace. To quantify the dynamic coupling between the internal residual $R^{k+1}|_{\mathcal{E}^k}$ and the perturbation $R^k|_{(\mathcal{E}^k)^c}$, we establish the following fundamental relations. Hereafter, we slightly extend the notation by letting $\hat{A} := A|_{\mathcal{E}^k}$ and $\tilde{A} := A|_{(\mathcal{E}^k)^c}$ to encompass the generalized subspace $\mathcal{E}^k$.
\begin{prop}\label{prop:E_Ec_couple}
Consider the sequence $\{x_{\text{ave}}^k\}$ generated by ASM-L1 with subspaces $\{\mathcal{E}^k\}$ and averaging factors $\{d^k\}$. If each support $E^k \subseteq \mathcal{E}^k$, then the residual sequence $\{R^k\}$ satisfies the following properties:
\begin{enumerate}
    \item $R^{k+1}|_{\mathcal{E}^k} - (1-d^k) R^k|_{\mathcal{E}^k} = \gamma \Delta \mu^{k}|_{\mathcal{E}^k} - v\lambda \Delta p^{k}|_{\mathcal{E}^k};$
    \item $\|R^{k+1}|_{\mathcal{E}^k} - (1-d^k) R^k|_{\mathcal{E}^k}\| \leq d^k(\|R^k|_{\mathcal{E}^k}\|^2-J^k/\hat v)^\frac12$, provided $v \geq \hat{v}$ or $\Delta p^{k}|_{\mathcal{E}^k} = 0$.
\end{enumerate}
\textcolor{black}{Here $J^k := J(\sum_{i\in \mE^k}A_iR^k_i, \sum_{j\in (\mE^k)^c}A_jR^k_j)$, where $A_i$ is the $i$th column of $A$, $\gamma := \hat v/(v+\hat v)$, and for any $u,w\in \mathbb{R}^M$, $J(u,w)$ is defined as follows, with $G := (I + \hat{v}\hat{A}\hat{A}^T)^{-1}$:
\begin{equation}
\begin{aligned}
    J(u,w) :=& \left\| [vw-(\hat v-v)u]/2 - G[(\hat{v}+v)u + vw] \right\|^2\\
             &- \left\| [vw-(\hat v-v)u]/2 \right\|^2.\label{eq:J_k}
\end{aligned}
\end{equation}}
\end{prop}
\begin{IEEEproof}
See Appendix~\ref{sec:algebraic_relationships_between_asm_iterations}.
\end{IEEEproof}

As demonstrated in the proof, Property 2) implies $\|R^{k+1}|_{\mathcal{E}^k}\|^2 \leq \|R^{k}|_{\mathcal{E}^k}\|^2 - (d^k/\hat v) J^k$. Crucially, $J(u,w)$ is a homogeneous quadratic form, implying that its non-negativity depends solely on the residual partition. Specifically, when $\hat v = v$, there exists a ratio $r_0$ such that whenever $\|\tilde A R^k|_{(\mathcal{E}^k)^c}\| \leq r_0 \|\hat A R^k|_{\mathcal{E}^k}\|$, we have $J^k \geq 0$.
Conversely, when $\|\hat A R^k|_{\mathcal{E}^k}\|$ is relatively small, although $J^k$ may become negative, the corresponding $R^k|_{(\mathcal{E}^k)^c}$ remains bounded away from zero. Since $R^{k+1}|_{(\mathcal{E}^k)^c}$ is guaranteed to be strictly contractive with respect to $R^{k}|_{(\mathcal{E}^k)^c}$, the overall sufficient descent can still be maintained, given that $\|J^k\| \sim \mathcal{O}(v^2 \|R^k|_{(\mathcal{E}^k)^c}\|^2)$ as $v \to 0$.

\begin{thm}[Global Convergence of ASM-L1]\label{thm:Ek_converge}
    Let $\{x_{\text{ave}}^k\}$ be the sequence generated by the ASM-L1 \eqref{eq:core_ASM}, with subspaces $\{\mathcal{E}_0^k\}$ given by \eqref{eq:ave_epsilon} and averaging factors $\{d^k\}$ satisfying the \textit{safe averaging rule}. Under the \textit{generic conditions}, there exists a constant $v_0 > 0$ such that for any $\hat{v} = v \leq v_0$, the sequence $\{x_{\text{ave}}^k\}$ converges to the LASSO optimal solution $x^*$. Moreover, if the parameter $\epsilon$ in \eqref{eq:ave_epsilon} satisfies $\epsilon < \omega_0$ (where $\omega_0$ is defined in the \textit{generic conditions}), then $\mathcal{E}_0^k = \mathcal{E}^*$ for all sufficiently large $k$.
\end{thm}
\begin{IEEEproof}
The proof is given in Appendix~\ref{app:converge_Ek_Ek0}.
\end{IEEEproof}

\textcolor{black}{
The theorem indicates that if ASM closely tracks the support evolution with $\mathcal{E}^k=E^k$, it imposes constraints on $v$ and $\hat{v}$.} Before presenting two strategies in Sec.~\ref{sec:practical_strategies_for_asm} for relaxing these, we remark that the requirement on $v_0$ is generally loose. First, the proof rests on a stringent assertion where sufficient descent holds even if the positive term in~\eqref{eq:J_k} vanishes. Second, even in this worst case, $v_0 \approx 1$ typically suffices. From the proof, $v_0$ is admissible if $\sqrt{v_0}\|\sum_{j \in (\mathcal{E}^k)^c} A_j R_j^k\| \leq 2\|\sum_{j \in (\mathcal{E}^k)^c} R_j^k\|$. For instance, if $\Delta \mathcal{E}^k := \mathcal{E}^{k-1} \setminus \mathcal{E}^{k}$ is the newly suppressed set, $A$ is column-normalized, and $R^k$ has vanished on $(\mathcal{E}^{k-1})^c$ by the contractility of ASM on it, the condition follows from the triangle inequality.

\section{Practical Strategies for ASM}\label{sec:practical_strategies_for_asm}

In this section, we introduce two strategies to enhance the efficiency of ASM: dynamic subspace updates using low-rank operations and adaptive step-size selection.

\textcolor{black}{
\subsection{Adaptive Subspace with Low-Rank Updates}\label{sec:Adap_sub_with_LR}
}

As established in Theorem~\ref{thm:Ek_converge}, closely tracking the support when $\mathcal{E}^k = E^k$ imposes a step-size constraint, as large $v$ may yield $J^k \leq 0$ and induce instability. Viewing $J^k$ (for fixed $R^k$) as a functional of $\mathcal{E}^k$, ASM represents the extreme $\mathcal{E}^k = E^k$, whereas ADMM occupies the opposite pole $\mathcal{E}^k = [N]$. In practice, we can balance efficiency and convergence by repartitioning $(E^k)^c$ to expand $\mathcal{E}^k$, ensuring $R^k|_{\mathcal{E}^k}$ dominates the remaining components and maintaining $J^k \geq 0$ for larger $v$.

At each iteration, we only select a subset $\Delta\mathcal{E}^k \subseteq \mathcal{E}^{k-1} \setminus E^k$ (e.g., based on the magnitudes of $|R_i^k|$) for exclusion, thereby sustaining the dominance of $\sum_{i \in \mathcal{E}^k} A_i R^k_i$. By controlling the cardinality of $\Delta\mathcal{E}^k$, we can efficiently perform a non-negativity check on $J^k$ prior to the actual removal. Since the matrix $G$ is already required in ASM to compute $\hat{\mathcal{L}}^k_{\hat{v}}$ in~\eqref{eq:mL_def}, this checking procedure is handled via efficient low-rank updates (\textcolor{black}{See SM~\ref{supple:Low-Rank_Updates}}). Let $\bar{E}^k := \{i : (x^k_{\text{ave}})_i \neq 0\} \cup \mathcal{E}_0^k$, with $\mathcal{E}_0^k$ defined as in \eqref{eq:ave_epsilon}.
\textcolor{black}{
We define the \textit{safeguarding rule} as follows:}

\begin{defi}[Safeguarding Rule]\label{defi:Safe_Rule}
\textcolor{black}{We say $\{x_{\text{ave}}^k\}$ generated by the ASM-L1 \eqref{eq:core_ASM} follows the \textit{safeguarding rule} if its subspaces $\{\mathcal{E}^k\}$ and averaging factors $\{d^k\}$ are determined as follows. Fix a sufficiently small $r>0$, a $d_2 \in (0,1)$ for averaging and $c_0 \in (0,1/2)$ for the \textit{non-monotonic descent condition}. At each iteration,} for any set $\Delta \mathcal{E}^k \subseteq \bar{E}^k \setminus E^k$, we define $\mathcal{E}_1^k := \bar{E}^k \setminus \Delta \mathcal{E}^k$ and evaluate the corresponding $J_1^k$:
\begin{enumerate}
    \item If $J_1^k \geq r\|R|_{\mE^k_1}\|^2$, set $\mathcal{E}^k = \mathcal{E}_1^k$. If $\|\mu(x^{k+1})|_{(\mE^k)^c}\|_\infty \leq v\lambda$, set $d^k=1$; otherwise, repeatedly halve $d^k$ starting from $1$ \textcolor{black}{until the \textit{non-monotonic descent condition} holds.}
    \item If $J_1^k < 0$, set $\mathcal{E}^k = \bar{E}^k$ and $d^k=d_2$.
\end{enumerate}
\end{defi}

Intuitively, \textcolor{black}{when $\Delta \mathcal{E}^k$ is stably screened out, $R^k|_{\Delta \mathcal{E}^k}$} decays rapidly due to the contractivity induced by averaging. In the safeguarding rule, we explicitly nullify the impact of $R^k|_{(\mathcal{E}^k)^c}$ by setting $d^k=1$, thereby ensuring it does not hinder the exclusion of remaining elements. \textcolor{black}{In Sec.~\ref{sec:experiment}, we choose $\Delta E^k:=\arg\min_{i\in\bar E^k\setminus \mathcal{E}^{k}_0}|(x_{\text{ave}}^k)_i|$; empirically, $\mathcal{E}^k\setminus\mathcal{E}^k_0 \to \emptyset$ without affecting the asymptotic acceleration of ASM. Theoretically, we can guarantee that $\mathcal{E}^k=E^k$ by temporarily increasing $\hat{v}$ under certain conditions (see Appendix~\ref{sec:convergence_of_asm_with_shrinkage_checks}).}

In practice, the \textit{safeguarding rule} is rather conservative, as setting $d^k=1$ is often restrictive and unnecessary for achieving \textit{non-monotonic descent} during initial stages. For the experiments in Sec.~\ref{sec:experiment} (termed as ASM-L1-LR), we adopt a two-stage procedure with conditional shrinkage checks, incorporating a pre-specified sequence $\{a_k\}$ and a constant $c_0\leq 1/2$ for the non-monotonic descent condition, as follows:

\begin{enumerate}
    \item \textcolor{black}{\textbf{Initial Stage}: Given small constants $\varepsilon_0, c_0 > 0$, initialize $\mE^k_{(1)} := \mE^k_0 \cup \{i : |(x_{\text{ave}}^k)_i| \geq \varepsilon_0\}$, with $\mE^k_0$ given by~\eqref{eq:ave_epsilon} and $d^k$ given by the \textit{Safe Averaging Rule}. If the \textit{non-monotonic descent} condition is satisfied, the ASM update in~\eqref{eq:core_ASM} is performed using $\mE^k_{(1)}$ and $d^k$. Otherwise, the algorithm permanently transitions to the Safeguarding Stage.}
    \item \textbf{Safeguarding Stage}: Initially, we expand $\mathcal{E}^{k}$ until the ASM iteration with $d^k=1$ ensures the \textit{non-monotonic descent condition}. The algorithm then follows the \textit{safeguarding rule}. \textcolor{black}{Specifically, the set $\Delta \mathcal{E}^k$ of a fixed size $K$ is selected corresponding to the indices with the smallest magnitudes of $|(x_{\text{ave}}^k)_i|$.}
    \textcolor{black}{\item \textbf{Shrinkage Checks}: During the Safeguarding Stage, we fix a small constant $r_1 \in (0,1)$. Whenever $|E^k| \leq M$ while $\mE^k \neq E^k$, we initialize $\tau = 1$. Every $K_1$ iterations thereafter, provided that $E^k$ and $\mE^k$ remain unchanged, we double $\tau$, compute $x^{k+1}$ with $\hat{v} \leftarrow \tau v$, and evaluate the following conditions: 
    (i) $\|R(x^{k+1})\| \leq (1-r_1)\|R(x_\text{ave}^k)\|$; 
    (ii) $\|\mu(x^{k+1})|_{(E^k)^c}\|_\infty \leq v\lambda$; and 
    (iii) $[p(x^{k+1})-p(x^k_{\text{ave}})]|_{E^k} = 0$. 
    If these conditions are jointly satisfied, we set $\mathcal{E}^k = E^k$ and $d^k = 1$; otherwise, we resume the \textit{Safeguarding} iterations with $\hat{v} = v$.}
\end{enumerate}

\textcolor{black}{This procedure also guarantees the convergence of ASM-L1, ensuring that $x_{\text{ave}}^k \to x^*$ and $\mathcal{E}^k \to \mathcal{E}^*$, provided $\epsilon < \omega_0$ (see Appendix~\ref{sec:convergence_of_asm_with_shrinkage_checks}). } Although the transition between the two stages requires subspace expansion and may incur computational overhead in the worst case, it occurs at most once. Regarding the shrinkage check, the checking interval $K_1$ can be increased to further relieve computational burden.

\color{black}

\subsection{Empirical Strategies for Adaptive Step-Size}\label{sec:ave_strategies}

\textcolor{black}{For ASM variants that closely track the support ($\mathcal{E}^k = E^k$), we propose an empirical schedule for the step sizes $v$ and $\hat{v}$. As implied in the proof of Theorem~\ref{thm:Local_geometric_convergence}, increasing $\hat{v}$ can reduce the contraction factor as $E^k$ approaches $\mathcal{E}^*$. Motivated by this, we initialize $v = \hat{v}$ at a conservative value to ensure early stability, and subsequently increase $\hat{v}$ to accelerate convergence. This strategy is detailed below and evaluated in Sec.~\ref{sec:experiment}.}

\begin{enumerate}
    \item Keep $\hat{v}^k$ and $v^k$ fixed as $\hat{v}$ and $v$ throughout the iteration.
    
    \item Keep $v^k$ fixed as $v$ but allow $\hat{v}^k$ to vary according to:
    \begin{equation}
        \hat{v}^{k} = \left[(\rho^k \cdot v + (1 - \rho^k) \cdot \bar{v}^k)^{-1} - v^{-1}\right]^{-1}. \label{eq:mix_hatv}
    \end{equation}
    Here $\bar{v}^k := v \cdot |E^k| / N$ is suggested by VAMP, and is mixed with $v$ to prevent $\hat{v}^k$ from increasing too rapidly.
    
    \item Adjust both $\hat{v}^k$ and $v^k$ according to subspace-restricted message passing:
    \begin{equation}
        v^{k+1/2} = |E^k|^{-1}\mathrm{tr}\big[((\hat{v}^k)^{-1} \cdot I + \hat{A}^T \hat{A})^{-1}\big], \label{eq:vmix_both1}
    \end{equation}
    then set $v^{k+1} := 1 / (1/v^{k+1/2} - 1/\hat{v}^k)$. For $\hat{v}^k$, we use:
    \begin{equation}
        \hat{v}^{k} = [\big(\mathrm{mean}_{i \in E^k}(v_{\text{post},i}^k) \cdot \alpha\big)^{-1} - (v^{k+1})^{-1}]^{-1}. \label{eq:vmix_both2}
    \end{equation}
\end{enumerate}

For ASM-L1, we recommend strategy 2). Specifically, we set $\rho^k$ as follows in our experiments:
\begin{equation}
\rho^k =
\begin{cases}
    \rho_0, & \text{if } |\bar{E}^{s,k}| > (1 + c) M \\
    |E^k| / (|\bar{E}^{s,k}| + \varepsilon), & \text{if } |\bar{E}^{s,k}| \le (1 + c) M
\end{cases}
\label{eq:rho_increase}
\end{equation}
where $\bar{E}^{s,k}:=\cup_{j=0}^s E^{k-j}$, and $\varepsilon > 0$ is a small stability constant. Heuristically, $\rho^k$ monitors the variation of $E^k$: when the support set becomes stationary, $\rho^k$ increases accordingly.

Strategy 3) is inspired by the variance update in Vector Approximate Message Passing (VAMP)~\cite{VAMP} and is adopted for the algorithms introduced in Sec.~\ref{sec:ASM_Structured}. Every strategy can be implemented by substituting $\hat{v}^k$ and $v^k$ in Algorithms~\ref{algo:ASframework}. From the VAMP perspective, \eqref{eq:vmix_both1} approximates the posterior variance $v^{k+1/2}$ through moment matching within the subspace, similar to the approach used for $x^{k}$ in~\eqref{eq:core_ASM}. Regarding~\eqref{eq:vmix_both2}, the posterior variance matched on $E^k$ is scaled by $\alpha \in (0,1)$ to mitigate degeneration inherent to the VAMP framework.

Although these strategies often yield commendable performance, as demonstrated in Sec.~\ref{sec:experiment}, we must emphasize their heuristic nature. We stress that while sophisticated $\hat{v}^k$ and $v^k$ tuning can improve ASM's performance, the core acceleration mechanism stems from the subspace framework in~\eqref{eq:core_ASM}, as explained in Sec.~\ref{sec:Sub4Accelerate}.

\section{ASM with General Denoiser}\label{sec:ASM_Structured}

The ASM framework, as illustrated in Algorithm~\ref{algo:ASframework}, can be enhanced by suitably designed denoisers $\mathcal{D}$. In this section, we adopt a Bayesian perspective to demonstrate how structured priors can be leveraged within ASM. Treating $x^\dagger$ as a random variable, we obtain the likelihood function $p(y|x) \propto \exp\left\{-\|y - A x\|_2^2 / (2\sigma_w^2)\right\}$. Given any prior distribution $p^{\text{pri}}(x)$, one may estimate $x^\dagger$ using either the maximum a posteriori (MAP) or minimum mean square error (MMSE) estimators:
\begin{equation}
    x_{\text{MAP}} := \mathop{\arg\max}_x p(x|y),\ \ x_{\text{MMSE}} := \mathbb{E}[X | Y = y].
\end{equation}

Various methods have been developed for this problem, including the recently proposed VAMP \cite{VAMP}, also known as OAMP \cite{OAMP} or Turbo CS \cite{TCS}. As demonstrated in~\cite{ECAI}, VAMP for $x_{\text{MAP}}$ under Laplace prior $p_\text{L}^{\text{pri}}(x) \propto \exp(-\lambda_0 \|x\|_1)$ produces iterations identical to ADMM, but interprets the ADMM parameter as the variance and updates it via a dedicated procedure.

Without delving into the complex derivations of VAMP, we directly present the prior-induced denoiser $\mathcal{D}$ as follows:
\begin{align}
    &\mathcal{D}_{\text{MAP}}(\mu, v; p^{\text{pri}}) = \mathop{\arg\min}_z \frac{1}{2} \|z - \mu\|^2 - v \log p^{\text{pri}}(z),\label{eq:DeMAP}\\
    &\mathcal{D}_{\text{MMSE}}(\mu, v; p^{\text{pri}}) = \mathbb{E} Z,\ \ p(z) \propto e^{-\frac{1}{2v}\|z - \mu\|^2} \cdot p^{\text{pri}}(z).\label{eq:DeMMSE}
\end{align}
To illustrate this, we implement ASM with a Bernoulli-Gaussian prior \eqref{eq:BG_pri} and a hidden Markov chain (HMC) modeled prior \eqref{eq:HMC_pri}. The resulting algorithms are designated as ASM-BG and ASM-HMC, respectively.

\subsection{ASM with Denoisers Induced by Posterior Mean}

If $p^{\rm pri}$ is a Bernoulli-Gaussian distribution defined as
\begin{equation}
X_i \sim \begin{cases}
0, & \text{with probability } 1 - \epsilon \\
\mathcal{N}(0, \sigma_0^2), & \text{with probability } \epsilon
\end{cases}\label{eq:BG_pri}
\end{equation}
then $x_{\text{MMSE}}$ is also essentially sparse. According to~\eqref{eq:DeMMSE}, this prior induces a denoiser $\mathcal{S}^{\text{BG}}$ of the form
\begin{align}
    &\mS^{\text{BG}}_{\sigma_{0},v^k}(\mu^k_i):=\mu^{k}_{i}\cdot\beta_i^{k}\cdot\frac{\sigma_{0}^2}{v^{k}+\sigma_{0}^2},\ \ \ \ \beta_i^{k}:=\frac1{1+\Gamma^{k}_i},\\
    &\Gamma^{k}_i:=\frac{1-\epsilon}{\epsilon}\sqrt{\frac{v^{k}+\sigma_{0}^2}{v^{k}}}\exp\left\{-\frac{v^{k}+\sigma_{0}^2}{2v^{k}\sigma_{0}^2}|q_i^k|^2 \right\}.\label{eq:BG_Gamma}
\end{align}
where $\mu^k$ is the input signal, $q_i^k$ is shorthand for $\mu_i^{k} \cdot \sigma_{0}^2 / (v^{k} + \sigma_{0}^2)$, and $v^k$ is the parameter to be tuned. This denoising operation is accomplished by multiplying the input $\mu_i^k$ by a shrinkage factor $\beta_i^k \cdot \sigma_0^2 / (v^k + \sigma_0^2)$, where $\beta_i^k \in (0,1)$ controls the sparsity level. Notably, the VAMP framework provides not only  structure of this denoiser but also a recommended update rule for $v^k$ as specified in \eqref{eq:vmix_both2} with $v_{\text{post},i}^{k}$ satisfying
\begin{equation}
\begin{aligned}
    &v_{\text{post},i}^{k}=\frac{v^{k}\sigma_{0}^2}{v^{k}+\sigma_{0}^2}\beta_i^{k}+\big|q^{k}_i\big|^2\cdot\beta_i^{k}(1-\beta_i^{k}).
\end{aligned}\label{eq:ZeeBetaGamma}
\end{equation}

To apply subspace fidelity, two issues remain: 1) Although the output $x^k$ of $\mathcal{S}^{\text{BG}}$ exhibits intrinsic sparsity, none of the elements $x_i^k$ will be exactly zero; 2) The operator $\partial \hat{g}$ in ASM-L1 algorithm~\eqref{eq:core_ASM} is induced by the subgradient of the $\ell_1$-norm in the LASSO problem, which lacks an explicit counterpart for the MMSE case (as opposed to MAP).

To address the first issue, one can manually threshold $x^k$ to select the support. While thresholding based on noise level is common, we observe for this specific denoiser that both $|x_i^k|$ and $v_{p,i}^k$ approach zero as $\beta_i$ tends to zero, as seen in \eqref{eq:ZeeBetaGamma}. Therefore, we define the support set as $E^k := \{i : \beta_i \ge c\}$ for some constant $c \in (0,1)$. If $x^k$ converges, a larger value of $c$ generally yields a sparser limit point. \textcolor{black}{In practice, a small threshold such as $c = 0.05$ is generally recommended. Empirically, we observe that the more structured and accurate the prior information captured by the denoiser, the less sensitive the ASM framework becomes to the choice of $c$.}

For the second issue, we choose to update $\hat{\nu}^k$ such that $\nu_{i}^k = x_i^k - (\hat{v}^k / v^k)(\mu_i^k - x_i^k)$ for all $i \in E^k$. For readers familiar with VAMP, this exactly follows from  the extrinsic information~\cite{TurboDecoding}. In fact, this can also be understood from an optimization perspective. When deriving $\nabla g$ in~\eqref{eq:cplx_L}, we utilized the property in~\eqref{eq:mul_equal}. Using the notations from Algorithm~\ref{algo:ASframework} and denoting $\mu^{k} := x^{k-1} + v^k s^{k-1}$, it follows from~\eqref{eq:mul_equal} that $\hat{\nu} = \hat{z}^k - v^k \hat{s}^k$ is exactly equivalent to the formula above.

\subsection{Structured Denoiser}

If the ground truth $x^\dagger$ is known to have its non-zero elements grouped into clusters, we consider ASM-HMC with the following structured prior \cite{STCS}:
\begin{equation}
\begin{aligned}
    &p^{\text{pri}}(x_i|s_i) = \delta_{\{s_i=0\}}\delta(x_i) + \delta_{\{s_i=1\}}\mathcal{CN}(0,\sigma^2_0),\\
    &p(s_i|s_{i-1}) = (1-p_{01})^{1-s_i}p_{01}^{s_i} \cdot \delta_{\{s_{i-1}=0\}} \\
    &\hspace*{1.5cm} + (1-p_{10})^{s_i}p_{10}^{1-s_i} \cdot \delta_{\{s_{i-1}=1\}},
\end{aligned}
\label{eq:HMC_pri}
\end{equation}
with $p(s_1) = \frac{p_{01}}{p_{01}+p_{10}} \delta_{\{s_1=1\}} + \frac{p_{10}}{p_{01}+p_{10}} \delta_{\{s_1=0\}}$, where $s_i \in \{0,1\}$ are hidden variables and $\delta(x)$ is the Dirac delta function. The hyperparameters $p_{01}$ and $p_{10}$ are transition probabilities that control the cluster size of the recovered signal, and $p_{01}/(p_{01}+p_{10})$ characterizes the sparsity.

By incorporating the alternating subspace strategy into structured turbo CS \cite{STCS}, the resulting iteration follows the ASM framework, and the corresponding denoiser $\mathcal{S}^{\text{HMC}}_{\sigma_0,v^k}$ takes the same form as $\mathcal{S}^{\text{BG}}_{\sigma_0,v^k}$, except that all instances of $\epsilon$ in \eqref{eq:BG_Gamma} are replaced by $\pi_i^k$, which is derived via message passing rules as described in \cite{STCS}. We employ the same subspace determination strategies as in ASM-BG. In Sec.~\ref{sec:experiment}, we use a channel estimation problem to evaluate the performance of ASM-BG and ASM-HMC.

\section{Numerical Experiments}\label{sec:experiment}

In this section, we present numerical experiments to evaluate the ASM algorithm for: 1) the LASSO problem, 2) a channel estimation problem, and 3) a dynamic compressed sensing problem. Experiment 1) demonstrates the convergence and acceleration properties of ASM; Experiment 2) illustrates the integration of general denoisers; and Experiment 3) showcases the performance gain of ASM in real-time systems leveraging historical information.

\subsection{LASSO Problem}\label{sec:LASSO_problem}

\subsubsection{System Settings}

We generate LASSO problems using model~\eqref{problem:LinearInverse}. Signals $x^\dagger$ are drawn from a Bernoulli-Gaussian distribution~\eqref{eq:BG_pri} with $\epsilon = 0.25$ and $N = 2M$ (and $\epsilon = 0.125$ and $0.0625$ for $N = 2M$ and $4M$, respectively). Noise level is controlled by setting the noise variance $\sigma^2_w$ such that $\text{SNR} = 10 \log_{10}(\|A x^\dagger\|^2 / (M \sigma_w^2))$(dB), and we set $\lambda=\sigma^2_w$. We emphasize that, while higher SNR typically improves estimation accuracy, it necessitates smaller regularization parameters $\lambda$ in LASSO, which in turn makes the problem more ill-conditioned and computationally challenging.

For measurement matrices, we use: 1) Gaussian matrix $A_G$ with components sampled from i.i.d. Gaussian as described in Sec.~\ref{sec:4A}. To demonstrate the general applicability of ASM, we supplement additional tests with the measurement matrices: 2) Row-orthogonal (R-O) matrix $A_O$ with rows sampled from an $N \times N$ orthogonalized i.i.d. Gaussian matrix. 3) An $N \times N$ Toeplitz matrix with elements decaying at correlation coefficient $\rho=0.97$; 4) Partial discrete cosine transformation (P-DCT) matrix with frequency sampling from an exponential distribution $\exp(-\gamma f/f_{\max})$ where $\gamma = 0.2$ controls the sampling concentration; 5) Bernoulli matrix with each element randomly drawn from $\{\pm1\}$ with equal probability.

\subsubsection{Algorithm Settings}

We compare ASM-L1 \textcolor{black}{and its variant ASM-L1-LR (where `LR' stands for low-rank updates) introduced in Sec.~\ref{sec:practical_strategies_for_asm}}, with VAMP~\cite{VAMP}, ADMM as defined in~\eqref{eq:DRS_ADMM}, and SSNAL~\cite{SSNAL}. Where applicable, we also include Diag-VAMP, a variant of VAMP where the covariance matrix is determined by diagonal moment matching. We select SSNAL as a benchmark, since it represents the state-of-the-art for the LASSO and has been shown to outperform leading alternatives including FPC\_AS~\cite{FPC_AS}, mfIPM~\cite{mfIPM} and LADMM~\cite{LADMM}. For ASM-L1, we fix the averaging factor at $d = 1/2$ and employ Strategy 2 from Sec.~\ref{sec:ave_strategies} for parameter adjustment. The parameters in~\eqref{eq:rho_increase} are set to $s = 4$, $c = 0.5$, and $\rho_0 = 0.7$.

For accuracy evaluation, we employ the relative KKT residual by terminating the iteration when $\mathrm{Res}_{\mathrm{KKT}}(x^k)$ falls below $10^{-6}$ or when the number of iterations reaches $10^4$ (extended to $10^5$ for the more challenging case SNR=50dB).

\subsubsection{Experiments in Different Settings}

\begin{figure}[t]
\centering
\includegraphics[width=0.45\textwidth]{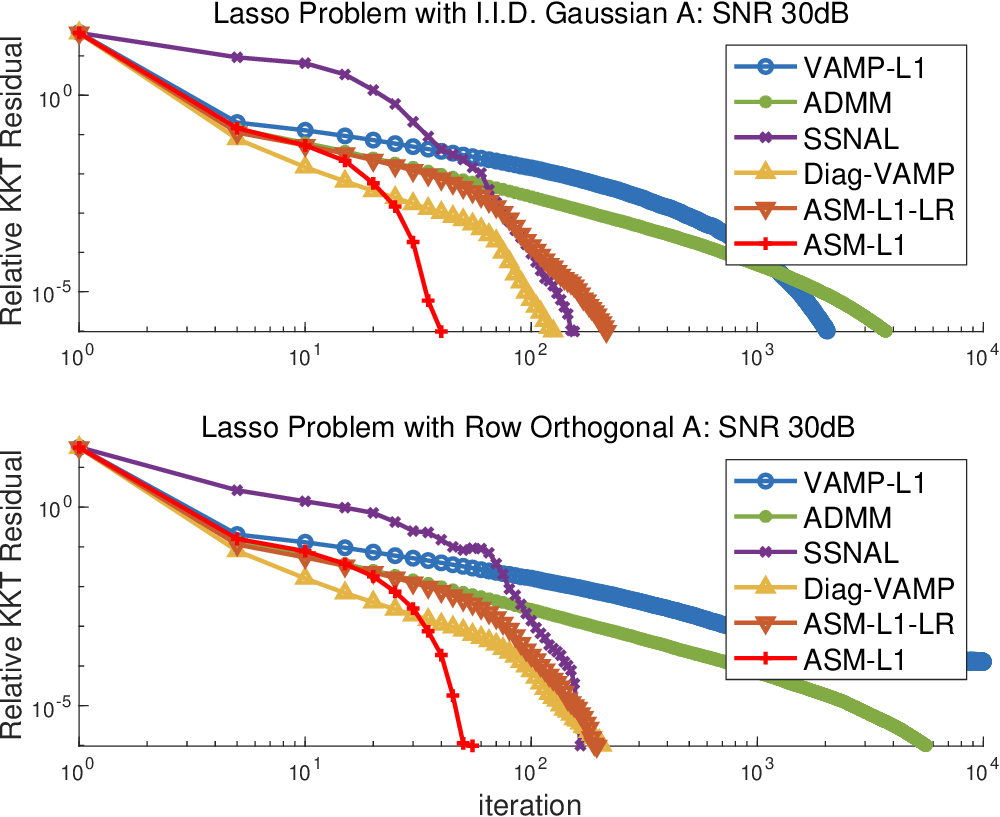}
\caption{Median relative KKT residual versus iteration count across 200 realizations for different measurement matrices. Here, $A \in \mathbb{R}^{200 \times 400}$ is I.I.D. Gaussian (top) or row-orthogonal (bottom) and SNR $= 30$ dB.}\label{fig:GaussOrth}
\end{figure}

\begin{figure}
\centering
\includegraphics[width=0.45\textwidth]{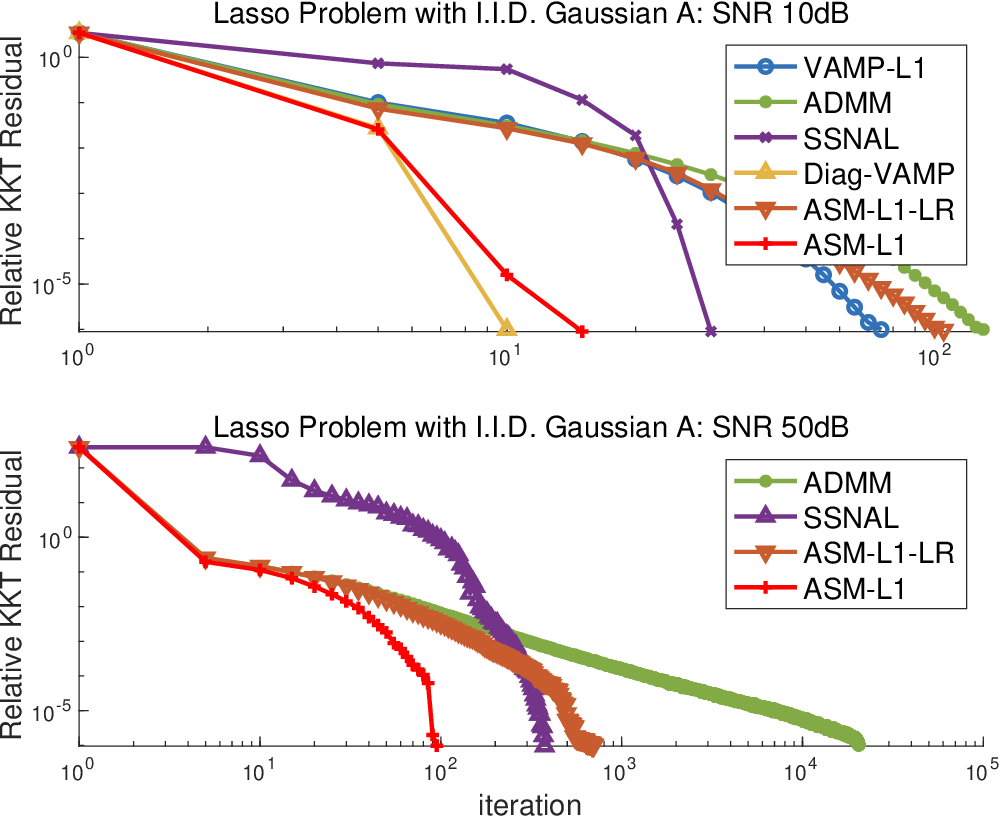}
\caption{The median relative KKT residual versus iteration over 200 realizations for different SNR. $A\in\mathbb R^{200\times 400}$ is I.I.D. Gaussian. SNR is $10$dB (Top) or $50$dB (Bottom).}\label{fig:GaussdB}
\end{figure}

\begin{figure}
\centering
\includegraphics[width=0.45\textwidth]{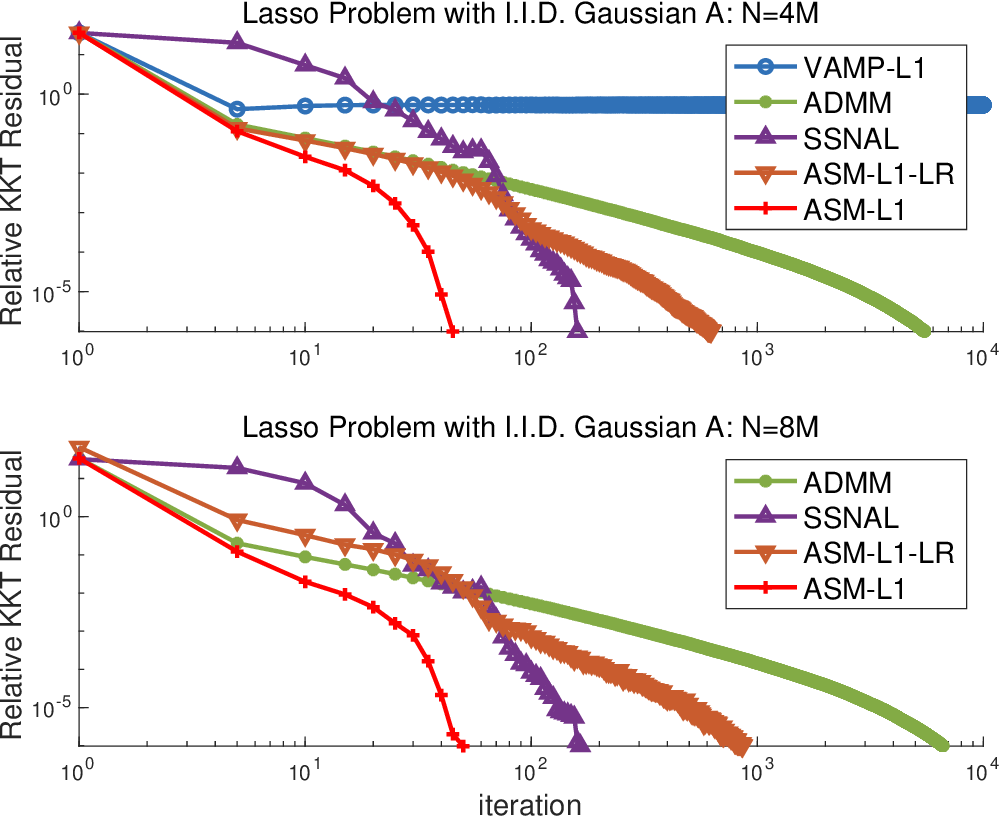}
\caption{The median relative KKT residual versus iteration over 200 realizations for different signal scales. I.I.D. Gaussian $A$ satisfies that $A\in\mathbb R^{200\times 800}$ (Top) or $A\in\mathbb R^{200\times 1600}$ (Bottom) and SNR =$30$dB.}\label{fig:GaussScale} 
\end{figure}

For clarity, we only present the convergence history for \eqref{problem:LASSO} with matrices $A_G$ (Figs.~\ref{fig:GaussOrth}, \ref{fig:GaussdB} and \ref{fig:GaussScale}) and $A_O$ (Fig.~\ref{fig:GaussOrth}). The convergence behavior for other measurement matrices are similar, and we only present their computational cost in Table~\ref{tab:time}. In Fig.~\ref{fig:GaussOrth}, each data point represents the minimum residual recorded per 5 iterations. It demonstrates that while all ADMM-like methods converge rapidly to a solution of moderate accuracy, only ASM and Diag-VAMP maintain acceleration to achieve high precision ($10^{-6}$). In contrast, SSNAL exhibits less efficiency in the early stages compared to ADMM-like methods.

\begin{table}[t!]
\caption{Median CPU time (s) over 200 realizations of the LASSO Problem in different settings.}
\label{tab:time}
\centering
\begin{tabular}{@{}>{\centering\arraybackslash}p{1.5cm} *{5}{c}@{}}
\toprule
Settings & VAMP & ADMM & SSNAL & ASM-L1-LR & ASM-L1 \\
\midrule
G.10dB          & \textbf{0.00369} & \textcolor{black}{0.00531} & 0.0188  & \textcolor{black}{0.0141}  & 0.00617 \\
G.30dB          & 0.0445           & \textcolor{black}{0.0740}  & 0.108   &  \textcolor{black}{0.0308}  & \textbf{0.0174} \\
G.50dB          & $\backslash$     & \textcolor{black}{0.395}& 0.259   & \textcolor{black}{0.0603} & \textbf{0.0400} \\
G. 4M           & $0.428^*$        & \textcolor{black}{0.225}   & 0.144   & \textcolor{black}{0.0992} & \textbf{0.0253} \\
G. 8M           & $\backslash$     & \textcolor{black}{0.978}   & 0.207   & \textcolor{black}{0.158} & \textbf{0.0334} \\
R-O\ \textsuperscript{1} & $0.280^*$ & \textcolor{black}{0.0955}  & 0.146   &  \textcolor{black}{0.0274}  & \textbf{0.0263} \\
Toeplitz        & $0.193^*$        & \textcolor{black}{0.00834}  & 0.0168  & \textcolor{black}{0.0270} & \textbf{0.00448} \\
P-DCT           & 0.0587           & \textcolor{black}{0.121}  & 0.111   &  \textcolor{black}{0.0374}  & \textbf{0.0240} \\
Bernoulli       & 0.0422           & \textcolor{black}{0.0727}  & 0.127   &   \textcolor{black}{0.0296}  & \textbf{0.0176} \\
\bottomrule
\end{tabular}

\begin{tablenotes}
\footnotesize
\item[1] Matrix settings: G. stands for i.i.d. Gaussian, with SNR 10/30/50 dB or dimension $N=4M/8M$; R-O (row orthogonal), Toeplitz, P-DCT (partial DCT), and Bernoulli ($\pm 1$ elements) are all tested at SNR=30dB. The symbol `$\backslash$' indicates algorithm divergence, while the superscript `$*$' denotes cases where the maximum iteration limit was reached without achieving the target accuracy. The entries corresponding to the minimal CPU time are emphasized in boldface in each row. \textcolor{black}{Since Diag-VAMP only converges under the G.10dB, G.30dB, R-O, and P-DCT settings (with CPU times of 0.0129 s, 0.159 s, 0.290 s, and 0.860 s, respectively), it is omitted from the table.}
\end{tablenotes}
\end{table}

It is essential to compare the computational cost across different methods (Table~\ref{tab:time}). All algorithms discussed here involve solving RLS problems, which we address via Cholesky decomposition. Although ADMM and VAMP require large-scale RLS solutions, their efficiency can be improved through warm-start strategies in this specific scenario. To ensure a fair comparison, we warm-start VAMP by precomputing the eigenvalue decomposition of $A A^H$ and ADMM by precomputing $\left(I + v A A^T\right)^{-1}$. Combined with the Sherman-Morrison-Woodbury (SMW) identity, this warm-start reduces the cost of RLS in both methods to several matrix-vector multiplications per iteration. In contrast, Diag-VAMP requires a matrix inversion at each step due to its diagonal moment-matching procedure, incurring significant computational overhead. We present the median CPU time for different methods and different measurement matrices in Table~\ref{tab:time}. With the chosen stopping criterion, we may consider the tested methods achieve the same precision finally.  For fairness, the time spent computing the KKT residual is excluded. Table~\ref{tab:time} clearly shows the efficiency of ASM-L1 over the other methods in different parameter setups. 

\textcolor{black}{
Specifically, the aforementioned warm-start strategy is not applied to SSNAL, as its restricted linear system changes dynamically, and its convergence relies on a penalty parameter that asymptotically approaches infinity and is directly coupled into the matrix inverse. In contrast, because ASM's convergence rigorously holds for a fixed step size, ASM-L1-LR can efficiently update the required inverse once the active subspace stabilizes.
}

\textcolor{black}{
Additionally, to ensure a fair comparison between ASM-L1-LR and ADMM, we assign both algorithms identical fixed parameters ($v = \hat{v} = \sigma^{-2}$). While this heuristic setting may not be theoretically optimal, it performs well empirically and is easy to implement. Under this setting, the initially identical performance and ASM's asymptotic acceleration become clearly evident. The only exception is the G.8M setting, where we increase the parameters for ASM-L1-LR to $v = \hat{v} = 4\sigma^{-2}$. We observe that a larger step size exerts a more profound impact in large-scale scenarios; thus, strictly enforcing identical parameters would preclude a fair assessment of ASM's potential (for context, applying this parameter to ADMM increases its CPU time to 1.36 s).
}

When the problem scale is small, ASM's superiority stems primarily from its fast convergence. This is particularly evident in ill-conditioned cases (e.g., SNR = 50 dB case in Fig.~\ref{fig:GaussdB}), where ASM converges within a hundred iterations while ADMM requires tens of thousands of iterations. Paradoxically, in simpler problems, the subspace restriction may render ASM less efficient. \textcolor{black}{In such cases, convergence occurs before adaptive tuning or low-rank updates can yield computational gains.} Consequently, ADMM may outperform ASM (e.g., SNR = 10 dB case in Fig.~\ref{fig:GaussdB}) despite requiring far more iterations. However, as the problem scale increases, the computational cost of ADMM grows dramatically, even though the number of iterations remains relatively stable (compare Figs.~\ref{fig:GaussOrth} and \ref{fig:GaussScale}). In contrast, ASM is much less affected due to the efficiency gained through subspace strategy.

\begin{figure}[t]
\centering
\includegraphics[width=0.45\textwidth]{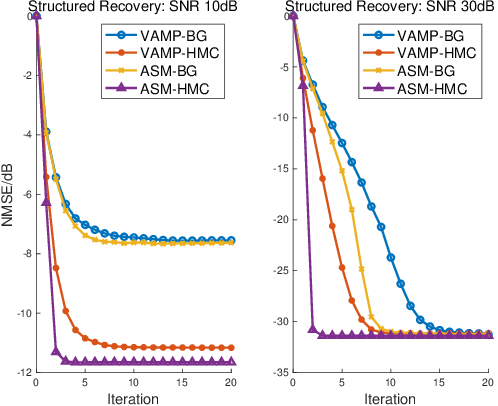}
\caption{Median NMSE versus iteration count across 200 realizations for different SNR settings: SNR = 10 dB (left) and SNR = 30 dB (right). The channel $h$ is simulated using~\eqref{eq:HMC_pri} with $p_{01} = 1/750$, $p_{10} = 1/250$, and $\sigma_0^2 = (p_{01} + p_{10}) / p_{01}$. Here, $M = 200$ and $N = 400$.}\label{fig:DFTdB}
\end{figure}

\subsection{Channel Estimation}

We evaluate the ASM-BG and ASM-HMC introduced in Sec.~\ref{sec:ASM_Structured} in the channel estimation problem. It is noteworthy that both $x^\dagger$, $w$, and $A$ are complex under this setting. What we need is simply to change $A^T$ to the Hermitian transpose $A^H$, take modulo instead of absolute value, and use $\mathcal{CN}(0,\sigma_0^2)$ in the Bernoulli-Gaussian prior \eqref{eq:BG_pri}.

\subsubsection{System Settings}
We consider a downlink massive MIMO channel system with one Base Station (BS) and a single user. The BS is equipped with a uniform linear array of $N$ antennas, and the user terminal has a single antenna. Given $M$ training sequences $\{z_t\}_{t=1}^M\ (z_t\in\mathbb C^{N})$, the user receives them as $y\ (y\in\mathbb C^M)$ through the channel $\tilde h\in\mathbb C^{N}$. Contaminated by noise $w\sim\mathcal{CN}(0,\sigma^2I_N)$, we have $y=Z^H \tilde h+w$, where $Z:=(z_1,\ldots,z_M)$.

Channel estimation seeks to recover $\tilde h$ by leveraging its sparsity in a transform domain. To be precise, if we use $F$ to denote the Discrete Fourier Transform (DFT) matrix and define $h:=F^H\tilde h$, then we can use model~\eqref{problem:LinearInverse} with $A:=Z^HF,\ x^\dagger:=h$ to reconstruct $\tilde h$. Since $h$ lies in the angular domain~\cite{booktse} and has non-zero elements clustered due to physical constraints, it is proper to further assume group sparsity. Specifically, we take $Z^H:=SFPF$ \cite{STCS} with $S$ being a selection matrix and $P$ being a random permutation matrix. We generate testing channels via the HMC model in \eqref{eq:HMC_pri} or by the WINNER2 model \cite{Winner2} using the default settings.

\subsubsection{Algorithm Settings}

Besides ASM-BG and ASM-HMC, we also implement the Turbo-CS from \cite{TCS} (denoted as VAMP-BG) and the Structured Turbo CS (denoted as VAMP-HMC) from \cite{STCS}. For ASM, we update $\hat v^k$ and $v^k$ according to \eqref{eq:vmix_both1} and set $\alpha=0.5$. The averaging strategy is applied not only to ASM but also to VAMP to mitigate variance degeneration. \textcolor{black}{To ensure stability within each VAMP iteration, we damp the potentially unstable denoiser output using the reliable RLS estimate.}

\begin{figure}[t]
\centering
\includegraphics[width=0.45\textwidth]{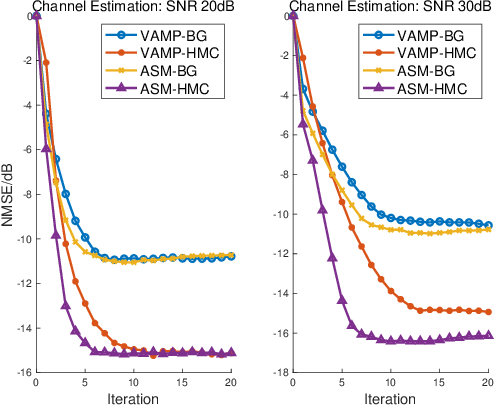}
\caption{Median NMSE versus iteration over 200 realizations for different SNR settings. SNR = 20 dB (Left) and SNR = 30 dB (Right). Channel $h$ is simulated using the WINNER2 model \cite{Winner2} with $M=50$ and $N=128$.}\label{fig:CEdB} 
\end{figure}

\subsubsection{Experiments with Different Settings}

When $h$ is simulated using \eqref{eq:HMC_pri}, the results are shown in Fig.~\ref{fig:DFTdB}. Since the prior of ASM-HMC coincides with the data distribution, it approaches the accuracy limit within two iterations. This suggests that a stronger denoiser may lead to further acceleration. For $h$ generated by the WINNER2 model (Fig.~\ref{fig:CEdB}), ASM-HMC can still reach the accuracy limit within six iterations.

Different from ASM-L1, the parameters $v^k$ and $\hat v^k$ in ASM-BG are not only relevant to the convergence rate but also to the final accuracy. Although we adopt averaging strategies and add the relaxation factor $\alpha$ to mitigate the degeneration, there is no guarantee of reaching the accuracy limit.

\subsection{Dynamic Compressed Sensing}

In this section, we use a time-division duplex (TDD) wireless system to illustrate that, when processing sequential data, historical information can provide an effective initial guess, allowing ASM to fully leverage its fast local convergence. For such problems, the Partial Laplacian dynamic compressed sensing (PLAY-CS) framework~\cite{PLAY-CS, RM-BPDN} employs the minimizer of \eqref{eq:LossPLAY} at each time interval, explicitly incorporating both historical channel gains and support sets as priors:
\begin{align}
    & x^{t|t-1} =\Psi_t(x^{t-1}), \label{eq:StateTrans}\\
%\begin{aligned}
    x^{t}=\mathop{\arg\min}_x&\frac12\|y_t-A_tx\|^2+\lambda_1\Big\|x_{S_{t|t-1}^c}-x^{t|t-1}_{S_{t|t-1}^c}\Big\|_1\nonumber \\ 
    &+\lambda_2\|x_{S_{t|t-1}}-x^{t|t-1}_{S_{t|t-1}}\|^2, \label{eq:LossPLAY}
%\end{aligned}
\end{align}
where $\Psi_t(\cdot)$ is the state transition function, $y_t$ and $A_t$ constitute the observation system at the $t$-th symbol, and $S_{t|t-1}$ is the support of $x^{t|t-1}$.

\subsubsection{System Settings}

Specific to the TDD system, we consider a single-input single-output configuration with orthogonal frequency-division multiplexing (OFDM) employing $30$\,kHz subcarrier spacing. The base station samples every 8th subcarrier across a total bandwidth of $408$ subcarriers in the frequency domain---termed the full band for reconstruction. During tracking, the BS receives training signals at 5\,ms intervals (every 10 time slots), with pilot symbols occupying $1/k$ ($k\in\mathbb{Z}_+$) of the full band at randomized locations. This yields the observation model $\tilde y_t=P_t\varPhi_t Fx^t+\tilde w_t$, where $P_t:=\mathrm{diag}(p_1^t,\ldots,p_M^t)$ is the pilot matrix satisfying $P^HP=I$, $\varPhi_t\in\mathbb C^{M\times kM}$ is the $0$-$1$ selection matrix, and $F\in\mathbb C^{kM\times N}$ is the super-resolution matrix mapping the delay domain to the full band.

Letting $y_t:=P_t^H\tilde y_t$ and $A_t:=\varPhi_t F$, we can apply \eqref{eq:LossPLAY} for TDD channel tracking. One remaining challenge arises when considering the channel prediction in \eqref{eq:StateTrans}, due to the presence of the unknown Doppler phase rotation \cite{Doppler, 2Stage}. Specifically, $\Psi_t(\cdot)$ in \eqref{eq:StateTrans} takes the form $\Psi_t(x^{t-1})=e^{\mathrm i \boldsymbol \varphi^{t-1}}\odot x^{t-1}$, where $\boldsymbol{\varphi}^t := (\varphi_1^t, \ldots, \varphi_N^t)$ denotes the Doppler phase shift. We estimate this shift via exponential smoothing, leveraging its slow variation:
\begin{equation}
    \hat{\boldsymbol \varphi}^t:=\angle(x^t\odot \overline{x^{t-1}}),\ \ \boldsymbol \varphi^t:= \alpha \hat{\boldsymbol \varphi}^t+(1-\alpha)\boldsymbol \varphi^{t-1}\label{eq:Doppler}
\end{equation}
with $\alpha=0.1$ and $\boldsymbol{\varphi}^0=0$. The tracking procedure requires an initial full-band observation to obtain $x^0$ and $N_S := |S^0_{1|0}|$. Contrary to the thresholding operation in~\cite{PLAY-CS}, which retains components based on energy thresholds, we instead preserve exactly $N_S$ highest-energy components of $x^t$ to enforce sparsity. This approach presumes limited channel variation; hence, all experiments track only 20 symbol intervals (100\,ms duration).

\subsubsection{Algorithm Settings}

Applying PLAY-CS means executing \eqref{eq:StateTrans}, \eqref{eq:LossPLAY}, and \eqref{eq:Doppler} iteratively. Among them, \eqref{eq:LossPLAY} involves sparse recovery on $S_{t|t-1}^c$. We test ASM, ADMM, and SSNAL on it. Additionally, we test the Accelerated Proximal Gradient method (APG) \cite{APG} and the Chambolle-Pock method~\cite{C-P}. APG is known as FISTA for the LASSO problem, and the Chambolle-Pock method is a primal-dual hybrid gradient method, which is the optimization analog of AMP \cite{AMP1}.

In practice, we let $y_t' := y_t - A_t x^{t|t-1}$ and solve the following problem with $x^t$ updated as $x^t := x^{t|t-1} + \delta x^t$:
\begin{equation}
\begin{aligned}
    \delta x^t&=\mathop{\arg\min}_{x'}\frac12\|y_t'-A_tx'\|^2+\lambda_1\|x_{S_{t|t-1}^c}'\|_1+\lambda_2\|x_{S_{t|t-1}}'\|^2.\label{eq:LossPLAY2}
\end{aligned}
\end{equation}
All methods are evaluated under 4 times super-resolution with a partial DFT matrix $F$. All algorithms are implemented similarly to the LASSO case, but replace the soft-thresholding operator $\mS_{\lambda}(\cdot)$ in \eqref{eq:STvOP} with a mixed operator. Specifically, ASM retains the update framework of Algorithm~\ref{algo:ASframework}, with operator $\mathcal{D}$ replaced by $\mathcal{S}'_{v\lambda_1,v\lambda_2}$ satisfying $\mathcal S'_{v\lambda_1,v\lambda_2}(z_i)=z_i/(1+v\lambda_2),$ for $i \in S_{t|t-1}$ and $\mathcal S'_{v\lambda_1,v\lambda_2}(z_i)=\mS_{v\lambda_1}(z_i),$ for $i \in S_{t|t-1}^c$.

We evaluate algorithm performance under 3GPP Clustered Delay Line-C (CDL-C) channel conditions with 2\,km/h terminal velocity, generated using MATLAB 5G Toolbox. For each observation, the algorithms iterate until either the modified KKT residual $\mathrm{Res}_\mathrm{KKT}'$ (computed via \eqref{eq:ResKKT} with $\mathcal{S}_{\lambda}$ replaced by $\mathcal{S}'_{\lambda_1,\lambda_2}$) falls below $10^{-3}$ or the iteration count reaches $10^3$. We set the regularization parameters as $\lambda_1=\sqrt{2\sigma_t^2\cdot N\log(N)/M}$ and $\lambda_2=0.1\lambda_1$, where $\sigma_t^2$ denotes the noise power.

\begin{table*}[t!]
\centering
\caption{Comparison of computational time (seconds) and time per unit precision (TPUP, in parentheses) across different algorithms under varying SNR conditions and sampling rates for dynamic channel tracking.}
\label{tab:time2}
\begin{tabular}{@{}>{\centering\arraybackslash}p{2cm} *{5}{c}@{}}
\toprule
\multicolumn{1}{c}{Settings} & \multicolumn{1}{c}{APG} & \multicolumn{1}{c}{Chambolle-Pock} & \multicolumn{1}{c}{ADMM} & \multicolumn{1}{c}{SSNAL} & \multicolumn{1}{c}{ASM} \\
\midrule
SNR = 20 dB & 0.402 (2.53$\times$10$^{-3}$) & 0.130 (8.20$\times$10$^{-4}$) & 0.165 (1.09$\times$10$^{-3}$) & 0.0619 (3.94$\times$10$^{-4}$) & \textbf{0.0329 (2.06$\times$10$^{-4}$)} \\
SNR = 25 dB & 0.763 (2.67$\times$10$^{-3}$) & 0.135 (4.91$\times$10$^{-4}$) & 0.219 (7.92$\times$10$^{-4}$) & 0.0931 (3.38$\times$10$^{-4}$) & \textbf{0.0347 (1.23$\times$10$^{-4}$)} \\
SNR = 30 dB & 0.839 (1.84$\times$10$^{-3}$) & 0.164 (3.55$\times$10$^{-4}$) & 0.287 (6.35$\times$10$^{-4}$) & 0.106 (2.37$\times$10$^{-4}$) & \textbf{0.0366 (8.15$\times$10$^{-5}$)} \\
SNR = 35 dB & 1.20 (2.16$\times$10$^{-3}$) & 0.222 (4.17$\times$10$^{-4}$) & 0.383 (7.56$\times$10$^{-4}$) & 0.132 (2.47$\times$10$^{-4}$) & \textbf{0.0433 (7.96$\times$10$^{-5}$)} \\
SNR = 40 dB & 1.63 (3.42$\times$10$^{-3}$) & 0.330 (6.80$\times$10$^{-4}$) & 0.539 (1.09$\times$10$^{-3}$) & 0.184 (3.86$\times$10$^{-4}$) & \textbf{0.0473 (9.84$\times$10$^{-5}$)} \\
\midrule
N = 6M & 0.856 (7.38$\times$10$^{-4}$) & 0.170 (1.51$\times$10$^{-4}$) & 0.276 (2.45$\times$10$^{-4}$) & 0.123 (1.13$\times$10$^{-4}$) & \textbf{0.0484 (4.25$\times$10$^{-5}$)} \\
N = 8M & 0.811 (1.01$\times$10$^{-3}$) & 0.165 (2.10$\times$10$^{-4}$) & 0.278 (3.60$\times$10$^{-4}$) & 0.113 (1.51$\times$10$^{-4}$) & \textbf{0.0428 (5.51$\times$10$^{-5}$)} \\
N = 17M & 0.813 (3.35$\times$10$^{-3}$) & 0.167 (6.68$\times$10$^{-4}$) & 0.298 (1.22$\times$10$^{-3}$) & 0.0973 (4.06$\times$10$^{-4}$) & \textbf{0.0327 (1.39$\times$10$^{-4}$)} \\
\bottomrule
\end{tabular}

\begin{tablenotes}
\footnotesize
\item The first five rows present results under varying SNR conditions with fixed sampling rate (N = 12M). The subsequent three rows show results under varying sampling rates with fixed SNR (30 dB). The baseline configuration (SNR = 30 dB, N = 12M) is included in the SNR variation experiment. The entries corresponding to the minimal CPU time and the minimal TPUP are emphasized in boldface in each row.
\end{tablenotes}
\end{table*}

\subsubsection{Experiments with Different Settings}

The computational cost comparison is summarized in Tables~\ref{tab:time2}, reporting the median CPU time and the time per unit precision (Time/NMSE$^{-1}$). The processing of the RLS step is the same as in Sec.~\ref{sec:LASSO_problem}. As the local convergence behavior of SSNAL is independent of line search, while it requires significant computational effort, we remove it in experiments for fairness.

By virtue of the quasi-static nature of the delay taps, the support set does not vary significantly across adjacent symbol intervals. This leads to exceptional sparsity in $\delta x^t_{S_{t|t-1}^c}$ and makes $x^{t|t-1}$ an effective initial guess for~\eqref{eq:LossPLAY}, enabling ASM to leverage its fast local convergence. This efficiency advantage becomes particularly pronounced in ill-conditioned scenarios (SNR = 40\,dB, $k=12$), where ASM achieves an 11.4$\times$ speed-up over ADMM and a 3.89$\times$ speed-up over SSNAL. In simpler cases (SNR = 20\,dB, $k=12$), ASM also maintains 5.02$\times$ and 1.88$\times$ speed-ups, respectively.

\section{Conclusion}

We proposed the Alternating Subspace Method (ASM) which accelerates ADMM-like methods by consistently restricting the data fidelity step to the support set subspace. Numerical experiments on the LASSO problem demonstrate that ASM retains the rapid initial convergence of ADMM while effectively achieving high accuracy. 
The subspace fidelity strategy significantly reduces computational cost, enhancing the efficiency of ASM for large-scale problems. Furthermore, its compatibility with general denoisers was confirmed through its application to channel estimation problems. \textcolor{black}{We also established the global convergence analysis of ASM for the LASSO problem.} Future topics include the development of adaptive parameter strategies with theoretical guarantees and studies on the convergence radius and step size, etc.

\appendices

\color{black}

\section{Proof of Theorem~\ref{thm:Local_geometric_convergence}}\label{app:Local_geometric_convergence}

\textcolor{black}{
We first verify that $\|\cdot\|_{\mE^*}$ is a valid norm. Only the positive definiteness deserves checking. $\|x\|_{\mE^*} = 0$ yields $\|x_{(\mE^*)^c}\| = 0$ and $\|(I - v\hat{A}^T\hat{A})x|_{\mE^*}\| = 0$, where $\hat{A} := A|_{\mE^*}$. Given the non-singularity assumption on $I - v\hat{A}^T\hat{A}$, we have $x = 0$. This proof applies to $\|\cdot\|_{\mE^*}$ without change. Hereafter, we generalize the RLS parameter to $\hat{v}$ (cf. Sec.~\ref{sec:averaging_and_the_proximal_residual_control}).}

\textcolor{black}{When $\mE^* \subseteq E^k$, the ASM acts as a fixed-point iteration on $x^*$. Below we set $\hat{A}:=A|_{E^k}$, $\mu^*:=\mu(x^*)$, and this rule applies to variables with similar notations. Let $H^k:=(I + \hat{v}\hat{A}^T\hat{A})^{-1}(I - v\hat{A}^T\hat{A})$. By uniformly defining $\hat{\mu} := \mu|_{E^k}$ and $\hat{\nu} := (1 + \hat{v}/v)z - (\hat{v}/v)\hat{\mu}$ for both the optimum ($z^* := \mathcal{S}_{v\lambda}(\hat{\mu}^*)$) and the $k$th iteration ($z^k := \hat{z}^k$), we explicitly obtain $\hat{\mu}^* = H^k \hat{\nu}^*$ alongside the auxiliary variables $\hat{\mu}^k$ and $\hat{\nu}^k$.
}

\begin{lemma}\label{lemma:Non-expan_mG}
    If $\mE^*\subseteq E^k$ and $\mu_i^k\cdot\mu_i^*>0$ for $i\in E^k$, then: 
    (i) $\|\hat{\nu}^k-\hat{\nu}^*\| \le \|\hat{\mu}^k-\hat{\mu}^*\|$ if $\mE^*= E^k$ or $v=\hat v$; 
    (ii) $\|\hat{\mu}^{k+1}-\hat{\mu}^*\| \leq \tilde C^k\|\hat{\nu}^k-\hat{\nu}^*\|$, where $\tilde C^k:=d^k \|H^k\|+(1-d^k)$.
\end{lemma}
\begin{IEEEproof}
    See SM \ref{SM:proof_of_lemma_ref_lemma_non_expan_mg}.
\end{IEEEproof}

\textcolor{black}{
Since $x_\text{ave}^{k+1}|_{(E^k)^c}=(1-d^k)x_\text{ave}^{k}|_{(E^k)^c}$, $x^*|_{(E^k)^c}=0$, and $\|(x_\text{ave}^{k}-x^*)|_{E^k}-v\hat{A}^TA(x_\text{ave}^{k}-x^*)\|=\|\hat{\mu}^k-\hat{\mu}^*\|$, we have
\begin{equation}
\begin{aligned}
    &\|\hat \mu^{k+1}-\hat\mu^*\|+\|(x^{k+1}_\text{ave}-x^*)|_{(E^k)^c}\| \\
    \leq& \tilde C^k\|\hat \mu^{k}-\hat\mu^*\|+(1-d^k)\|(x^{k}_\text{ave}-x^*)|_{(E^k)^c}\|
\end{aligned}
\end{equation}
leading to $\|x^{k+1}_\text{ave}-x^*\|_{E^k}\leq \tilde C^k\|x^{k}_\text{ave}-x^*\|_{E^k}$.
}

\textcolor{black}{
Finally, since $x_{\text{ave}}^k \to x^*$ and the mapping $\mu(\cdot)$ is Lipschitz continuous, there exists an integer $k_0 > 0$ such that for all $k \geq k_0$, $\|\mu(x_{\text{ave}}^k) - \mu^*\| \leq \min_i \big| |\mu_i^*| - v\lambda \big| / 2$. This guarantees the exact identification of the optimal support, i.e., $E^k = E^{k+1} = \mE^*$, and $\mu_i^k\cdot\mu_i^*>0$ for $i\in \mE^k$. Consequently, the aforementioned contraction relationship strictly holds for all $k \geq k_0$ under the fixed support $\mE^*$, completing the proof.
}

\section{Proof of Proposition~\ref{prop:ADMM}}\label{app:prop_ADMM}

Before proceeding, we establish a few algebraic properties.
\begin{lemma}\label{lem:algebraic_facts}
    For any $a,b\in\mathbb R$, the following properties hold:
    \begin{enumerate}
        \item $(a-b)^2=a^2-(2a-b)b$;
        \item If $ab\geq 0$ and $2|a|\geq|b|$ (or $b=0$), then $|a-b|\leq |a|$;
        \item For any $c>0$, if $b_1=\mathcal P_{[-c,c]}(a_1)$ and $b_2=\mathcal P_{[-c,c]}(a_2)$, then $(a_1-a_2)(b_1-b_2)\geq 0$ and $|b_1-b_2| \leq |a_1-a_2|$.
    \end{enumerate}
\end{lemma}

Properties 1) and 2) follow from direct expansion, while 3) is a standard monotonic and non-expansive property of the projection operator (see, e.g., Lemma 3.1 in~\cite{2008-Elaine}). 

\subsection{The Non-expansiveness of $\|R^k\|$}\label{sub:Properties_ADMM}

Define the scaled variables $\mu^k_\gamma:=\gamma\mu^k$ and $p^k_\lambda:=v\lambda p^k$, where $\gamma:=v/(\hat v+v)$. Applying Lemma~\ref{lem:algebraic_facts}-1) yields:
\begin{equation}
    \| \Delta \mu^k_\gamma- \Delta p^k_\lambda\|^2 = \| \Delta \mu^k_\gamma\|^2-(2\Delta \mu^k_\gamma- \Delta p^k_\lambda)^T\Delta p^k_\lambda.
    \label{eq:mu_p_descent_ADMM}
\end{equation}
Furthermore, using the fact that $p^k_\lambda=\mathcal P_{[-v\lambda,v\lambda]}(\mu^k)$, the vectors $\Delta \mu^k_\gamma$ and $\Delta p^k_\lambda$ satisfy the conditions of Lemma~\ref{lem:algebraic_facts}-2) coordinately when $v\geq \hat v$ (which ensures $2\gamma\geq 1$). Consequently, $\| \Delta \mu^k_\gamma- \Delta p^k_\lambda\|\leq \| \Delta \mu^k_\gamma\|$. By reorganizing the iterations of ASM-L1 in~\eqref{eq:core_ASM} in terms of $R^k$, $\mu^k$, and $p^k$, the following relations can be verified (detailed in SM~\ref{sub:the_identities_in_appendix_admm}):
%\begin{enumerate}
\begin{equation}\label{Eq:Non-Expansive-Rk-Relations}
\begin{aligned}
& 1)\ R^{k+1}-(1-d)R^k=\Delta \mu^k_\gamma-\Delta p^{k}_\lambda,\\
& 2)\ \Delta\mu^k_\gamma=d\cdot (I+\hat vA^TA)^{-1}(I-vA^TA)R^k. 
\end{aligned}
\end{equation}
%\end{enumerate}
Sequentially applying the relations above, we have
\begin{equation}
\begin{aligned}
    &\|R^{k+1}-(1-d)R^{k}\|=\|\Delta \mu^k_\gamma-\Delta p^{k}_\lambda\|\leq \|\Delta \mu^k_\gamma\| \\
    \leq&  d \|(I+\hat vA^TA)^{-1}(I-vA^TA)\|_2\cdot \|R^k\|\leq d\|R^k\|\label{eq:Ineq_ADMM1}
\end{aligned}
\end{equation}
where $\|(I+\hat vA^TA)^{-1}(I-vA^TA)\|_2\leq 1$ since $v=\hat v$ (via the diagonalization technique). Then by the triangle inequality and \eqref{eq:Ineq_ADMM1} we obtain
\begin{equation*}
\|R^{k+1}\|\leq (1-d)\|R^{k}\|+\|R^{k+1}-(1-d)R^{k}\|\leq \|R^k\|.
\end{equation*}

\subsection{Strict Descent of the Proximal Residual}
\label{sub:Plateau_Conditions_ADMM}

Denote $x=x^{(0)}$ and $x'=x^{(1)}$ for ease of notation.
\begin{defi}[Plateau Conditions for ADMM]\label{def:plateau_conditions_ADMM}
    Given $x\in\mathbb R^N$, define $S_0:=\{i:v\lambda[p_i(x)-p_i(x')]=\mu_i(x)-\mu_i(x')\}$ and $S_1:=\{i\in S_0^c:p_i(x)-p_i(x')=0\}$. We say $x$ satisfies the Plateau Conditions if all the following hold: (i) $AR(x)=0$; (ii) $(S_1\cup S_0)^c=\emptyset$; (iii) $R(x)|_{S_0}=0$; (iv) $|S_1|> M$.
\end{defi}
%By examining the conditions for the equality in \eqref{eq:Ineq_ADMM1} and \eqref{eq:Ineq_ADMM2}, we establish the following result.
\begin{lemma}\label{lemma:Violation_plateau_implication}
Under generic conditions and the condition $\hat v=v$ (implying $2\gamma=1$), the violation of any of the  \textit{Plateau Conditions} implies either $R(x)=0$ or the existence of a constant $\delta_x>0$ such that $\|R(x')\|\leq \|R(x)\|-\delta_x$.    
\end{lemma} 
\begin{IEEEproof}
The proof is given in SM \ref{sub:violating_the_plateau_condition_for_admm}.
\end{IEEEproof}

Violating the Plateau Conditions directly yields Proposition~\ref{prop:ADMM} with $n_s=1$. Conversely, if all conditions hold, utilizing relations 1) and 2) from Appendix~\ref{sub:Properties_ADMM}, 
%alongside the definitions of $S_1$ and $S_0$, 
we have:
\begin{enumerate}
    \item $\mu(x')|_{S_1}=\mu(x)|_{S_1}+\gamma^{-1}d\cdot R(x)|_{S_1}$;
    \item $\mu(x')|_{S_0}=\mu(x)|_{S_0}$;
    \item $R(x')|_{S_1}=R(x)|_{S_1}$,\quad $R(x')|_{S_0}=0$.
\end{enumerate}

If $|\mu_i(x^{(2)})| \geq v\lambda$ continues to hold for all $i\in S_1$, then $x'$ also satisfies the Plateau Conditions, and the iterative relations above apply similarly to the transition from $x'$ to $x^{(2)}$. Therefore, for the ideal ADMM iteration starting from $x$, breaking the plateau stage can occur only on the set $S_{-}:=\{i\in S_1:R_i(x)\cdot \mu_i(x)<0 \}$. Indeed, we have:
\begin{claim}\label{claim:S_minus_non_empty}
    Under generic conditions, if $x$ satisfies the Plateau Conditions, then $S_{-}\neq \emptyset$.
\end{claim}

See SM~\ref{sec:proof_of_claim_s_minus_non_empty} for a proof. Consequently, we define
\begin{equation}
    \mathcal F(x, i):=\gamma(|\mu_{i}(x)|-v\lambda)/|d\cdot R_{i}(x)|
\end{equation}
and $i_s:=\arg\min_{i\in S_{-}}\mathcal F(x, i)$, $n_s:=\lfloor\mathcal F(x, i_s)\rfloor+1$. Then the Plateau Conditions hold for all $x^{(n)}$ ($n\leq n_s-1$). For $n=n_s$, $\mu_{i_s}(x^{(n_s)})=\mu_{i_s}(x)+n_s\gamma^{-1}d\cdot R_{i_s}(x)$ and $|n_s\gamma^{-1}d\cdot R_{i_s}(x)|>|\mu_{i_s}(x)|-v\lambda$. This implies $p_{i_s}(x^{(n_s)})-p_{i_s}(x^{(n_s-1)})\neq 0$, meaning there exists a constant $\delta_x>0$ such that $\|R(x^{(n_s)})\|\leq \|R(x^{(n_s-1)})\|-\delta_x=\|R(x)\|-\delta_x$ by~\eqref{eq:mu_p_descent_ADMM}.

%\section{Convergence of ASM}
\section{Proof of Theorem~\ref{thm:convergence_concern12}}
\label{sec:convergence_of_asm}

In this section, we establish a stronger version of the theorem by replacing the original conditions 1) and 2) with a generalized \textit{non-monotonic descent} condition:
\begin{equation}\label{eq:non-monotonic}
\begin{aligned}
    \|&R^{k+1}\|^2 \leq (1+a_k)\big[\|R^k\|^2 - c_0 d^k \delta^k_c \big], \\
    &\delta^k_c := \|R^{k+1}|_{(\mE^k)^c}\|^2 + \|R^k|_{(\mE^k)^c}\|^2,
\end{aligned}
\end{equation}
where $c_0 > 0$ is a predefined constant and $\{a_k\}$ satisfies $a_k \geq 0$ and $\sum_{j=1}^\infty a_j < \infty$. It is straightforward to verify that the original conditions 1) and 2) ensure the condition above by setting $c_0=1$ and $a_k \equiv 0$. As $d^k$ is presumed to satisfy $d^k \geq d_1$, we introduce a lumped constant $d_0 := d_1 c_0$, which simplifies the descent inequality to $\|R^{k+1}\|^2 \leq (1+a_k)[\|R^k\|^2 - d_0 \delta^k_c]$.

We first establish the uniform boundedness of the sequence $\{x^k_\text{ave}\}$, as formalized in the following lemma.
\begin{lemma}\label{lemma:ASM_bounded}
    Let $\{x^k_\text{ave}\}$ be generated by the ASM iteration with subspaces $\{\mE^k\}$ and averaging factors $\{d^k\}$. Suppose that $\mE^k_0 \subseteq \mE^k$ for all $k$, the matrix $I - v(A|_{\mE^k})^T A|_{\mE^k}$ is non-singular for each $\mE^k$, and $\|R^{k+1}\|^2 \leq (1+a_k)\|R^k\|^2$ for some sequence $a_k \geq 0$ satisfying $\sum a_k < \infty$. Then, there exist constants $C_\text{shrink}, C_\text{step} > 0$, independent of any specific choice of $\mE^k$ and $d^k$, such that for all $k$: 1) $\|x^{k+1}_\text{ave}\| \leq \|x^{k}_\text{ave}\| + C_\text{step}$; 2) whenever $\|x^k_\text{ave}\| \geq C_\text{shrink}$, it holds that $\|x^{k+1}_\text{ave}\| \leq \|x^{k}_\text{ave}\|$.
\end{lemma}
\begin{IEEEproof}
    The proof is given in SM \ref{SM:ASM_bounded}.
\end{IEEEproof}

Since the sequence $\{x_\text{ave}^k\}$ is uniformly bounded by $C_\text{step}+C_\text{shrink}$ and the residual mapping $R(\cdot)$ is continuous, there exists an accumulation point $x^*$ achieving $\|R^*\| = \liminf_{k \to \infty} \|R^k\|$, where $R^* := R(x^*)$. Our goal is to prove $R^* = 0$. For the subsequent local sequence analysis, we identify this limit point as the base state $x^{(0)} := x^*$. Because the algorithmic parameters $d^k$ and $\mE^k$ take values from finite sets, we can extract a subsequence $\{x_\text{ave}^{k_l}\}$ converging to $x^*$ along which these parameters remain strictly constant, naturally denoted by $d^{(0)}$ and $\mE^{(0)}$ respectively. Let $\mathcal T^{(0)}$ represent the single-step ASM operator associated with $d^{(0)}$ and $\mE^{(0)}$, which generates the ideal next iterate $x^{(1)} := \mathcal T^{(0)}(x^{(0)})$. Furthermore, since $\mu(\cdot)$ is continuous and the support condition $E^{k_l} \subseteq \mE^{k_l} \equiv \mE^{(0)}$ holds along the subsequence, it inherently follows that  $\{i : |\mu_i(x^{(0)})| > v\lambda\} \subseteq \mE^{(0)}$.

The subsequent proof closely parallels Proposition~\ref{prop:ADMM}. We will show that a non-optimal $x^*$ yields $\|R(x^{(n_s)})\| < \|R^*\|$ for some $n_s \geq 1$. Provided that these $n_s$ iterations constitute a continuous mapping, this strict decrease contradicts the limit inferior property of $R^*$. To establish this, we consider two main cases, just as in Proposition~\ref{prop:ADMM}, and introduce the following auxiliary definition:
\begin{defi}[Plateau Conditions]
    We denote $x' := x^{(1)}$, $S_0 := \{i\in\mE^{(0)} : 2\gamma(\mu_i(x')-\mu_i(x^*))=v\lambda (p(x'_i)-p(x^*_i))\}$ and $S_1:=S_0^c\cap \mE^{(0)}$. If $AR^*=0$, $[p(x')-p(x^*)]|_{S_1}=0$, $|S_1|>M$, and $R(x^*)|_{(\mE^{(0)})^c}=R(x')|_{(\mE^{(0)})^c}=0$, we say that the \textit{Plateau Conditions} hold.
\end{defi}
\begin{claim}\label{lemma:Ec_zero_delta_x}
    If $R^*\neq 0$, one of the following two cases must occurs:  (i) there exists a constant $\delta_x > 0$ such that $\|R(x')\| \leq \|R(x^*)\| - \delta_x$; or (ii) the \textit{Plateau Conditions} hold.
\end{claim}
\begin{IEEEproof}
    The proof is given in SM \ref{SM:proof_of_claim_ref_lemma_ec_zero_delta_x}.
\end{IEEEproof}

As the term \textit{Plateau Conditions} suggests, and analogous to Proposition~\ref{prop:ADMM}, the residual remains constant for several iterations before strictly decreasing at some step $n_s$. Before formalizing this in Claim~\ref{lemma:Plateau_delta}, we introduce a safe iteration horizon $n_0 := \min_{i\in S_{-}} \lfloor \gamma(|\mu_{i}(x^*)|-v\lambda)/|d_0 R_{i}^*| \rfloor$, where $S_{-}:=\{i\in S_1 : R_i^*\cdot \mu_i(x^*) < 0\}$ is non-empty when the \textit{Plateau Conditions} hold (cf. the proof of Proposition~\ref{prop:ADMM}). Because $\{\mE^k\}$ and $\{d^k\}$ are taken from finite sets, we can extract a further subsequence, still denoted by $\{x_\text{ave}^{k_l}\}$ for simplicity, along which the sequence of subspaces and averaging factors $\{\mE^{k_l+n}, d^{k_l+n}\}_{n=0}^{n_0}$ is identical for all $l$. We denote these invariant sequences as $\{\mE^{(n)}, d^{(n)}\}_{n=0}^{n_0}$, and let $\mathcal T^{(n)}$ be the single-step ASM operator associated with $(\mE^{(n)}, d^{(n)})$. The subsequent ideal iterates are then recursively defined as $x^{(n+1)} := \mathcal T^{(n)}(x^{(n)})$ starting from $x^{(0)} := x^*$. Similarly, for the actual subsequence, we define $x^{(n+1, l)} := \mathcal T^{(n)}(x^{(n, l)})$ initialized at $x^{(0, l)} := x^{k_l}$.
\begin{claim}\label{lemma:Plateau_delta}
    If $R^* \neq 0$, then under the \textit{Plateau Conditions}, there exist an index $n_s$ $(1 \leq n_s \leq n_0)$ and a constant $\delta_x > 0$ such that $\|R(x^{(n_s)})\| \leq \|R^*\| - \delta_x$.
\end{claim}
\begin{IEEEproof}
    The proof is given in SM \ref{SM:proof_of_claim_ref_lemma_ec_zero_delta_x}.
\end{IEEEproof}

Since the composite mapping $\mathcal{T}^{(n_s-1)}\circ\cdots\circ \mathcal{T}^{(0)}(\cdot)$ is continuous given that $\{\mE^{(n)}, d^{(n)}\}$ are strictly fixed for $n \leq n_0$, taking the limit over the subsequence $\{x_\text{ave}^{k_l}\}$ yields $\limsup_{l \to \infty} \|R(x_\text{ave}^{k_l+n_s})\| \leq \|R^*\| - \delta_x$, which contradicts the limit inferior property of $\|R^*\|$. Consequently, we must have $R^* = 0$. The global convergence of the entire sequence then directly follows from the lemma below:

\begin{lemma}\label{lemma:Rx_zero_x_converge}
    Let $\{u^k\}$ be any bounded sequence satisfying $\|R(u^{k+1})\|^2 \leq (1+a_k)\|R(u^k)\|^2$ for some sequence $a_k \geq 0$ such that $\sum a_k < \infty$. If there exists a subsequence $\{u^{k_l}\}$ such that $\lim_{l\to\infty} \|R(u^{k_l})\| = 0$, then $\{u^k\}$ converges to the unique LASSO optimum under the \textit{generic conditions}.
\end{lemma}
\begin{IEEEproof}
    The proof is given in SM \ref{SM:proof_of_lemma_ref_lemma_rx_zero_x_converge}.
\end{IEEEproof}

\section{Proof of Proposition \ref{prop:E_Ec_couple}} \label{sec:algebraic_relationships_between_asm_iterations}

For brevity, we introduce the following notations with respect to $\mE^k$: $\hat R^{k+1} := R^{k+1}|_{\mE^k}$, $\hat R^k := R^k|_{\mE^k}$, $\tilde R^k := R^k|_{(\mE^k)^c}$, $\hat x_\text{ave}^k := x^k_\text{ave}|_{\mE^k}$, $\hat \mu^k_\gamma := \gamma\mu^k|_{\mE^k}$, $\hat p^k_\lambda := v\lambda p^k|_{\mE^k}$, $\hat A := A|_{\mE^k}$, $\tilde A := A|_{(\mE^k)^c}$, and $\hat h(x) := -\hat A^T(y-Ax)$. 

The derivation of 1) parallels that of Eq.~\eqref{eq:R_mu_p_ADMM}. Despite the restriction to $\mE^k$, the fact $x^{k+1}|_{(\mE^k)^c} = 0$ (given $E^k \subseteq \mE^k$) naturally induces a final expression with no explicit out-of-subspace quantities. The detailed verifications are deferred to SM~\ref{SM:algebraic_relationships_between_asm_iterations}.
Similarly, following the rationale of Eqs.~\eqref{eq:mu_p_descent_ADMM} and \eqref{eq:Delta_mu_Rk_ADMM}, and using the fact that $\tilde R^k = -x_\text{ave}^k|_{(\mE^k)^c}$, we establish the following identities (cf. SM~\ref{SM:algebraic_relationships_between_asm_iterations}):
\begin{equation}
    \| \Delta\hat \mu^k_\gamma - \Delta\hat p^k_\lambda \|^2 = \| \Delta\hat \mu^k_\gamma\|^2 - \left(2\Delta\hat \mu^k_\gamma - \Delta\hat p^k_\lambda\right)^T\Delta\hat p^k_\lambda, \label{eq:mu_p_descent}
\end{equation}
\begin{equation}
    \Delta\hat \mu^k_\gamma = d^k (I+\hat v\hat A^T\hat A)^{-1}[(I-v\hat A^T\hat A)\hat R^k - v\hat A^T\tilde A\tilde R^k]. \label{eq:Delta_mu_Rk}    
\end{equation}

By invoking the properties established in Sec.~\ref{sub:Properties_ADMM} and using the relation in 1), we obtain the inequality $\|\hat R^{k+1}-(1-d^k)\hat R^k\| \leq \|\Delta\hat \mu^k_\gamma\|$. Defining $J^k = \hat v\big(\|R^k|_{\mathcal{E}^k}\|^2 - \|\Delta\hat \mu^k_\gamma /d^k\|^2\big)$ and substituting $\Delta\hat \mu^k_\gamma$ via Eq.~\eqref{eq:Delta_mu_Rk}, it is straightforward to verify that $J^k$ coincides with Eq.~\eqref{eq:J_k} (cf. SM~\ref{SM:algebraic_relationships_between_asm_iterations}), which concludes the proof of 2).

Furthermore, we remark that the triangle inequality yields
\begin{align*}
\|\hat R^{k+1}\| \leq (1-d^k)\|\hat R^{k}\| + d^k\big(\|\hat{R}^k|\|^2 - J^k/\hat v\big)^{1/2}
    %&\leq (1-d^k)\|\hat R^{k}\| + \|\hat R^{k+1}-(1-d^k)\hat R^k\| \\  
\end{align*}
from Property 2), and we obtain $\|\hat R^{k+1}\|^2 \leq \|\hat{R}^k\|^2 - (d^k/\hat v)J^k$ by Jensen's inequality.

\section{Convergence of ASM with $\mE^k=\mE^k_0$}\label{app:converge_Ek_Ek0}

\begin{claim}\label{claim:convergence_Ek_Ek0}
    If $v=\hat v\leq 4/(\max_{\mE}\|A|_{\mE}\|_2^2)$, then $\{R^k\}$ satisfies the \textit{non-monotonic descent condition} with $c_0=1/2$.
\end{claim}

See SM~\ref{sec:convergence_Ek_Ek0} for a proof. Since $\epsilon<\omega_0$ and $\mu^k\to\mu^*$ as $x^k\to x^*$, there is a $k_0$ such that for all $k\geq k_0$, $\|\mu^k-\mu^*\|_\infty\leq v\lambda|\omega_0-\epsilon|/2$ and $\|x^k_\text{ave}|_{(\mE^*)^c}\|_\infty\leq \varepsilon_0$, yielding $\mE^k_0=\mE^*$.

\section{Convergence of ASM under Safeguarding Rule with Shrinkage Checks}\label{sec:convergence_of_asm_with_shrinkage_checks}

\textbf{Scenario 1:} We first consider the scenario where the algorithm remains in the Initial Stage. Since the \textit{non-monotonic descent condition} holds throughout and simultaneously ensures, via Lemma~\ref{lemma:dk_guarantee}, that $d^k$ satisfies the conditions of Theorem~\ref{thm:convergence_concern12}, it follows that $x_{\text{ave}}^k \to x^*$. Given that $\mu(\cdot)$ is continuous and $\epsilon \leq \omega_0$, there exists a $k_0$ such that $\mE^k = E^k = \mE_0^k = \mE^*$ for all $k \geq k_0$. Furthermore, as $(x_{\text{ave}}^k)|_{(\mE^*)^c} \to 0$, we asymptotically obtain $\|(x_{\text{ave}}^k)|_{(\mE^*)^c}\|_\infty \leq \varepsilon_0/2$, and hence $\mE^k_{(1)} \subseteq \mE^*$.

\textbf{Scenario 2:} Alternatively, if the algorithm switches to the Safeguarding Stage at some iteration, we have:

\begin{claim}\label{claim:safe_shrink}
    Suppose that (i) $\epsilon \leq \omega_0$, (ii) the \textit{non-singular condition} holds, and (iii) $\{\mE^k\}$ and $\{d^k\}$ follow the \textit{safeguarding rule} with \textit{shrinkage checks}. Then: 
    \begin{enumerate}
        \item If $x_{\text{ave}}^k \to x^*$, there exists an integer $k_1$ such that $\mE^k = \bar{E}^k = \mE^k_0 = E^k = \mE^*$ for all $k \geq k_1$.
        \item If the \textit{shrinkage checks} and $d^k=1$ are triggered only finitely many times, then $x_{\text{ave}}^k \to x^*$.
        \item Both of the \textit{shrinkage checks} and the operation $d^k=1$ cannot be triggered infinitely many times.
    \end{enumerate}
\end{claim}

For 1), $\bar E^{k+1}\subseteq \bar E^k$ persists unless $E^{k+1}\setminus E^k\neq\emptyset$. As $\mu^k \to \mu^*$, this re-emergence terminates in finitely many iterations, and the \textit{shrinkage checks} ensure $\mE^k \to \mE^*$ (cf. SM \ref{SM:proof_of_claim_ref_claim_safe_shrink}).

For 2), we can verify that the \textit{safeguarding rule} ensures $J^k \geq 0$ at each iteration and guarantees the existence of a valid $\{d^k\}$ as in the \textit{safe averaging rule} (cf. SM \ref{SM:proof_of_claim_ref_claim_safe_shrink}). 

For 3), we first introduce the following lemma:
\begin{lemma} \label{lema:residual_zero_bounded}
    Consider the LASSO objective $F$ defined on $\mathbb{R}^N$. Suppose $r_1 \in (0,1)$ and a sequence $\{a_k\}$ satisfies $a_k \geq 0$ with $\sum a_k < \infty$. Then, for any sequence $\{u^k\}$ in $\mathbb{R}^N$:
    \begin{enumerate}
        \item If $\|R(u^{k+1})\|^2 \leq (1+a_k)\|R(u^k)\|^2$ for all $k$, and there exists a subsequence $\{u^{k_l}\}_{l=1}^\infty$ satisfying $\|R(u^{k_l+1})\| \leq (1-r_1)\|R(u^{k_l})\|$ for all $l$, then $R(u^k) \to 0$.
        \item If $R(u^k) \to 0$, then the sequence $\{u^k\}$ is bounded.
    \end{enumerate}
\end{lemma}
\begin{IEEEproof}
    See SM \ref{SM:residual_zero_bounded}.
\end{IEEEproof}

By Proposition~\ref{prop:E_Ec_couple}, given $d^k=1$, the condition $J^k\geq r\|R^k|_{\mE^k}\|^2$ implies $\|R^{k+1}\|\leq (1-r')\|R^k\|$ for some $r'>0$. If the \textit{shrinkage checks} or the operation $d^k=1$ are triggered infinitely often, applying Lemma~\ref{lema:residual_zero_bounded} to $\{x_{\text{ave}}^k\}$ yields $R^k \to 0$ and that $\{x_{\text{ave}}^k\}$ is bounded. Then, by Lemma~\ref{lemma:Rx_zero_x_converge} and assertion 1) of the claim, $x_{\text{ave}}^k \to x^*$ and $\mE^k = \mE^*$ eventually, yielding a contradiction. 

\color{black}

\section*{Acknowledgment}

The authors would like to thank Profs. Yong Xia and An Liu for stimulating discussions.  They also acknowledge the support from National Key R\&D Program of China under grant 2021YFA1003301, and National Science Foundation of China under grant 12288101.

% \clearpage

\bibliographystyle{IEEEtran}
\bibliography{IEEEabrv, ASM_bibfile}

\vfill

\clearpage

\makeatletter
% 调用 IEEEtran 内部备份的原始 section 命令
\let\section\@IEEEappendixsavesection
\makeatother

\setcounter{section}{0}
\setcounter{equation}{0}
\setcounter{figure}{0}
\setcounter{table}{0}
\setcounter{lemma}{0}
\setcounter{thm}{0}
\setcounter{prop}{0}
\setcounter{coro}{0}
\setcounter{claim}{0}
\setcounter{defi}{0}

\renewcommand{\thesection}{S.\arabic{section}}
\renewcommand{\thesectiondis}{S.\arabic{section}.}
\renewcommand{\theequation}{S\arabic{equation}}

\twocolumn[{
\begin{center}
    \vspace*{2em}
    {\Huge Supplementary Material} \\[1em]
    {\LARGE for ``\textit{Alternating Subspace Method for Sparse Recovery of Signals}"} \\
    \vspace*{2em}
\end{center}
}]

% \onecolumn
% \begin{center}
%     \vspace*{2em}
%     {\Huge Supplementary Material} \\[1em]
%     {\LARGE of ``\textit{Alternating Subspace Method for Sparse Recovery of Signals}"} \\
%     \vspace*{2em}
% \end{center}
% \twocolumn

\section{Equivalent Expressions of the ADMM Iteration}\label{sec:equivalent_expressions_of_the_admm_iteration}

Here we prove the equivalent ADMM formulation
\begin{equation}
    z^k - v\nabla h(x^{k+1}) = {x}_\text{ave}^{k+1} - v\nabla h({x}_\text{ave}^{k+1}) \label{eq:admm_equiv_claim}
\end{equation}
by induction with initializations $z^1=vA^Ty$ and $s^1=x^1_{\text{ave}}=0$. First, consider $k=1$. Since $x^2_\text{ave}=(x^2+x^1_\text{ave})/2$ and $x^1_{\text{ave}}=0$, the affinity of $\nabla h$ implies that Eq.~\eqref{eq:admm_equiv_claim} is equivalent to $z^1=(x^2+v\nabla h(x^2))/2+vA^Ty/2$. By Eq.~\eqref{eq:1st_OPT_L}, we have $x^2+v\nabla h(x^2)=z^1-vs^1$. Thus, the equivalence holds for $k=1$ since $z^1=vA^Ty$ and $s^1=0$.

Now assume Eq.~\eqref{eq:admm_equiv_claim} holds for $k=n-1$. For $k=n$, using $x^{n+1}_\text{ave}=(x^{n+1}+x^n_\text{ave})/2$ and the affinity of $\nabla h$, proving Eq.~\eqref{eq:admm_equiv_claim} is equivalent to showing
\begin{equation*}
    z^n=\frac{1}{2}(x^{n+1}+v\nabla h(x^{n+1}))+\frac{1}{2}(x^{n}_\text{ave}-v\nabla h(x^{n}_\text{ave})).
\end{equation*}
Substituting $x^{n+1}+v\nabla h(x^{n+1})=z^n-vs^n$ from Eq.~\eqref{eq:1st_OPT_L} and applying the induction hypothesis to the second term, this condition reduces to $z^n=(z^n-vs^n)/2+(z^{n-1}-v\nabla h(x^n))/2$, or equivalently, $z^n+vs^n=z^{n-1}-v\nabla h(x^n)$. This final equality holds identically since $z^n+vs^n=x^n+vs^{n-1}$ by Eq.~\eqref{eq:DRS_ADMM} and $z^{n-1}-v\nabla h(x^n)=x^n+vs^{n-1}$ by Eq.~\eqref{eq:1st_OPT_L}.

\section{Derivations on Null Space Error Evolvement}\label{app:derivation_null_space}

Here we present the derivation of \eqref{eq:eNull_ISTA}. Without loss of generality, we assume that $A$ is partially orthogonal satisfying $AA^T=I_M$, so that the null space projection can be simplified as $e^k_{\text{Null}}=\mathcal P_{\text{Null}}(e^k):=[I_N-A^TA](x^k-x^*).$ For ISTA, note that $\mathcal S_{\lambda v}(x^k-v\nabla h(x^k))=x^k-v\nabla h(x^k)-\lambda v p^k$, where $p^k=p(x^k-v\nabla h(x^k))$, then we have
\begin{equation*}
\begin{aligned}
    e^{k+1}_{\text{Null}}&= \mathcal P_{\text{Null}}\left(x^k+vA^T(y-Ax^k)-\lambda v p^k-x^* \right)\\
    &=e^{k}_{\text{Null}}-\lambda v\cdot\mathcal P_{\text{Null}}(p^k).
\end{aligned}
\end{equation*}
For ADMM, we have $(I+vA^TA)\big(I-\frac v{1+v}A^TA\big)=I$ by $AA^T=I_M$, which means $\mathcal P_{\text{Null}}\left(\mathcal L_v(u)-u\right)=0$ for any $u\in\mathbb R^N$ by~\eqref{eq:V_inverse}, and $\mathcal P_{\text{Null}}\big(\big(I_N-\frac{v}{1+v}A^TA\big)(u+vA^Ty)  \big)=\mathcal P_{\text{Null}}(u).$
Remind that $\lambda\nabla g(\mathcal S_{\lambda v}(x^k))$ denotes the multiplier $s^k$, one can verify that $s^k=\lambda p^k=s^{k-1}+(x^k-z^k)/v$. Since the averaging step in~\eqref{eq:cplx_L} is omitted, it follows from ISTA that $\mathcal P_{\text{Null}}(z^k-x^*)=e^{k}_{\text{Null}}-\lambda v\cdot\mathcal P_{\text{Null}}(p^k)$, leading to:
\begin{equation*}
\begin{aligned}
    &e^{k+1}_{\text{Null}}=\mathcal P_{\text{Null}}\left(\mathcal L_v\left(z^k-vs^k  \right)-x^*\right)\\
    &=\mathcal P_{\text{Null}}\left(z^k-\lambda v p^k -x^*\right)=e^{k}_{\text{Null}}-2\lambda v \mathcal P_{\text{Null}}\left(p^k \right).
\end{aligned}
\end{equation*}

\section{Proof of Lemma~\ref{lemma:Non-expan_mG}}\label{SM:proof_of_lemma_ref_lemma_non_expan_mg}

\setcounter{lemma}{\numexpr\getrefnumber{lemma:Non-expan_mG}-1\relax}
\begin{lemma}
    If $\mE^*\subseteq E^k$ and $\mu_i^k\cdot\mu_i^*>0$ for $i\in E^k$, then: 
    (i) $\|\hat{\nu}^k-\hat{\nu}^*\| \le \|\hat{\mu}^k-\hat{\mu}^*\|$ if $\mE^*= E^k$ or $v=\hat v$; 
    (ii) $\|\hat{\mu}^{k+1}-\hat{\mu}^*\| \leq \tilde C^k\|\hat{\mu}^k-\hat{\mu}^*\|$, where $\tilde C^k:=d^k \|H^k\|+(1-d^k)$.
\end{lemma}

\begin{IEEEproof}
For (i), it suffices to show $|\nu_i^k-\nu_i^*|\le |\mu^k_i-\mu^*_i|$ for all $i\in E^k$. For $i\in E^k\cap\mE^*$, since $z_i^k-\mu_i^k=-v\lambda \mathrm{sgn}(\mu_i^k)=-v\lambda \mathrm{sgn}(\mu_i^*)$, we trivially have $|\nu_i^k-\nu_i^*|=|\mu^k_i-\mu^*_i|$. For $i\in E^k\cap(\mE^*)^c$, assume $\mu_i^k>0$ without loss of generality, yielding $z_i^*=0$ and $\nu_i^k-\nu_i^*=\mu^k_i+(\hat{v}/v)\mu^*_i-\lambda(v+\hat{v})$. We can bound this from above: 
\begin{equation*}
    \nu_i^k-\nu_i^* = \mu_i^k-\mu_i^*-(1+\hat{v}/v)(v\lambda-\mu_i^*)\le \mu_i^k-\mu_i^*.
\end{equation*}
For the lower bound, we retain the generalized $\hat{v}$:
\begin{equation*}
\begin{aligned}
    \nu_i^k-\nu_i^* &= -(\mu_i^k-\mu_i^*) + 2(\mu_i^k-v\lambda) + (1-\hat{v}/v)(v\lambda-\mu_i^*) \\
    &= -(\mu_i^k-\mu_i^*) + 2(\mu_i^k-v\lambda) \ge -(\mu_i^k-\mu_i^*),
\end{aligned}
\end{equation*}
where we use the fact that $\hat{v}=v$ when $E^k\cap(\mE^*)^c\neq\emptyset$. For (ii), verifying $\mu(x^{k+1})|_{E^k}-\hat{\mu}^*=H^k(\hat{\nu}^{k}-\hat{\nu}^*)$ is straightforward. Combining this with $\hat{\mu}^{k+1}=d^k\mu(x^{k+1})|_{E^k}+(1-d^k)\hat{\mu}^{k}$ and part (i) finishes the proof.
\end{IEEEproof}

\section{Proof of Eq.~\eqref{Eq:Non-Expansive-Rk-Relations} in Appendix~\ref{sub:Properties_ADMM}}\label{sub:the_identities_in_appendix_admm}

Viewing the RLS step in~\eqref{eq:cplx_L} as an implicit gradient descent step, we have
\begin{equation}
    x^{k+1}=z^{k}-\hat v\nabla h(x^{k+1})-\hat vp^{k}. \label{eq:xk1_zk}
\end{equation}
Substituting $z^k=x^k_{\text{ave}}-v \nabla h(x^{k}_{\text{ave}})-v\lambda p^{k}$ yields
\begin{equation}
    x^{k+1}= x^k_\text{ave}-v\nabla h(x^{k}_\text{ave})- \hat v\nabla h(x^{k+1})-(\hat v+v)\lambda p^{k}. \label{eq:x_evo_ADMM}
\end{equation}
Using $x^{k+1}=(x^{k+1}_{\text{ave}}-(1-d)x^{k}_{\text{ave}})/d$, we obtain
\begin{equation*}
    x^{k+1}_{\text{ave}} = x^k_{\text{ave}} - \alpha\nabla h(x^{k}_{\text{ave}}) -\hat v \nabla h(x^{k+1}_{\text{ave}})-d (\hat v+v)\lambda p^{k},
\end{equation*}
where $\alpha:=vd-\hat v(1-d)$. Since $R^k=z^k-x_{\text{ave}}^k=-v\nabla h(x_{\text{ave}}^k)-v\lambda p^k$, rearranging terms directly leads to
\begin{equation}
    R^{k+1}-(1-d)R^k=\gamma\Delta\mu^k-v\lambda \Delta p^{k}, \label{eq:R_mu_p_ADMM}
\end{equation}
where $\gamma:=v/(\hat v+v)$. Furthermore, \eqref{eq:xk1_zk} implies
\begin{equation*}
    x^{k+1}=(I+\hat vA^TA)^{-1}(z^k+\hat vA^Ty-\hat vp^k).
\end{equation*}
Since $x^k_\text{ave}=(I+\hat vA^TA)^{-1}(I+\hat vA^TA)x^k_\text{ave}$, subtracting this from $x^{k+1}$ gives $x^{k+1}-x^k_\text{ave}=\gamma^{-1}(I+\hat vA^TA)^{-1}R^k$. Multiplying both sides by $(I-vA^TA)$ and using $\mu(x^{k+1})-\mu^k=(I-vA^TA)(x^{k+1}-x^k_\text{ave})$ yields
\begin{equation*}
    \mu(x^{k+1})-\mu^k=\gamma^{-1}(I-vA^TA)(I+\hat vA^TA)^{-1}R^k.
\end{equation*}
By the commutativity of the coefficient matrices and $\mu^{k+1}=d\cdot \mu(x^{k+1})+(1-d)\mu^k$, we conclude
\begin{equation}
    \gamma \Delta\mu^k=d\cdot (I+\hat vA^TA)^{-1}(I-vA^TA)R^k. \label{eq:Delta_mu_Rk_ADMM}
\end{equation}

\section{Proof of Lemma~\ref{lemma:Violation_plateau_implication} in Appendix~\ref{sub:Plateau_Conditions_ADMM}}\label{sub:violating_the_plateau_condition_for_admm}

Since $\|R(x)\|=0$ if and only if $x=x^*$ under the \textit{generic conditions}, it suffices to show $\|R(x')\|<\|R(x)\|$ to ensure $\delta_x>0$. Given $\hat v=v$ (so $\gamma=1/2$), if $(S_0\cup S_1)^c\neq \emptyset$ or $AR(x)\neq 0$, then $\|R(x')\|<\|R(x)\|$ by \eqref{eq:mu_p_descent_ADMM} and \eqref{eq:Ineq_ADMM1}.

Assuming $AR(x)=0$, we have $\gamma[\mu(x')-\mu(x)]=d\cdot R(x)$ and $R(x')=R(x)-v\lambda (p(x')-p(x))$ by Eq.~\eqref{eq:R_mu_p_ADMM}. By the definition of $S_0$, for $i\in S_0$,
\begin{equation*}
    R_i(x')=R_i(x)-[\mu_i(x')-\mu_i(x)]=(1-\gamma^{-1}d)\cdot R_i(x).
\end{equation*}
If $R_i(x)\neq 0$ for some $i\in S_0$, then by $(S_1\cup S_0)^c=\emptyset$, $\|R(x')\|<\|R(x)\|$ follows strictly from $1-\gamma^{-1}d\in(-1,1)$. 

If further $R(x)|_{S_0}=0$, combined with $AR(x)=0$, we have $A|_{S_1}R(x)|_{S_1}=0$. If $|S_1|\leq M$, generic condition 3) enforces $R(x)=0$. In conclusion, violating any of the Plateau Conditions yields either $\|R(x')\|<\|R(x)\|$ or $R(x)=0$, concluding the proof.

\section{Proof of Claim~\ref{claim:S_minus_non_empty} in Appendix~\ref{sub:Plateau_Conditions_ADMM}}\label{sec:proof_of_claim_s_minus_non_empty}

\setcounter{claim}{\numexpr\getrefnumber{claim:S_minus_non_empty}-1\relax}
\begin{claim}
    Under generic conditions, if $x$ satisfies the Plateau Conditions, then $S_{-}:=\{i\in S_1:R_i(x)\cdot \mu_i(x)<0 \}\neq \emptyset$.
\end{claim}

\begin{IEEEproof}
Since $AR(x)=0$ and $R(x)|_{S_0}=0$, we have $A|_{S_1}R(x)|_{S_1}=0$, i.e., $R(x)|_{S_1}\in \mathrm{Null}(A|_{S_1})$. By the relation
\begin{equation*}
\begin{aligned}
    R(x)|_{S_1}&=-v\nabla h(x)|_{S_1}-v\lambda p(x)|_{S_1}\\
    &=v(A|_{S_1})^T(y-Ax) -v\lambda p(x)|_{S_1},
\end{aligned}
\end{equation*}
projecting onto the null space of $A|_{S_1}$ yields $R(x)|_{S_1}=-v\lambda \mathcal P_{\mathrm{Null}}(p(x)|_{S_1})$. Consequently,
\begin{equation*}
    (R(x)|_{S_1})^Tp(x)|_{S_1} =-v\lambda \|\mathcal P_{\mathrm{Null}}(p(x)|_{S_1})\|^2<0
\end{equation*}
by generic condition 3). Thus, there exists an $i\in S_1$ such that $R_{i}(x)\cdot p_{i}(x)<0$. As $\mu_i(x)\cdot p_i(x)>0$ on $S_1$, we have $R_i(x)\cdot \mu_i(x)<0$, which concludes $S_{-}\neq\emptyset$.
\end{IEEEproof}

\section{Derivation for Appendix ~\ref{sec:algebraic_relationships_between_asm_iterations}}\label{SM:algebraic_relationships_between_asm_iterations}

For brevity, we introduce the following notations with respect to $\mE^k$: $\hat R^{k+1} := R^{k+1}|_{\mE^k}$, $\hat R^k := R^k|_{\mE^k}$, $\tilde R^k := R^k|_{(\mE^k)^c}$, $\hat x_\text{ave}^k := x^k_\text{ave}|_{\mE^k}$, $\tilde x_\text{ave}^k := x^k_\text{ave}|_{(\mE^k)^c}$, $\hat \mu^k_\gamma := \gamma\mu^k|_{\mE^k}$, $\hat p^k_\lambda := v\lambda p^k|_{\mE^k}$, $\hat A := A|_{\mE^k}$, $\tilde A := A|_{(\mE^k)^c}$, $\hat h(x) := -\hat A^T(y-Ax)$, and $G:=(I+\hat v\hat A\hat A^T)^{-1}$, where $\gamma:=v/(\hat v+v)$.

\subsection{The relationship between $R^{k+1}-(1-d^k)R^k$ and $\Delta\mu^k$}

Since $x^{k+1}|_{(\mE^k)^c}=z^k|_{(\mE^k)^c}=0$ (given $E^k\subseteq\mE^k$), the update for $\hat x^{k+1}$ follows similarly to Proposition~\ref{prop:ADMM}, yielding
\begin{equation}
    \hat x^{k+1}=\hat x^k_\text{ave}-v\nabla\hat h(x^{k}_\text{ave})-\hat v\nabla\hat h(x^{k+1})-(\hat v+v)\lambda\hat p^{k}.\label{eq:x_ave_evo}
\end{equation}
Substituting $x^{k+1}=(x^{k+1}_\text{ave}-(1-d^k)x^{k}_\text{ave})/d^k$ into the above, we have
\begin{equation}
    \hat x^{k+1}_\text{ave}=\hat x^k_\text{ave} -\alpha^k\nabla\hat h(x^{k}_\text{ave})-\hat v\nabla\hat h(x^{k+1}_\text{ave})-d^k (\hat v+v)\lambda\hat p^{k},\label{eq:x_hat_ave_evo}
\end{equation}
where $\alpha^k:=[vd^k-\hat v(1-d^k)]$. Since $R^k=-v\nabla h(x_\text{ave}^k)-v\lambda p^k$, $\mu^k=x^k_\text{ave}-v\nabla h(x^k_\text{ave})$, rearranging the iteration yields
\begin{equation}
    \hat R^{k+1}-(1-d^k)\hat R^k=\Delta\hat \mu^k_\gamma-\Delta\hat p^{k}_\lambda,\label{eq:R_mu_p}
\end{equation}
Note the following algebraic identity:
\begin{equation}
    \|  \Delta\hat \mu^k_\gamma- \Delta\hat p^k_\lambda\|^2=\|\Delta\hat \mu^k_\gamma\|^2-\left(2\gamma\Delta\hat \mu^k-\Delta\hat p^k_\lambda\right)^T\Delta\hat p^k_\lambda.
\end{equation}
Since $(p_\lambda^k)_i=\mathcal P_{[-v\lambda,v\lambda]}(\mu_i^k)$, by Lemma 3.1 in~\cite{2008-Elaine}, we know that $\Delta\mu_i^k-\Delta (p^k_\lambda)_i$, $\Delta\mu_i^k$, and $\Delta (p^k_\lambda)_i$ share the same sign (i.e., their pairwise products are non-negative) for each $i$. If $v\geq \hat v$ such that $2\gamma\geq 1$, or if $\Delta p^k=0$, it follows that $\left(2\gamma\Delta\hat \mu^k-\Delta\hat p^k_\lambda\right)^T\Delta\hat p^k_\lambda\geq 0$, which ensures
\begin{equation}
    \|\hat R^{k+1}-(1-d^k)\hat R^k\|\leq \|  \Delta\hat \mu^k_\gamma- \Delta\hat p^k_\lambda\|\leq\|\Delta\hat \mu^k_\gamma\|.
\end{equation}

\subsection{The relationship between $\Delta\mu^k$ and $R^k$}

To establish the relationship between $\Delta\hat \mu^k_\gamma$ and $d^k\cdot R^k$, we multiply both sides of Eq.~\eqref{eq:x_ave_evo} by $\hat A^T\hat A$, yielding
\begin{equation}
\begin{aligned}
    &\hat A^T\hat A(\hat x^{k+1}-\hat x^k_\text{ave})\\
    =&-\hat A^T\hat A[v\nabla\hat h(x^{k}_\text{ave})+\hat v\nabla\hat h(x^{k+1})+(\hat v+v)\lambda\hat p^{k}].
\end{aligned}\label{eq:ATA_x_h_p}
\end{equation}
By definition, we have
\begin{equation*}
    \hat A^T\hat A(\hat x^{k+1}-\hat x^k_\text{ave})=\nabla\hat h(x^{k+1})-\nabla\hat h(x^{k}_\text{ave})+\hat A^T\tilde A\tilde x^{k}_\text{ave},
\end{equation*}
and thus Eq.~\eqref{eq:ATA_x_h_p} can be reformulated as
\begin{equation*}
\begin{aligned}
    &(I-v\hat A^T\hat A)[-v\nabla\hat h(x^{k}_\text{ave})-v\lambda\hat p^{k}]+v\hat A^T\tilde A\tilde x^k_\text{ave}\\
    =&(I+\hat v\hat A^T\hat A)[ -v\nabla\hat h(x^{k+1})-v\lambda\hat p^{k} ]\\
    =&(I+\hat v\hat A^T\hat A) \gamma[ \mu(x^{k+1})|_{\mE^k}-\hat \mu^k ].
\end{aligned}
\end{equation*}
The second equality follows from Eq.~\eqref{eq:x_ave_evo}. Combining this with the relation $\hat \mu^{k+1}=d^k\mu(x^{k+1})|_{\mE^k}+(1-d^k)\hat \mu^k$, and recalling that $\hat R^k=-v\nabla\hat h(x^{k}_\text{ave})-v\lambda\hat p^{k}$ and $\tilde R^k=-\tilde x_\text{ave}^k$, we finally obtain
\begin{equation}
    \Delta\hat \mu^k_\gamma=d^k\cdot (I+\hat v\hat A^T\hat A)^{-1}[(I-v\hat A^T\hat A)\hat R^k-v\hat A^T\tilde A\tilde R^k].\label{eq:Delta_mu_Rk}
\end{equation}

\subsection{The relationship between $R^{k+1}$ and $J^k$}

Here, we verify the relation $J^k=\hat v\|\hat R^k\|^2-\hat v\|\Delta\hat \mu^k_\gamma/d^k\|^2$. From Eq.~\eqref{eq:Delta_mu_Rk} and the Sherman-Morrison-Woodbury identity, we have
\begin{equation}
    \Delta\hat \mu^k_\gamma/d^k=\hat R^k-(\hat v+v)\hat A^TG\hat A\hat R^k-v\hat A^TG\tilde A\tilde R^k.
\end{equation}
Denoting $u:=\hat A\hat R^k$ and $w:=\tilde A\tilde R^k$, expanding the squared norm yields
\begin{equation}
\begin{aligned}
    \|\Delta&\hat \mu_\gamma^k/d^k\|^2=\|\hat R^k\|^2+(\hat v+v)^2u^\top G^\top (\hat A\hat A^\top)Gu\\
    &+v^2 w^\top G^\top (\hat A\hat A^\top)Gw-2(\hat v+v)u^\top Gu\\
    &-2vu^\top Gw+2v(\hat v+v)u^\top G^\top(\hat A\hat A^\top)Gw.
\end{aligned}
\end{equation}
Using the identity $G^\top (\hat A\hat A^\top)G=[G-G^2]/\hat v$, we obtain
\begin{equation}
\begin{aligned}
    \|\Delta&\hat \mu_\gamma^k/d^k\|^2=\|\hat R^k\|^2+[(v^2-\hat v^2)u^\top Gu\\
    &-(\hat v+v)^2u^\top G^2u-2v(\hat v+v)u^\top G^2w\\
    &+2v^2u^\top Gw+v^2w^\top G w-v^2w^\top G^2w]/\hat v\\
    &:=\|\hat R^k\|^2-J(u,w)/\hat v.
\end{aligned}
\end{equation}
It is then straightforward to verify that $J(u,w)$ coincides with the definition in Eq.~\eqref{eq:J_k}.

\section{Proof of Lemma~\ref{lemma:ASM_bounded}}\label{SM:ASM_bounded}

\setcounter{lemma}{\numexpr\getrefnumber{lemma:ASM_bounded}-1\relax}
\begin{lemma}%\label{lemma:ASM_bounded}
    Let $\{x^k_\text{ave}\}$ be generated by the ASM iteration with subspaces $\{\mE^k\}$ and averaging factors $\{d^k\}$. Suppose that $\mE^k_0 \subseteq \mE^k$ for all $k$, the matrix $I - v(A|_{\mE^k})^T A|_{\mE^k}$ is non-singular for each $\mE^k$, and $\|R^{k+1}\|^2 \leq (1+a_k)\|R^k\|^2$ for some sequence $a_k \geq 0$ satisfying $\sum a_k < \infty$. Then, there exist constants $C_\text{shrink}, C_\text{step} > 0$, independent of any specific choice of $\mE^k$ and $d^k$, such that for all $k$: 1) $\|x^{k+1}_\text{ave}\| \leq \|x^{k}_\text{ave}\| + C_\text{step}$; 2) whenever $\|x^k_\text{ave}\| \geq C_\text{shrink}$, it holds that $\|x^{k+1}_\text{ave}\| \leq \|x^{k}_\text{ave}\|$.
\end{lemma}

\begin{IEEEproof}
    We observe that, by definition, $x^{k+1}_\text{ave} = d^k x^{k+1} + (1-d^k)x^k_\text{ave}$ is a convex combination for any $d^k \in [0, 1]$. Therefore, if we can establish that \textbf{(i)} $\|x^{k+1}\| \leq \|x^{k}_\text{ave}\| + C_\text{step}$, and \textbf{(ii)} $\|x^{k+1}\| \leq \|x^{k}_\text{ave}\|$ whenever $\|x^k_\text{ave}\| \geq C_\text{shrink}$, then the original claims 1) and 2) follow immediately.

    {First, we establish uniform bounds on the global projections of the iterates.} Since $\sum_{j=1}^\infty a_j<\infty$, the sequence $\{R^k\}$ is uniformly bounded by $\|R^k\|^2\leq \|R^1\|^2\prod_{j=1}^\infty(1+a_j)$. Because $R^k = v A^T(y - A x^k_\text{ave}) - v\lambda p(x^k_\text{ave})$ with $\|p(x^k_\text{ave})\|_\infty \leq 1$, the row-space projection $\|\mathcal P_{\text{Row}(A)}(x_\text{ave}^k)\|$ is bounded by a constant $C_{\text{Row}} > 0$.

    {Second, we show that at any iteration $k$, the projection} $\|\mathcal P_{\text{Row}(A|_{\mathcal{E}^k})}(x_\text{ave}^k|_{\mathcal{E}^k})\|$ {is also bounded by a uniform constant} that is independent of the specific subspace $\mathcal{E}^k$. To see this, first recall that the algorithmic truncation (induced by $\mE^k_0\subseteq\mE^k$) guarantees $\|x_\text{ave}^k|_{(\mathcal{E}^k)^c}\| \leq C\sqrt{N} := C_{\text{out}}$. Next, denoting $\hat A := A|_{\mathcal{E}^k}$ and $\hat x^k_\text{ave} := x^k_\text{ave}|_{\mathcal{E}^k}$, we use the algebraic splitting $A x^k_\text{ave} = \hat A \hat x^k_\text{ave} + A|_{(\mathcal{E}^k)^c} x^k_\text{ave}|_{(\mathcal{E}^k)^c}$. Multiplying both sides by $\hat A^T$ and noting that $A x^k_\text{ave} = A \mathcal P_{\text{Row}(A)}(x^k_\text{ave})$, we obtain the splitting:
    \begin{equation*}
        \hat A^T \hat A \hat x^k_\text{ave} = \hat A^T A \mathcal P_{\text{Row}(A)}(x^k_\text{ave}) - \hat A^T A|_{(\mathcal{E}^k)^c} x^k_\text{ave}|_{(\mathcal{E}^k)^c}.
    \end{equation*}
    By definition, the boundedness of the projection $\|\mathcal P_{\text{Row}(\hat A)}(\hat x^k_\text{ave})\|$ is equivalent to that of $\|\hat A^T \hat A \hat x^k_\text{ave}\|$. From the right-hand side of the splitting above, the latter is clearly bounded by a constant $C(\mathcal{E}^k)$ depending exclusively on the choice of $\mathcal{E}^k$. Since the possible choices for $\mathcal{E}^k$ are combinatorially finite, this proves the existence of a global uniform constant $C_{\text{RowSub}} > 0$ such that $\|\mathcal P_{\text{Row}(A|_{\mathcal{E}^k})}( x^k_\text{ave}|_{\mathcal{E}^k})\| \leq C_{\text{RowSub}}$.

    Third, we show that the same assertion applies also to $x^{k+1}$. To see this, we multiply both sides of Eq.~\eqref{eq:x_ave_evo} by $\hat A$, which yields
    \begin{equation}
    \begin{aligned}
        &W^k A(x^{k+1}-x^{k}_\text{ave}) \\
        =&\gamma^{-1} \hat A[\gamma (\mu(x^{k+1})|_{\mE^k}-\hat \mu^k)]-A|_{(\mE^k)^c} x_\text{ave}^k|_{(\mE^k)^c},
    \end{aligned}
    \end{equation}
    where we utilized the fact that $x^{k+1}|_{(\mE^k)^c}=0$, and introduced the matrix $W^k := I-v\hat A\hat A^T$ while inheriting the remaining notations from Eq.~\eqref{eq:x_ave_evo}. Furthermore, evaluating Eq.~\eqref{eq:Delta_mu_Rk} at $d^k=1$ provides that:
    \begin{equation*}
    \begin{aligned}
        &\gamma (\mu(x^{k+1})|_{\mE^k}-\hat \mu^k) \\
        =&(I+\hat v\hat A^T\hat A)^{-1}[(I-v\hat A^T\hat A)\hat R^k+vA|_{(\mE^k)^c} x_\text{ave}^k|_{(\mE^k)^c}],
    \end{aligned}
    \end{equation*}
    where we used the fact that $R^k|_{(\mE^k)^c}=-x^k_\text{ave}|_{(\mE^k)^c}$. Consequently, since $W^k$ is invertible, $\|x_\text{ave}^k|_{(\mathcal{E}^k)^c}\| \leq C_{\text{out}}$, and $\|\hat R^k\|\leq \|R^k\|$ is uniformly bounded, substituting the latter equation into the former and taking the supremum over all possible choices of $\mE^k$ demonstrates that $\|A(x^{k+1}-x^{k}_\text{ave})\|$ is uniformly bounded. This guarantees the existence of a constant $C'_\text{Row}$ such that $\|\mathcal P_{\text{Row}(A)}(x^{k+1})\|\leq C'_{\text{Row}}$. Following the exact same argument as for $x^k_\text{ave}$, there uniformly exists a constant $C'_\text{RowSub}$ such that $\|\mathcal P_{\text{Row}(A|_{\mathcal{E}^k})}(x^{k+1}|_{\mathcal{E}^k})\|\leq C'_{\text{RowSub}}$.

    \textbf{To prove statement (i),} our previous analysis implies that it suffices to establish the existence of a constant $C_\text{step} > 0$ such that $\|\hat x^{k+1}\| \leq \|\hat x^k_\text{ave}\| + C_\text{step}$, where we denote $\hat {x}^{k+1} := x^{k+1}|_{\mathcal{E}^k}$ and $\hat {x}_\text{ave}^{k} := x_\text{ave}^{k}|_{\mathcal{E}^k}$ for brevity. From Eq.~\eqref{eq:x_ave_evo}, the subspace update takes the form:
    \begin{equation}
        \hat x^{k+1} = \hat x_\text{ave}^{k} + \hat A^T u^k - \eta \hat p^k,\label{eq:dif_xk1_xk_ave}
    \end{equation}
    where $u^k := \hat v(y - Ax^{k+1}) + v(y - Ax^{k}_\text{ave})$ and $\eta := \hat v+v$. Since the projections of $x^{k+1}$ and $x^{k}_\text{ave}$ onto $\text{Row}(A)$ are uniformly bounded, and the parameters $\hat v, v$ are fixed and finite, both $u^k$ and $\eta$ are rigorously bounded. Consequently, the update increments $\hat A^T u^k$ and $\eta \hat p^k$ are globally bounded, which directly guarantees the existence of $C_\text{step}$.

    \textbf{To prove statement (ii),} we decompose $\|\hat x^{k+1}\|^2$ using orthogonal projections onto $\text{Row}(\hat A)$ and $\text{Null}(\hat A)$:
    \begin{equation*}
    \begin{aligned}
        \|\hat x^{k+1}&\|^2=\|\mathcal P_{\text{Row}(\hat A)}(\hat x^{k+1})\|^2+\|\mathcal P_{\text{Null}(\hat A)}(\hat x^{k+1}-\hat x_\text{ave}^{k})\|^2\\
        &+\|\mathcal P_{\text{Null}(\hat A)}(\hat x_\text{ave}^{k})\|^2+2(\hat x^{k+1}-\hat x_\text{ave}^{k})^T\mathcal P_{\text{Null}(\hat A)}(\hat x_\text{ave}^{k})
    \end{aligned}
    \end{equation*}
    We first demonstrate that for any arbitrarily large constant $C_0 > 0$, there exists a threshold $C_\text{shrink}'$ such that whenever $\|\hat x_\text{ave}^{k}\| \geq C_\text{shrink}'$, the cross term on the right-hand side is strictly negative and its absolute value exceeds $C_0$.

    From Eq.~\eqref{eq:dif_xk1_xk_ave}, we know
    $$
    \begin{aligned}
        (&\hat x^{k+1}-\hat x_\text{ave}^{k})^T\mathcal P_{\text{Null}(\hat A)}(\hat x_\text{ave}^{k})=-\eta(\hat p^k)^T\mathcal P_{\text{Null}(\hat A)}(\hat x_\text{ave}^{k})\\
        &=-\eta (\hat p^k)^T \hat x_\text{ave}^{k} + \eta (\hat p^k)^T \mathcal P_{\text{Row}(\hat A)}(\hat x_\text{ave}^{k})
    \end{aligned}
    $$
    Since $\mathcal P_{\text{Row}(\hat A)}(\hat x_\text{ave}^{k})$ and $\hat p^k$ are uniformly bounded, the asymptotic behavior of the cross term is dominated by $-\eta(\hat p^k)^T \hat x_\text{ave}^{k}$ as $\|\hat x_\text{ave}^{k}\|\to\infty$. {For any coordinate $i$ where the sign aligns} (i.e., $(x_\text{ave}^k)_i \cdot p_i^k > 0$ and $|p_i^k|=1$), the product $-\eta(x_\text{ave}^k)_i \cdot p_i^k \to -\infty$ as $|(\hat x_\text{ave}^{k})_i|\to\infty$, given that such unbounded coordinates must exist if the overall norm grows, while the remaining misaligned elements are strictly bounded. In fact, {for any coordinate $i$ that does not align} (i.e., $(x_\text{ave}^k)_i \cdot p_i^k \leq 0$ or $|p_i^k| < 1$), the definition of $p^k$ implies that $(x_\text{ave}^k)_i+v A_i^T A(y-Ax_\text{ave}^k)$ is either suppressed below the threshold $v\lambda$ or reversed in direction relative to $(x_\text{ave}^k)_i$ and in either case, $|(x_\text{ave}^k)_i|\leq v|A_i^T A(y-Ax_\text{ave}^k)|+v\lambda$. Because the right-hand side is uniformly bounded (since $\|\mathcal P_{\text{Row}(A)}(x^{k}_\text{ave})\|\leq C_{\text{Row}}$), these unaligned components cannot grow unbounded. Consequently, $-\eta (\hat p^k)^T \hat x_\text{ave}^{k}\to-\infty$ as $\|\hat x_\text{ave}^{k}\|\to\infty$, and the assertion holds.

    Next, we note that: {(a)} $\|\mathcal P_{\text{Row}(\hat A)}(\hat x^{k+1})\| \leq C_\text{RowSub}'$; {(b)} $\|\mathcal P_{\text{Null}(\hat A)}(\hat x^{k+1} - \hat x_\text{ave}^{k})\| = \eta\|\mathcal P_{\text{Null}(\hat A)}(\hat p^k)\| \leq \eta\sqrt{N}$. By setting the target constant to $C_0 := (C_\text{RowSub}')^2 + \eta^2 N$, we ensure that whenever $\|\hat x_\text{ave}^{k}\| \geq C_\text{shrink}'$, the negative cross term strictly offsets the terms in (a) and (b). Consequently, this yields: $\|\hat x^{k+1}\|^2 \leq \|\mathcal P_{\text{Null}(\hat A)}(\hat x_\text{ave}^{k})\|^2 \leq \|\hat x_\text{ave}^{k}\|^2$.  

    Finally, by setting $C_\text{shrink} := C_\text{shrink}' + C_{\text{out}}$, any condition $\|x_\text{ave}^k\| \geq C_\text{shrink}$ naturally enforces $\|\hat x_\text{ave}^{k}\| \geq C_\text{shrink}'$. Since $x^{k+1}|_{(\mE^k)^c} = 0$, this ensures $\|x^{k+1}\| \leq \|\hat x_\text{ave}^{k}\| \leq \|x_\text{ave}^k\|$, completing the proof of statement (ii).
\end{IEEEproof}

\section{Proof of Claims \ref{lemma:Ec_zero_delta_x} and \ref{lemma:Plateau_delta} in Appendix \ref{sec:convergence_of_asm}}\label{SM:proof_of_claim_ref_lemma_ec_zero_delta_x}

\setcounter{claim}{\numexpr\getrefnumber{lemma:Ec_zero_delta_x}-1\relax}
\begin{claim}
    If $R^*\neq 0$, one of the following two cases must occurs:  (i) there exists a constant $\delta_x > 0$ such that $\|R(x')\| \leq \|R(x^*)\| - \delta_x$; or (ii) the \textit{Plateau Conditions} hold.
\end{claim}
\begin{IEEEproof}
    Since $\mE^{(0)}$ and $d^{(0)}$ are fixed, $\mathcal T^{(0)}$ is continuous. 

    If $\delta_0 := \|R(x^*)|_{(\mE^{(0)})^c}\|^2+\|R(x')|_{(\mE^{(0)})^c}\|^2 \neq 0$, then, 
    \begin{equation}
    \begin{aligned}
        &\|R'\|^2=\lim_{l\to\infty}\|R^{k_l+1}\|^2\\
        \leq& \lim_{l\to\infty}(1+a^{k_l})[\|R^{k_l}\|^2-d_0\delta^{k_l}_0]=\|R^*\|^2-d_0\delta_0,
    \end{aligned}
    \end{equation}
    where $\delta^{k_l}_0:=\|R^{k_l+1}|_{(\mE^{(0)})^c}\|^2 +\|R^{k_l}|_{(\mE^{(0)})^c}\|^2$, and thus a $\delta_x>0$ exists. 

    Otherwise, $\delta_0=0$, which implies $R(x')|_{(\mE^{(0)})^c}=R(x^*)|_{(\mE^{(0)})^c}=0$. This means that $x^*|_{(\mE^{(0)})^c}=0$, and once there exists a $\delta_x'>0$ such that $\|R(x')|_{\mE^{(0)}}\|=\|R(x^*)|_{\mE^{(0)}}\|-\delta_x'$, then setting $\delta_x=\delta_x'$ meets condition (i). Note that $x^*|_{(\mE^{(0)})^c}=0$, and $\mathcal T^{(0)}$ indeed induces an ADMM iteration from $\mE^{(0)}$ to $\mE^{(0)}$. From Proposition~\ref{prop:ADMM}, $\delta_x'>0$ exists unless the \textit{Plateau Conditions} hold.
\end{IEEEproof}

\setcounter{claim}{\numexpr\getrefnumber{lemma:Plateau_delta}-1\relax}
\begin{claim}
    If $R^* \neq 0$, then under the \textit{Plateau Conditions}, there exist an index $n_s$ $(1 \leq n_s \leq n_0)$ and a constant $\delta_x > 0$ such that $\|R(x^{(n_s)})\| \leq \|R^*\| - \delta_x$.
\end{claim}
\begin{IEEEproof}
    For brevity, we denote $R^{(n)}:=R(x^{(n)})$, $\mu^{(n)}:=\mu(x^{(n)})$, and $p^{(n)}:=p(x^{(n)})$.

    First, suppose there exists an $n\in\{1,\ldots,n_0-1\}$ such that $\delta_c^{(n)}:=\|R^{(n+1)}|_{(\mE^{(n)})^c}\|^2+\|R^{(n)}|_{(\mE^{(n)})^c}\|^2\neq 0$. Then
    \begin{equation*}
    \begin{aligned}
        &\|R(x^{(n+1,l)})\|^2 \leq (1+a^{k_l+n}) \big[ \|R(x^{(n,l)})\|^2-d_0\delta_c^{k_l+n} \big]\\
        &\leq (1+a^{k_l+n})\bigg[ \|R(x^{k_l})\|^2\prod_{j=0}^{n-1}(1+a^{k_l+j})-d_0\delta_c^{k_l+n} \bigg].
    \end{aligned}
    \end{equation*}
    By the continuity of $\mathcal T^{(n)}\circ\cdots\circ \mathcal T^{(0)}$ and $R(\cdot)$, taking the limit $l\to\infty$ yields $\|R^{(n+1)}\|^2\leq \|R^*\|^2-d_0\delta_c^{(n)}$. Setting $\delta_x:=\delta_c^{(n)}$ and $n_s:=n+1$ concludes this case.

    Otherwise, $\delta_c^{(n)}=0$ for all $n\leq n_0-1$. Recall that $\|R^{(n)}|_{(\mE^{(n)})^c}\|=0$ ensures $z^{(n)}:=\mS_{v\lambda}(\mu^{(n)})|_{(\mE^{(n)})^c}=0$, which guarantees that $\mathcal T^{(n)}$ is a valid ASM iteration on $x^{(n)}$ (i.e., the support of $z^{(n)}$ is covered by $\mE^{(n)}$). Consequently, all properties established in SM.~\ref{SM:algebraic_relationships_between_asm_iterations} hold for each $\mathcal T^{(n)}$.

    Second, we claim that for any set $S$, if $|\mu_i^{(n)}|\geq v\lambda$ for all $i\in S$, then $S\subseteq \mE^{(n)}$. Since $\mathcal V:=\mu\circ \mathcal T^{(n)}\circ\cdots\circ \mathcal T^{(0)}$ is continuous, there exists an $l_0$ such that for all $l\geq l_0$, $\|\mathcal V(x^*)-\mathcal V(x^{k_l})\|\leq v\lambda\epsilon/2$, where $\epsilon$ is defined in~\eqref{eq:ave_epsilon} for $\mE^k_0$. Then 
    \begin{equation*}
    \begin{aligned}
        |\mathcal V_i(x^{k_l})|&=|\mu_i(x^{(n,l)})|\geq |\mu_i(x^{(n)})|-v\lambda\epsilon/2\\
        &\geq v\lambda(1-\epsilon/2)>v\lambda(1-\epsilon).
    \end{aligned}
    \end{equation*}
    Hence, $i\in \mE_0^{k_l+n}\subseteq \mE^{k_l+n}=\mE^{(n)}$.

    Third, we define $S_{-}:=\{i\in S_1: \mu_i(x^*)\cdot R_i^*>0\}$. Following reasoning analogous to Claim~\ref{claim:S_minus_non_empty}, $S_{-}\neq \emptyset$ under the Plateau Conditions and generic condition 3). We define $i_s:=\mathop{\arg\min}_{i\in S_{-}} (|\mu_i(x^*)|-v\lambda)/|R_i^*|$. Then, there exists an $n_s\geq 2$ (since $S_{-}\subseteq \{i:p^{(1)}_i=p^*_i\}$) such that
    \begin{equation}
        \sum_{j=0}^{n_s-2}d^{(j)}|R_{i_s}^*| \leq \gamma(|\mu_{i_s}(x^*)|-v\lambda) < \sum_{j=0}^{n_s-1}d^{(j)}|R_{i_s}^*|\label{eq:ns_defi}
    \end{equation}
    % , it follows that $n_s\geq 2$ and $|\mu_i(x^*)|>v\lambda$ for all $i\in S_{-}$.

    With this preparation, we show by induction that for every $n\leq n_s-1$:
    \begin{enumerate}
        \item $\mu_i^{(n)}\cdot p_i(x^*)\geq v\lambda\ (\forall i\in S_1)$;
        \item $A|_{S_1}R^{(n)}|_{S_1}=0$ and $R^{(n)}|_{S_1^c}=0$.
    \end{enumerate}

    For the base case $n=0$, $\|R^*|_{S_1^c}\|=0$ and $A|_{S_1}R^*|_{S_1}=AR^*=0$ hold by the Plateau Conditions. By the definition of $S_1$, $\mu_i(x^*)\cdot p_i(x^*)=|\mu_i(x^*)|\geq v\lambda\ (\forall i\in S_1)$. Next, for a fixed $n\leq n_s-1$, we assume that $\mu_i^{(r)}\cdot p_i(x^*)\geq v\lambda\ (\forall i\in S_1)$, $A|_{S_1}R^{(r)}|_{S_1}=0$ and $R^{(r)}|_{S_1^c}=0$ for all $r\leq n-1\leq n_s-2$.

    % Then we prove that the assertion holds for $j=n\leq n_s-1$. From the induction assumptions, we claim that
    % \begin{enumerate}
    %     \item $[\gamma(\mu^{(r+1)}-\mu^{(r)})]|_{\mE^{(r)}}=d^{(r)}R^{(r)}$,
    %     \item $[R^{(r+1)}-R^{(r)}]|_{\mE^{(r)}}=-v\lambda[p^{(r+1)}-p^{(r)}]|_{\mE^{(r)}}$,
    %     \item $[p^{(r)}-p^{(r-1)}]|_{S_1}=0$
    % \end{enumerate}
    % hold for all $r\leq n-1$. Here 1) and 2) hold by Eq.~\eqref{eq:R_mu_p}, ~\eqref{eq:Delta_mu_Rk} and the induction assumption that $A|_{\mE^{(r)}}R^{(r)}|_{\mE^{(r)}}=0$. As for 3), Since $|\mu_i^{(r)}|=|\mu_i^{(r)}p(x_i^*)|\geq v\lambda$ and $p_i^{(r)}\cdot p(x_i^*)\geq0$ for all $r\leq n-1$, we know $|p_i^{(r)}|=1$ for every $i\in S_1$ and thus 3) holds.

    We now prove that the assertions hold for step $n$. From the induction assumptions, we claim the following hold for all $r\leq n-1$:
    \begin{enumerate}
        \item $[\gamma(\mu^{(r+1)}-\mu^{(r)})]|_{\mE^{(r)}}=d^{(r)}R^{(r)}$,
        \item $[R^{(r+1)}-R^{(r)}]|_{\mE^{(r)}}=-v\lambda[p^{(r+1)}-p^{(r)}]|_{\mE^{(r)}}$,
        \item $[p^{(r)}-p^{(r-1)}]|_{S_1}=0$.
    \end{enumerate}
    Here, 1) and 2) follow from \eqref{eq:R_mu_p}, \eqref{eq:Delta_mu_Rk}, and the induction assumption that $A|_{\mE^{(r)}}R^{(r)}|_{\mE^{(r)}}=0$. For 3), since $\mu_i^{(r)}\cdot p_i(x^*)\geq v\lambda$ implies $|\mu_i^{(r)}|\geq v\lambda$ and $p_i^{(r)}$ shares the sign of $p_i(x^*)$ for all $r\leq n-1$, it follows that $p_i^{(r)}=p_i(x^*)$ for every $i\in S_1$, and thus 3) holds.

    Consequently, $R^{(n)}|_{S_1}=\cdots=R^{(0)}|_{S_1}=R^*|_{S_1}$. For any $i\in S_1$, we have
\begin{equation}
\begin{aligned}
    \mu_i^{(n)}&=\mu_i(x^*)+\gamma^{-1}\sum_{r=0}^{n-1}d^{(r)}R_i^{(r)}\\
    &=\mu_i(x^*)+\gamma^{-1}\sum_{r=0}^{n-1}d^{(r)}R_i^*.
\end{aligned}
\end{equation}
    \begin{enumerate}
        \item To verify $\mu_i^{(n)}\cdot p_i(x^*)\geq v\lambda$ for all $i\in S_1$: (i) For $i\in S_{-}^c\cap S_1$, we have $\mu_i(x^*)\cdot R_{i}^*\geq 0$, which implies $|\mu_i^{(n)}|\geq |\mu_i(x^*)|\geq v\lambda$. (ii) For $i\in S_{-}$,
        \begin{equation*}
        \begin{aligned}
            &\mu_i^{(n)}p_i^*=\mu_i(x^*)p_i^*+\gamma^{-1}\sum_{r=0}^{n-1}d^{(r)}R_{i}^*p_i^*\\
            &=v\lambda+\bigg[(|\mu_i(x^*)|-v\lambda)/|R_{i}^*|-\gamma^{-1}\sum_{r=0}^{n-1}d^{(r)}\bigg]|R_{i}^*|\\
            &\geq v\lambda+\bigg[(|\mu_{i_s}(x^*)|-v\lambda)/|R_{i_s}^*|-\gamma^{-1}\sum_{r=0}^{n_s-2}d^{(r)}\bigg]|R_{i}^*|,
        \end{aligned}
        \end{equation*}
        where $p_i^*:=p_i(x^*)$. This strictly leads to $\mu_i^{(n)}\cdot p_i(x^*)\geq v\lambda$ for all $i\in S_1$ by \eqref{eq:ns_defi}.
        \item To verify $R^{(n)}|_{S_1^c}=0$: (i) For $i\in\mE^{(n)}\cap S_1^c$, we have $\mu_i^{(n)}-\mu_i^{(n-1)}=\gamma^{-1}d^{(n)}R_i^{(n-1)}=0$, which implies $p_i^{(n)}=p_i^{(n-1)}$ and $R_i^{(n)}=R_i^{(n-1)}=0$; (ii) For $i\in(\mE^{(n)})^c$, the condition $\delta_c^{(n)}=0$ ensures $R_i^{(n)}=0$.
        \item To verify $A|_{S_1}R^{(n)}|_{S_1}=0$: Since $\mu_i^{(n)}\cdot p_i(x^*)\geq v\lambda$ and $\mu_i^{(n-1)}\cdot p_i(x^*)\geq v\lambda$ for $i\in S_1$, we have $[p^{(n)}-p^{(n-1)}]|_{S_1}=0$ and $R^{(n)}|_{S_1}=R^{(n-1)}|_{S_1}$. Consequently, $A|_{S_1}R^{(n)}|_{S_1}=A|_{S_1}R^{(n-1)}|_{S_1}=0$.
    \end{enumerate}
    This completes the induction.

    Thus far, we have established that $S_1\subseteq \mE^{(n)}$ and $[p^{(n)}-p^{(n-1)}]|_{S_1}=0$ for all $n\leq n_s-1$. By the definitions of $n_s$ and $i_s$, we have $\mu_{i_s}^{(n_s)}\cdot p_{i_s}(x^*)< v\lambda$, and thus $p_{i_s}^{(n_s)}-p_{i_s}^{(n_s-1)}\neq 0$. Since $i_s\in S_{-}\subseteq S_1\subseteq S_0^c$, there exists a constant $\delta_x>0$ such that $\|R^{(n_s)}|_{S_1}\|\leq \|R^{(n_s-1)}|_{S_1}\|-\delta_x$. Following the induction logic, $R^{(n_s)}|_{S_1^c}=0$ is guaranteed by $R^{(n_s-1)}|_{S_1^c}=0$ and $A|_{S_1}R^{(n_s-1)}|_{S_1}=0$. Finally, since $\|R^{(n_s)}|_{S_1^c}\|=\|R^*|_{S_1^c}\|=0$ and $\|R^{(n_s-1)}|_{S_1}\|=\|R^*|_{S_1}\|$, we conclude that $\|R^{(n_s)}\|\leq \|R^*\|-\delta_x$.

\end{IEEEproof}

\section{Proof of Lemma~\ref{lemma:Rx_zero_x_converge}}\label{SM:proof_of_lemma_ref_lemma_rx_zero_x_converge}

\setcounter{lemma}{\numexpr\getrefnumber{lemma:Rx_zero_x_converge}-1\relax}

\begin{lemma}
    Let $\{u^k\}$ be any bounded sequence satisfying $\|R(u^{k+1})\|^2 \leq (1+a_k)\|R(u^k)\|^2$ for some sequence $a_k \geq 0$ such that $\sum a_k < \infty$. If there exists a subsequence $\{u^{k_l}\}$ such that $\lim_{l\to\infty} \|R(u^{k_l})\| = 0$, then $\{u^k\}$ converges to the unique LASSO optimum under the \textit{generic conditions}.
\end{lemma}

\begin{IEEEproof}
By assumption, for any $\varepsilon>0$, there exists a $l_0$ such that for any $l\geq l_0$, $\|R(u^{k_l})\|\leq \varepsilon/\prod_{j=1}^\infty(1+a^j)$. For any $k\geq k_{l_0}$, we have $\|R(u^k)\|\leq \|R(u^{k_{l_0}})\|\prod_{j=k_{l_0}}^k(1+a^j)\leq \varepsilon$. Since the sequence $\{u^k\}$ is bounded and $\lim_{k \to \infty} \|R(u^k)\| = 0$, 
any accumulation point $\tilde{u}$ of the sequence must satisfy $R(\tilde{u}) = 0$ due to the continuity of $R(\cdot)$. Under the assumption that the optimal solution is unique, $u^*$ is the unique accumulation point of $\{u^k\}$ and $\lim_{k \to \infty} u^k = u^*$.
\end{IEEEproof}

\section{Proof of Claim~\ref{claim:convergence_Ek_Ek0} in Appendix \ref{app:converge_Ek_Ek0}}\label{sec:convergence_Ek_Ek0}

\setcounter{claim}{\numexpr\getrefnumber{claim:convergence_Ek_Ek0}-1\relax}

\begin{claim}
    If $v=\hat v\leq 4/(\max_{\mE}\|A|_{\mE}\|_2^2)$, then $\{R^k\}$ satisfies the \textit{non-monotonic descent condition} with $c_0=1/2$.
\end{claim}

\begin{IEEEproof}
Adopting the notations from Appendix~\ref{sec:algebraic_relationships_between_asm_iterations}, we denote $v_0:=4/(\max_{\mE}\|A|_{\mE}\|^2)$. The safe averaging rule yields $R^{k+1}|_{(\mE^k)^c}=(1-d^k)\tilde R^{k}$.

Using the decompositions $\|R^{k+1}\|^2=\|\hat R^{k+1}\|^2+(1-d^k)^2\|\tilde R^k\|^2$ and $\|R^{k}\|^2=\|\hat R^{k}\|^2+\|\tilde R^k\|^2$, proving
\begin{equation*}
    \|R^{k+1}\|^2\leq \|R^k\|^2-(d^k/2)[\|R^{k+1}|_{(\mE^k)^c}\|^2+\|\tilde R^k\|^2]
\end{equation*}
is equivalent to showing $\|\hat R^{k+1}\|^2\leq\|\hat R^{k}\|^2+d^k(1+d^k-(d^k)^2/2)\|\tilde R^k\|^2$. Since $d^k-(d^k)^2/2\geq 0$, it suffices to establish
\begin{equation}
    \|\hat R^{k+1}\|^2\leq\|\hat R^{k}\|^2+d^k\|\tilde R^k\|^2. \label{eq:Rk1_Rk_dk}
\end{equation}
From the proof of Proposition~\ref{prop:E_Ec_couple}, we obtain
\begin{equation}
\begin{aligned}
    \|\hat R^{k+1}\|^2&\leq \|\hat R^k\|^2-(d^k/\hat v) J^k\\
    &\leq \|\hat R^k\|^2+(d^k/\hat v) \|v\tilde A\tilde R^k/2\|^2\\
    &\leq \|\hat R^k\|^2+d^k v^2/(\hat v v_0)\|\tilde R^k\|^2.
\end{aligned}
\end{equation}
For $\hat v=v\leq v_0$, this directly yields \eqref{eq:Rk1_Rk_dk}.
\end{IEEEproof}

\section{Proof of Assertions 1) and 2) of Claim~\ref{claim:safe_shrink} in Appendix \ref{sec:convergence_of_asm_with_shrinkage_checks}}\label{SM:proof_of_claim_ref_claim_safe_shrink}

\setcounter{claim}{\numexpr\getrefnumber{claim:safe_shrink}-1\relax}

\begin{claim}
    Suppose that (i) $\epsilon \leq \omega_0$, (ii) the \textit{non-singular condition} holds, and (iii) $\{\mE^k\}$ and $\{d^k\}$ follow the \textit{safeguarding rule} with \textit{shrinkage checks}. Then: 
    \begin{enumerate}
        \item If $x_{\text{ave}}^k \to x^*$, there exists an integer $k_1$ such that $\mE^k = \bar{E}^k = \mE^k_0 = E^k = \mE^*$ for all $k \geq k_1$.
        \item If the \textit{shrinkage checks} are triggered only finitely many times, then $x_{\text{ave}}^k \to x^*$.
        % \item The \textit{shrinkage checks} cannot be triggered infinitely many times.
    \end{enumerate}
\end{claim}
\subsection{Proof of Assertion 1) in Claim~\ref{claim:safe_shrink}}

Since $x^k_{\text{ave}} \to x^*$, it follows from the continuity of $\mu(\cdot)$ and $\epsilon \leq \omega_0$ that there exists $k_0$ such that for all $k \geq k_0$, $\|(\mu^k-\mu^*)|_{(\mE^*)^c}\|_\infty \leq v\lambda(1-\epsilon/2)$ and $(p^{k}-p^*)|_{\mE^*}=0$. This implies $\mE^k_0 = E^k = \mE^*$ and that $\{i : |(x^k_{\text{ave}})_i| \neq 0\}$ is non-increasing by the \textit{safeguarding rule}. Consequently, $\bar{E}^{k+1} \subseteq \bar{E}^k$ for $k \geq k_0$. We assert that there exists $k_1 \geq k_0$ such that either the \textit{safeguarding rule} directly shrinks $\bar{E}^{k_1}$ to $E^{k_1}$, or the \textit{shrinkage check} is triggered to achieve the same reduction.

Suppose otherwise. By the monotonic shrinkage of $\bar{E}^k$, there must exist $k_1$ and a constant subset $\bar{E}$ such that $\bar{E}^k = \bar{E} \supsetneq \mE^*$ for all $k \geq k_1$. We now show that the conditions of the \textit{shrinkage check} will inevitably be satisfied. Let $L$ be the Lipschitz constant of the mapping $\mu(\cdot)$. From the proof of Theorem~\ref{thm:Local_geometric_convergence} and $E^k=\mE^*$, we obtain
\begin{equation*}
\begin{aligned}
    &\|\mu(x^{k+1})-\mu(x^*)\|_2 \leq L\|x^{k+1}-x^*\|\\
    =& L\|(x^{k+1}-x^*)|_{E^k}\| \leq L(1+\hat{v}\tau_{\min})^{-1}\|x^k_{\text{ave}}-x^*\|_{\mE^*},
\end{aligned}
\end{equation*}
where $\tau_{\min{}}:=\mathrm{eig}_{\min{}}(\hat A^T\hat A)$ with $\hat{A} := A|_{\mE^*}$.

Consequently, there exists a sufficiently large $k_2 \geq k_1$ such that for any $\hat{v} \geq v$, conditions (ii) and (iii) of the \textit{shrinkage check} hold for all $k \geq k_2$. Regarding condition (i), note that for $k \geq k_2$, conditions (ii) and (iii), together with \eqref{eq:R_mu_p} and \eqref{eq:Delta_mu_Rk}, ensure that if we select $\mE^{k_3}=E^{k_3}$ for some $k_3 \geq k_2$, we have
\begin{equation*}
\begin{aligned}
    &\|R(x^{k_3+1})\| = \|R(x^{k_3+1})|_{\mE^{k_3}}\|= \|[\mu(x^{k_3+1})-\mu^{k_3}]|_{\mE^{k_3}}\|\\
    &\leq \|(I+\hat{v}\hat{A}^T\hat{A})^{-1}(I-v\hat{A}^T\hat{A})\|_2 \cdot \|R^{k_3}|_{\mE^{k_3}}\|\\
    &\quad + \|v(I+\hat{v}\hat{A}^T\hat{A})^{-1}\hat{A}^TA|_{(\mE^{k_3})^c}\| \cdot \|R^{k_3}|_{(\mE^{k_3})^c}\|.
\end{aligned}
\end{equation*}
Because $|\mE^{k_3}| = |\mE^*| \leq M$ and $\hat{A}^T\hat{A}$ is strictly positive definite by the \textit{generic conditions}, for any given $r_1 \in (0,1)$, there exists $\tau_0$ such that for $\hat{v} \geq \tau_0 v$:
\begin{equation*}
\begin{aligned}
    &\|(I+\hat{v}\hat{A}^T\hat{A})^{-1}(I-v\hat{A}^T\hat{A})\|_2 \leq (1-r_1)/2,\\
    &\|v(I+\hat{v}\hat{A}^T\hat{A})^{-1}\hat{A}^TA|_{(\mE^{k_3})^c}\|_2 \leq (1-r_1)/2.
\end{aligned}
\end{equation*}
This implies $\|R(x^{k_3+1})\| \leq (1-r_1)\|R^{k_3}\|$. Hence, condition (i) is guaranteed to be triggered.

\subsection{Proof of Assertion 2) in Claim~\ref{claim:safe_shrink}}

Since the \textit{shrinkage checks} and the operation $d^k=1$ are triggered only finitely many times, there exists a $k_1$ such that for all $k \geq k_1$, the algorithm strictly follows the \textit{safeguarding rule} and the $J_1^k\geq r\|R|_{\mE_1^k}\|^2$ case is triggered with $d^k\leq 1/2$.

For $J_1^k\geq r\|R|_{\mE_1^k}\|^2$: Following reasoning analogous to the \textit{safe averaging rule}, there exists a $d_0>0$ such that for any $d^k\leq d_0$, the corresponding $R^{k+1}$ satisfies $R^{k+1}|_{(\mE^k)^c} = (1-d^k_1)R^{k}|_{(\mE^k)^c}$ and, given $J_1^k \geq 0$ and $\hat{v}=v$, $\|R^{k+1}|_{\mE^k}\| \leq \|R^{k}|_{\mE^k}\|$ by Proposition~\ref{prop:E_Ec_couple}. Consequently, the backtracking procedure for determining $d^k$ is guaranteed to terminate with $d_0/2\leq d^k\leq 1/2$ and $\mathrm{card}(\{d^k\}) < \infty$.

For $J_1^k < 0$, since $\bar{E}^k \subseteq \mE^k$, we have $R^k|_{(\mE^k)^c}=0$, so that the conditions in Theorem~\ref{thm:convergence_concern12} are naturally satisfied. Combining both cases, we conclude that $x^k_{\text{ave}} \to x^*$ by Theorem~\ref{thm:convergence_concern12}.

For $J_1^k < 0$, since $\bar{E}^k \subseteq \mE^k$, we have $R^k|_{(\mE^k)^c}=0$, so that the conditions in Theorem~\ref{thm:convergence_concern12} are naturally satisfied. Combining both cases, we conclude that $x^k_{\text{ave}} \to x^*$ by Theorem~\ref{thm:convergence_concern12}.

\section{Proof of Lemma~\ref{lema:residual_zero_bounded}}\label{SM:residual_zero_bounded}

\setcounter{lemma}{\numexpr\getrefnumber{lema:residual_zero_bounded}-1\relax}
\begin{lemma}
    Consider the LASSO objective $F$ defined on $\mathbb{R}^N$. Suppose $r_1 \in (0,1)$ and a sequence $\{a_k\}$ satisfies $a_k \geq 0$ with $\sum a_k < \infty$. Then, for any sequence $\{u^k\}$ in $\mathbb{R}^N$:
    \begin{enumerate}
        \item If $\|R(u^{k+1})\|^2 \leq (1+a_k)\|R(u^k)\|^2$ for all $k$, and there exists a subsequence $\{u^{k_l}\}_{l=1}^\infty$ satisfying $\|R(u^{k_l+1})\| \leq (1-r_1)\|R(u^{k_l})\|$ for all $l$, then $R(u^k) \to 0$.
        \item If $R(u^k) \to 0$, then the sequence $\{u^k\}$ is bounded.
    \end{enumerate}
\end{lemma}

\subsection{Proof of Assertion 1) in Lemma \ref{lema:residual_zero_bounded}}

As $\sum a_k < \infty$, there exists an integer $l_0 \geq 0$ such that $(1-r_1)\prod_{j=k_{l_0}}^\infty (1+a_j)^{\frac12} \leq 1 - r_1/2$. Letting $R^k := R(u^k)$, for any $l \geq l_0$ and $k \in \{k_l+1, \ldots, k_{l+1}\}$, we have
\begin{equation*}
    \|R^{k}\| \leq (1-r_1)\|R^{k_{l}}\|\prod_{j=k_l}^{k-1} (1+a_j)^{1/2} \leq \left(1-\frac{r_1}{2}\right)\|R^{k_l}\|.
\end{equation*}
This yields the subsequence contraction $\|R^{k_{l+1}}\| \leq (1-r_1/2)\|R^{k_l}\|$ and the intermediate bound $\|R^k\| \leq \|R^{k_l}\|$. Consequently, $\|R^{k_l}\| \to 0$, which ensures that the entire sequence $\|R^k\| \to 0$ as $k \to \infty$.

\subsection{Proof of Assertion 2) in Lemma \ref{lema:residual_zero_bounded}}

We proceed by contradiction. Suppose that $\{u^k\}$ is unbounded. Let $u^{k+} = u^k + R(u^k)$. Since $R(u^k) \to 0$, the sequence $\{u^{k+}\}$ must also be unbounded. Thus, there exists a subsequence such that $\|u^{k+}\|_1 \to \infty$. By the optimality condition of the proximal operator, we have $-v^{-1}R(u^k) \in \nabla h(u^k) + \partial g(u^{k+})$. Thus, a subgradient $\xi^k \in \partial F(u^{k+})$ can be explicitly constructed as $\xi^k := -v^{-1}R(u^k) + \nabla h(u^{k+}) - \nabla h(u^k)$, where $F(u) = \frac{1}{2}\|Au-y\|_2^2 + \lambda\|u\|_1$. Since $\nabla h$ is $L$-Lipschitz continuous, we can bound the norm of $\xi^k$ by $\|\xi^k\|_2 \leq (v^{-1}+L)\|R(u^k)\|_2$. Given $\lim_{k \to \infty} \|R(u^k)\|_2 = 0$, it immediately follows that $\|\xi^k\|_2 \to 0$. By the convexity of $F$, the subgradient inequality yields
$$
\begin{aligned}
    F(u^{k+}) &\leq F(u^*) + \langle \xi^k, u^{k+} - u^* \rangle\\
    &\leq F(u^*) + \|\xi^k\|_2 \|u^{k+} - u^*\|_2.
\end{aligned}
$$
Because $F(x) \geq \lambda\|x\|_1$, we obtain $\lambda\|u^{k+}\|_1 \leq F(u^*) + \|\xi^k\|_2 \|u^{k+} - u^*\|_2$. Note that $\|x\|_2 \leq \|x\|_1$ for any $x$. Thus, we can further bound the right-hand side: $\lambda\|u^{k+}\|_1 \leq F(u^*) + \|\xi^k\|_2 (\|u^{k+}\|_1 + \|u^*\|_1)$. Dividing both sides by $\|u^{k+}\|_1$ yields
$$
\lambda \leq \frac{F(u^*) + \|\xi^k\|_2 \|u^*\|_1}{\|u^{k+}\|_1} + \|\xi^k\|_2.
$$
Taking the limit strictly along the aforementioned subsequence where $\|u^{k+}\|_1 \to \infty$, and noting that $\|\xi^k\|_2 \to 0$, the right-hand side converges to $0$, which implies $\lambda \leq 0$. This fundamentally contradicts the premise that $\lambda > 0$. Therefore, the sequence $\{u^k\}$ must be bounded.

\section{The Low-Rank Updates of ASM-L1-LR}\label{supple:Low-Rank_Updates}

To efficiently execute the RLS step involving $(I+A|_{\mE^k}^TA|_{\mE^k})^{-1}$ (setting $v=1$ w.l.o.g.) when $\hat{v}=v$, we adaptively evaluate its action without strictly forming the large explicit inverse. Let $\Delta \mE^k := \Delta\mE^k_+ \cup \Delta\mE^k_-$ denote the support changes, where $\Delta\mE^k_+ := \mE^{k+1} \setminus \mE^{k}$ and $\Delta\mE^k_- := \mE^{k} \setminus \mE^{k+1}$. Given a threshold $\alpha \in (0,1)$ and a capacity $C \in \mathbb{Z}_+$, we adopt three scenarios: (i) When $|\mE^{k}| \leq (1+\alpha)M$, we directly compute and store $(I+A|_{\mE^k}^TA|_{\mE^k})^{-1}$ for maximal numerical precision, leveraging the inherent sparsity and stability of the support in this regime. (ii) When $|\mE^{k}| > (1+\alpha)M$ and $|\Delta \mE^k| > C$, we avoid forming the large primal inverse. Instead, we explicitly compute the $M \times M$ dual inverse $G^k := (I + A|_{\mE^k}(A|_{\mE^k})^T)^{-1}$ from scratch and apply the Sherman-Morrison-Woodbury (SMW) identity purely via sequential matrix-vector multiplications. (iii) When $|\mE^{k}| > (1+\alpha)M$ but $|\Delta \mE^k| \leq C$, we recursively update $G^{k+1}$ from $G^k$ via a low-rank SMW strategy. Specifically, letting $U^k := [A|_{\Delta\mE^k_+}, A|_{\Delta\mE^k_-}]$ and $D := \mathrm{diag}(I_{|\Delta\mE^k_+|}, -I_{|\Delta\mE^k_-|})$, we have $(G^{k+1})^{-1} = (G^{k})^{-1} + U^k D (U^k)^T$, we obtain:
\begin{equation}
\begin{aligned}
    G^{k+1} &= G^k - G^k U^k (D + (U^k)^T G^k U^k)^{-1} (U^k)^T G^k \\
    &= G^k - (W^k)^T H^k W^k,\label{eq:low_rank}
\end{aligned}
\end{equation}
where we denote $W^k := (U^k)^T G^k$ and $H^k := (D + (U^k)^T G^k U^k)^{-1}$. Since the inner matrix $(D + (U^k)^T G^k U^k)$ is at most of order $C \times C$, its inversion is computationally trivial, drastically reducing the overall complexity.

This perfectly complements Sec.~\ref{sec:Adap_sub_with_LR}, where Scenario (iii) dominates the Initial Stage. Although $W^k$ is low-rank, explicitly forming the dense $M \times M$ update $(W^k)^T H^k W^k$ at each step introduces substantial overhead. To circumvent this, we accumulate $(W^t)^T$ and $H^t W^t$ ($k_0\le t \le k$) to form block matrices $P^k_{k_0}$ and $Q^k_{k_0}$, respectively. This allows evaluating $G^k = G^{k_0} - P^k_{k_0} Q^k_{k_0}$ purely via fast matrix-vector products. We periodically recompute $G^k$ to clear the buffers and mitigate numerical round-off errors.

During the Safeguarding Stage, evaluating $J^k$ can similarly exploit low-rank updates for efficiency. As indicated by Eq.~\eqref{eq:J_k}, this evaluation requires tentatively removing $\Delta \mE^k$ to compute the corresponding $G^k$, which constitutes the primary computational overhead. Given that $|\Delta \mE^k|$ is extremely small, we again leverage Eq.~\eqref{eq:low_rank} to extract $W^k$ and $H^k$. The update is then evaluated strictly as a linear operator via sequential matrix-vector products, completely bypassing any explicit matrix inversion.

\vfill

\end{document}